\documentclass[11pt]{article}
\usepackage[T1]{fontenc}
\usepackage[a4paper,margin=28mm]{geometry}
\usepackage{amsmath,amssymb,amsthm}
\usepackage{tikz}
\usetikzlibrary{arrows.meta,positioning}
\usetikzlibrary{calc,decorations.pathreplacing,fit,backgrounds}
\definecolor{figA}{HTML}{C8553D}
\definecolor{figB}{HTML}{2A6F97}
\definecolor{figC}{HTML}{3F8F5F}
\definecolor{figgray}{HTML}{8A8A8A}
\tikzset{
  fv/.style={circle,draw,fill=white,inner sep=0pt,minimum size=4.5pt},
  fvb/.style={fv,fill=black},
  fvg/.style={fv,fill=figgray!45},
  fterm/.style={rectangle,draw,fill=white,inner sep=0pt,minimum size=4.5pt},
  flab/.style={font=\footnotesize},
  fbox/.style={draw=hypothesisink!45,fill=hypothesisfill,rounded corners=2pt},
  fnode/.style={fbox,font=\footnotesize,inner sep=4pt,align=center},
  farc/.style={-{Stealth[length=4.5pt]}},
  fimp/.style={-{Implies[]},double,double distance=1.7pt},
  fiff/.style={{Implies[]}-{Implies[]},double,double distance=1.7pt},
  hiA/.style={draw=figA,line width=1.2pt},
  hiB/.style={draw=figB,line width=1.2pt},
  hiC/.style={draw=figC,line width=1.2pt},
  fbrace/.style={decorate,decoration={brace,amplitude=4pt}},
}
\usepackage{array,booktabs}
\usepackage[hidelinks]{hyperref}
\usepackage{cleveref}
\usepackage{todonotes}

\definecolor{hypothesisink}{HTML}{24465A}
\definecolor{hypothesisfill}{HTML}{F2F7FA}
\newcommand{\hypothesisbox}[2]{%
  \par\medskip\noindent
  \begingroup
  \setlength{\fboxsep}{8pt}%
  \setlength{\fboxrule}{0.45pt}%
  \fcolorbox{hypothesisink!45}{hypothesisfill}{%
    \begin{minipage}{\dimexpr\linewidth-2\fboxsep-2\fboxrule\relax}
      {\color{hypothesisink}\bfseries\boldmath #1}\par\smallskip
      #2
    \end{minipage}}%
  \endgroup\par\medskip}

\newcommand{\problembox}[4]{%
  \hypothesisbox{#1}{\textbf{Input.} #2\par
    \textbf{Question.} #3\par
    \textbf{Parameter.} #4}}
\DeclareFontShape{T1}{cmr}{m}{scit}{<->ssub*cmr/m/sc}{}
\newtheorem{theorem}{Theorem}
\newtheorem{lemma}{Lemma}
\newtheorem{proposition}{Proposition}

\DeclareRobustCommand{\problem}[1]{\texorpdfstring{\textsc{#1}}{#1}}
\title{Running-Time Lower Bounds under \texorpdfstring{\(\mathrm{FPT}\ne\mathrm{W}[1]\)}{FPT distinct from W[1]}}
\author{Hendrik Molter\thanks{Supported by the Deutsche Forschungsgemeinschaft (DFG, German Research Foundation) -- 565381415.}\\Hasso Plattner Institute\\\texttt{hendrik.molter@hpi.de}}
\date{}
\begin{document}
\maketitle

\begin{abstract}
We revisit running-time lower bounds historically based on the
Exponential Time Hypothesis. We survey bounds that can be recovered
from earlier work under \(\mathrm{FPT}\ne\mathrm{W}[1]\) and prove
seven lower bounds for graph problems and set systems under the same
assumption. The new proofs encode clique choices in few bits and
control the target parameter in terms of the encoding. This lets us
turn a sufficiently fast algorithm for the target problem into a
fixed-parameter algorithm for \problem{Clique} or
\problem{Multicolored Clique}. In several cases, the new bound matches
the function of the parameter in the historical exponent, with the
additional requirement that the cutoff for any prescribed saving in
the exponent be computable.
\end{abstract}

\section{Introduction}

Finding fast algorithms is one of the central goals of computer
science. For a graph problem, an improvement can come from a better
dependence on the number of vertices or on a parameter such as the
size of the solution. In parameterized algorithms, the focus lies on the latter. We illustrate this with the two graph problems \problem{Vertex Cover} and \problem{Clique}.
For \problem{Vertex Cover} with budget \(k\), branching on the two endpoints
of an uncovered edge gives a \(2^k n^{O(1)}\)-time algorithm on an
\(n\)-vertex graph. The natural target is a subexponential dependence on \(k\).
For \problem{Clique}, enumerating sets of \(k\) vertices takes \(n^{O(k)}\)
time. Here the corresponding target is an \(o(k)\) exponent of~\(n\),
allowing a separate factor that depends only on \(k\).

The Exponential Time Hypothesis (ETH), proposed by Impagliazzo and
Paturi in 1999~\cite{ip1999}, ruled out both improvements conditionally. In each case, a reduction shows
that an algorithm with the proposed improvement would also solve
\problem{3-SAT} in subexponential time~\cite{chkx}, \cite[Theorems~4.1 and~5.1]{lms-eth}.
On October 26, 2026, OpenAI released a preprint reporting a
deterministic subexponential algorithm for \problem{3-SAT}~\cite{ethfalse}.
Anthropic independently reported such an algorithm later that
month~\cite{anthropic}. Both algorithms count satisfying assignments,
which also decides whether an assignment exists. If the reported
bounds hold, each algorithm refutes ETH, including the two uniform
versions discussed below. This is the motivation for revisiting
earlier lower bounds. In this work, we aim to explore the following question: which exclusions can we establish using
\(\mathrm{FPT}\ne\mathrm{W}[1]\) instead?

\subsection{From ETH to other assumptions}

Let \(n\) denote the number of variables in a 3-CNF formula and \(L\)
its encoding length. In its original form, ETH asserts that the infimum
of the constants \(c>0\) for which \problem{3-SAT} has a deterministic
\(2^{cn}L^{O(1)}\)-time algorithm is positive~\cite{ipz}.

\hypothesisbox{Exponential Time Hypothesis (ETH)}{There exists a
constant \(c>0\) such that \problem{3-SAT} has no deterministic algorithm running
in time \(2^{cn}L^{O(1)}\).}

The constant \(c\) is the same for every algorithm. To refute this
formulation, it suffices to show that for each \(c>0\) there exists a
\problem{3-SAT} algorithm running in time \(2^{cn}L^{O(1)}\). For example, a
lower-bound proof can fix \(c\) first and then tailor its reduction
to that choice. If a fast algorithm for the target problem yields
the required \problem{3-SAT} algorithm for every \(c\), the contradiction
follows. This argument need not give a procedure that takes \(c\)
as input and produces the corresponding algorithm.

We can also ask for \emph{one} algorithm that achieves a sublinear exponent
as the input grows. Chen, Eickmeyer, and Flum~\cite{cef} distinguish
this single-algorithm formulation, called \emph{uniform} ETH, from the
original formulation, which they call nonuniform ETH.

\hypothesisbox{Uniform ETH}{There is no deterministic \problem{3-SAT} algorithm
running in time \(2^{o(n)}L^{O(1)}\).}

A single subexponential algorithm also runs in time \(2^{cn}L^{O(1)}\)
for every fixed \(c>0\), so the original ETH implies uniform ETH.
In the opposite direction, the existence of an algorithm for each
fixed \(c\) does not tell us how to assemble them into one
subexponential algorithm.

The Satisfiability Hypothesis predates ETH. In 1990, Stearns and Hunt
asked how small a power of the input length can appear in the exponent
of a satisfiability algorithm~\cite{stearns-hunt}. Their formulation measures
the bit length of an unrestricted CNF formula.

\hypothesisbox{Satisfiability Hypothesis (SH)}{For every fixed
\(0<\varepsilon<1\), there is no deterministic
\(O(2^{L^{1-\varepsilon}})\)-time algorithm for \problem{CNF-SAT}
on inputs of bit length \(L\)~\cite{sh}.}

The MIT Hardness Group shows that SH is equivalent to Weak ETH\footnote{The name “Weak ETH” has also been used for other
hypotheses~\cite{chikk,lok-weth}. The formulation used here is also
known as “Almost-ETH”~\cite[2026 revision]{man-almosteth}.},
which rules out \(2^{O(n^{1-\varepsilon})}L^{O(1)}\) time for
\problem{3-SAT} for every fixed \(0<\varepsilon<1\), where \(n\) is
the number of variables~\cite[Theorem~2.4]{sh}.
Historical ETH lower bounds used sparsification to reduce to 3-CNF
formulas with linearly many clauses~\cite{ipz}. The proof of the
equivalence between SH and Weak ETH uses a quantitative version of sparsification
to preserve a fixed power saving~\cite[Lemma~2.2 and Theorem~2.4]{sh}.
Lower bounds on the running time of intermediate problems such
as \problem{Clique} could then be carried over to other problems by
reductions~\cite{chkx,lms-eth,lms}.
In particular, for reductions from
\problem{3-SAT}, one can equivalently use the number \(s=n+m\) of
variables plus clauses in place of the CNF input length, where \(m\)
is the number of clauses~\cite[Lemma~2.1]{sh}. In this formulation,
SH rules out an exponent such as \(s^{0.99}\) but allows \(s/\log s\).
The reported subexponential SAT algorithms~\cite{anthropic,ethfalse} do not save a fixed power
of \(s\). Hence, SH remains a possible premise. However, some historical bounds
distinguished between exponents differing by only a logarithmic
factor. To retain those distinctions, we can start from a problem
for which the reported algorithms leave the historical bound untouched.

One such problem is \(k\times k\) \problem{Clique}. The vertices
are arranged in a \(k\times k\) grid, and we ask whether there is a
clique with one vertex from each row. Enumerating the \(k^k\) ways to
choose one vertex per row gives a \(2^{O(k\log k)}\)-time algorithm. ETH ruled out \(2^{o(k\log k)}\) time through the reduction
of Lokshtanov, Marx, and Saurabh~\cite[Proposition~2.1 and
Theorem~2.4]{lms}. The same reduction under SH gives a fixed-power
exclusion of \(2^{O(k^{1-\varepsilon})}\) for every fixed
\(0<\varepsilon<1\)~\cite[Theorem~3.1]{sh}. In particular, it leaves a single-exponential improvement open.

One way to use the reported \problem{3-SAT} algorithms for \(k\times k\)
\problem{Clique} is to encode the instance as a SAT formula. The direct
encodings either use clauses of growing length or introduce too many
variables for the reported bounds to yield a \(2^{o(k\log k)}\)-time
algorithm. Szeider also discusses applying the tensor-network method
directly~\cite[Section~4, following Proposition~3]{square}. Conditional on OpenAI's Theorem~D.1,
this gives a \(2^{o(k\log k)}\)-time algorithm when only
\(o(k\sqrt{\log k})\) pairs of rows contain a nonedge. The reported
algorithms are not known to give such a running time for general
instances. Szeider proposes retaining the historical exclusion as a
hypothesis in its own right~\cite{square}.

\hypothesisbox{\(k\times k\) \problem{Clique} Hypothesis}{There is no
deterministic \(2^{o(k\log k)}\)-time algorithm for deciding whether
a graph on a \(k\times k\) grid has a clique with one vertex from each row.}

Szeider shows that this hypothesis recovers several historical ETH
bounds, including the \(d\log d\) exponent for \problem{Line Distortion}
by distortion \(d\) and the \(p\log p\) exponents for
\problem{Cycle Packing} and \problem{Vertex-Disjoint Paths} by
width of the supplied path decomposition \(p\). It also yields an exclusion
of \(F(k)n^{o(k)}\) time for \problem{Clique} by requested
size \(k\)~\cite[Table~1 and Proposition~1]{square}.

These examples show how a hypothesis about one source problem can
recover bounds for several others. Our starting point is
the standard intractability assumption of
parameterized complexity. The hypothesis that \(\mathrm{FPT}\ne\mathrm{W}[1]\)
plays a fundamental role in parameterized complexity theory, analogous to \(\mathrm{P}\ne\mathrm{NP}\): it rules out
fixed-parameter tractability for \(\mathrm{W}[1]\)-hard problems.
We use it to ask which running-time improvements can be excluded
without assuming a specific exponent for the source problem.

\hypothesisbox{\(\mathrm{FPT}\ne\mathrm{W}[1]\)}{There is no deterministic
algorithm for \problem{Clique} on \(n\) vertices, parameterized by the requested
clique size \(k\), with running time \(F(k)n^c\) for a computable
function \(F\) and a constant \(c\)~\cite{df}.}

The connection with ETH has a long history. ETH was known to imply
\(\mathrm{FPT}\ne\mathrm{W}[1]\), and researchers asked whether the
converse might hold. The miniaturization class \(\mathrm{M}[1]\)
captures one part of that connection~\cite{fgmini,fg}. 
Miniaturization asks whether a SAT instance becomes tractable when
its number of variables is small relative to the logarithm of the
available input length, which can be increased by padding.
Chen and Grohe~\cite{chen-grohe-mini} showed that miniaturization
preserves reductions and gives a correspondence between
subexponential-time and parameterized complexity.
The miniaturized problem is complete for \(\mathrm{M}[1]\)
under fixed-parameter Turing reductions. Strongly uniform ETH\footnote{This
version rules out a single subexponential \problem{3-SAT} algorithm
with computable cutoffs for its exponent. We define it precisely in
Section~\ref{sec:transfer}.} is
equivalent to \(\mathrm{FPT}\ne\mathrm{M}[1]\), and
\(\mathrm{M}[1]\subseteq\mathrm{W}[1]\)~\cite{cef,fg}. The
equivalence for \(\mathrm{M}[1]\) therefore gave one direction of
the proposed connection with \(\mathrm{W}[1]\).

Using a parameterized complexity class separation to exclude faster algorithms for
problems in FPT is an established idea. In their 2003 journal article,
Cai and Juedes~\cite{cj} gave, for each level \(\mathrm{W}[t]\), a
problem in FPT for which a subexponential dependence on the parameter
would imply \(\mathrm{FPT}=\mathrm{W}[t]\). Pilipczuk and
Wrochna~\cite{pw-space} used the same reparameterization idea to study
algorithms on graph decompositions.
We summarize specific lower bounds that follow from earlier work in
\Cref{sec:contribution} and give their derivations in
\Cref{sec:prior-bound-derivations}.

\subsection{How the running-time transfer works}\label{sec:transfer}

A parameterized reduction can transfer a running-time bound as well
as a hardness classification. A standard fixed-parameter reduction
bounds the target parameter by a computable function of the source
parameter alone. To exclude a specific running time, we can allow
the target parameter to grow with the source input, provided that
the hypothetical solver is still fast enough on the instances we
produce.

For example, suppose that in time \(F(k)n^{O(1)}\) we can reduce
\problem{Clique} on \(n\) vertices with requested size \(k\) to
a target instance with parameter \(p\le g(k)\log_2 n\), for a
computable positive integer function \(g\). A \(2^{cp}\)-time solver
with a fixed polynomial input factor would contribute at most
\(n^{c g(k)}\) to the running time. This fails to give an FPT
bound when the exponent of \(n\) grows with \(k\).

Now suppose the solver takes \(2^{h(p)}\) times a fixed polynomial
and there is a computable function \(T\) such that \(h(p)\le p/j\)
whenever \(j\ge1\) and \(p\ge T(j)\). Choose \(j=g(k)\).
For a target instance with \(p\ge T(j)\), we have
\[
 h(p)\le \frac{p}{j}
 \le \frac{g(k)\log_2 n}{g(k)}
 = \log_2 n.
\]
Thus the solver's exponential factor is at most \(n\).

\begin{figure}[t]
\centering
\begin{tikzpicture}[x=1cm,y=1cm]
  \fill[hypothesisfill] (0,0) rectangle (4,4.3);
  \fill[figC!10] (4,0) rectangle (10,4.3);
  \draw[farc] (0,0) -- (10.9,0) node[flab,right] {\(p\)};
  \draw[farc] (0,0) -- (0,4.7) node[flab,above] {exponent};
  \draw[thick] (0,0) -- (10,3.2) node[flab,right] {\(p/j\)};
  \draw[hiA] plot[smooth] coordinates {(0,1.1) (0.6,2.4) (1.3,1.7) (2.1,2.9) (2.9,1.9) (3.5,1.6) (4,1.28) (5,1.33) (6.5,1.5) (8,1.62) (10,1.75)}
      node[flab,right,text=figA] {\(h(p)\)};
  \draw[dashed] (4,0) -- (4,4.3);
  \node[flab,below] at (4,0) {\(T(j)\)};
  \draw[dashed] (10,0) -- (10,4.3);
  \node[flab,below] at (10,0) {\(g(k)\log_2 n\)};
  \draw[dotted] (0,3.2) -- (10,3.2);
  \node[flab,left] at (0,3.2) {\(\log_2 n\)};
  \node[flab,align=center] at (2,3.75) {\(h(p)\le\max\{c_0,T(g(k))\}\)};
  \node[flab,align=center] at (6.2,3.75) {\(h(p)\le p/j\le\log_2 n\)};
  \node[flab,align=center,text=hypothesisink] at (2,-1.0) {exponential factor of the solver\\depends on \(k\) only};
  \node[flab,align=center,text=figC!70!black] at (7,-1.0) {exponential factor of the solver\\is at most \(n\)};
\end{tikzpicture}
\caption{The two cases of the running-time transfer. Below the cutoff, the solver's exponential factor is bounded by a function of the requested clique size alone. Beyond the cutoff, it is at most the number of vertices in the source graph. The thresholds and the curve are schematic. The interval beyond the cutoff can be empty for small source graphs.}
\label{fig:transfer}
\end{figure}
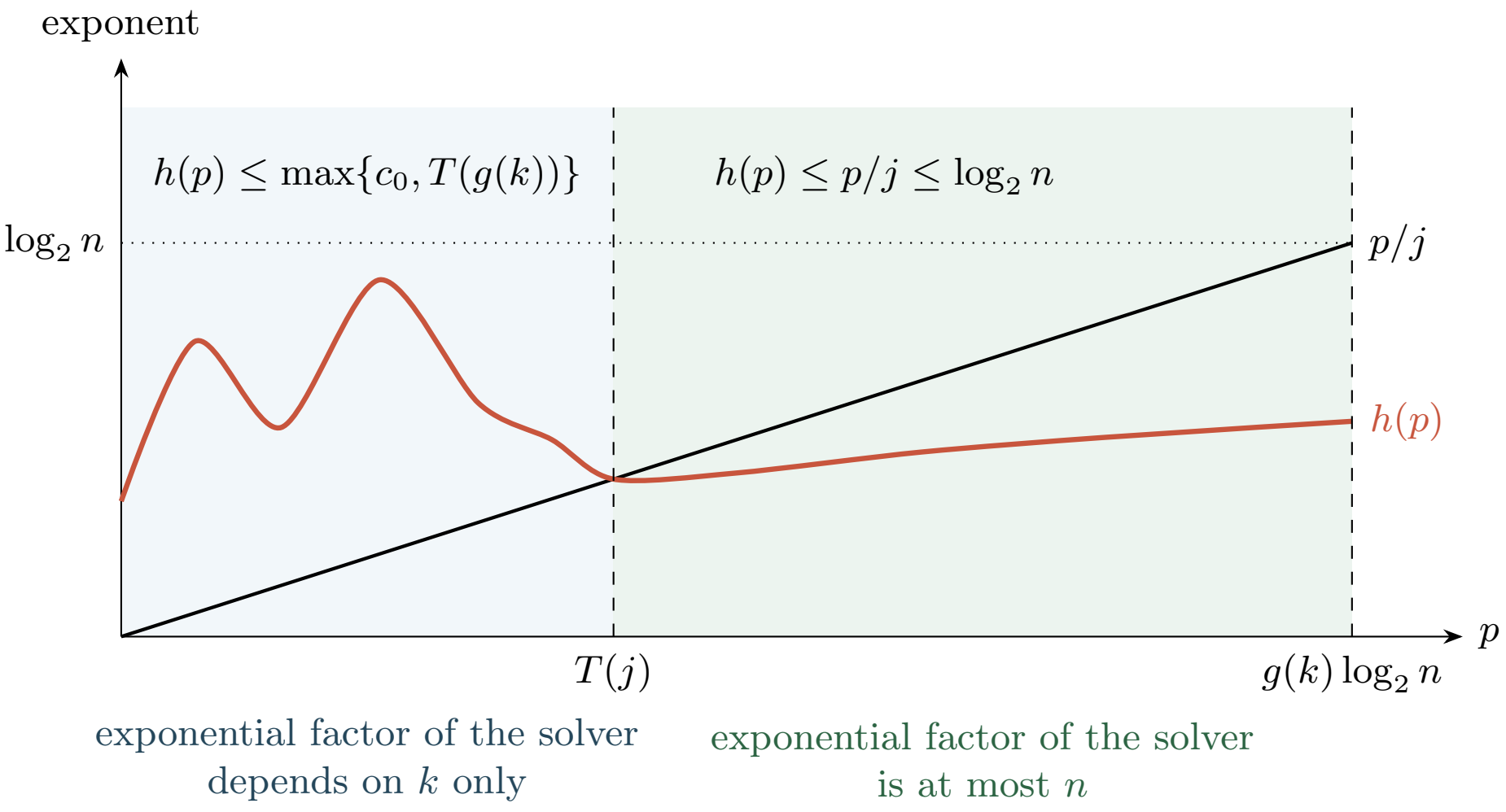

The same choice \(j=g(k)\) also sets the boundary for the
remaining case, \(p<T(g(k))\). Here we bound the solver's exponent
using that cutoff, which depends only on \(k\). Let \(c_0\) be an
integer bounding the finitely many values of \(h(p)\) with
\(p<T(1)\). For \(p\ge T(1)\), the cutoff for \(j=1\) gives
\(h(p)\le p\). Hence for \(p<T(g(k))\),
\[
 h(p)\le
 \begin{cases}
 c_0 & \text{if } p<T(1),\\
 p & \text{if } p\ge T(1).
 \end{cases}
\]
Thus \(h(p)\le\max\{c_0,T(g(k))\}\) below the cutoff. The solver's
exponential factor is at most \(n\) at or above the cutoff and is
bounded by a computable function of \(k\) below it. The polynomial input
factor and the reduction time then give an FPT algorithm for \problem{Clique} parameterized by
the requested size. \Cref{fig:transfer} illustrates the two cases.

With ordinary little-\(o\), the cutoff for the saving \(1/g(k)\)
still exists for each requested clique size \(k\). Our running-time
argument uses that cutoff to bound the solver's time on the target
instances below it. Without a way to compute a cutoff from \(k\),
this argument does not give a computable bound on that contribution
as a function of \(k\), as required for an FPT running time.

The same calculation applies to other exponents. For example, to
exclude an effective \(o(p\log p)\) exponent, the reduction needs a
bound of the form \(p\log p\le g(k)\log_2 n\). We then choose
the cutoff that guarantees \(h(p)\le p\log p/g(k)\).
The calculation also applies when a reduction produces several
target instances and accepts if any answer is yes. The construction time bounds both the number
of queries and the length of each. The parameter bound applies to
each query, so multiplying the maximum time for a solver call by
the number of queries still gives an FPT bound.

The need to compute a cutoff from the requested clique size motivates
a closer look at the third version of ETH. Uniform
ETH rules out a single subexponential \problem{3-SAT} algorithm. Strongly
uniform ETH rules out the more specific kind of algorithm for which
we can compute how large an instance must be before the exponent
is at most a prescribed fraction of \(n\)~\cite{cef}.

\hypothesisbox{Strongly uniform ETH}{There is no deterministic \problem{3-SAT}
algorithm with running time \(2^{h(n)}L^{O(1)}\) for which a computable
function \(T\) guarantees \(h(n)\le n/j\) for every positive integer
\(j\) and every \(n\ge T(j)\).}

Uniform ETH implies strongly uniform ETH. In this paper we call
\(T(j)\) a \emph{computable cutoff}. The corresponding asymptotic
notation is \emph{effective little-\(o\)}~\cite{cef}. We write
\(h=o_{\mathrm{eff}}(\varphi)\) if a computable cutoff \(T(j)\)
ensures \(h(p)\le \varphi(p)/j\) for every positive integer \(j\) and
all \(p\ge T(j)\).

\subsection{How the hypotheses are related}

\begin{figure}[t]
\centering
\begin{tikzpicture}[x=1cm,y=1cm]
  \node[fnode] (eth)  at (0,0)     {ETH};
  \node[fnode] (ueth) at (3.1,0)   {Uniform ETH};
  \node[fnode] (seth) at (7.1,0)   {Strongly uniform ETH};
  \node[fnode] (m1)   at (11.6,0)  {\(\mathrm{FPT}\ne\mathrm{M}[1]\)};
  \begin{scope}[on background layer]
    \node[draw=figA,dashed,rounded corners=3pt,inner xsep=7pt,inner ysep=9pt,
          fit=(eth)(m1)] (ref) {};
  \end{scope}
  \node[flab,text=figA,anchor=south west] at (ref.north west)
       {refuted if the reported bounds hold~\cite{anthropic,ethfalse}};
  \node[fnode] (kk)   at (1.6,-2.5) {\(k\times k\) \problem{Clique}\\Hypothesis};
  \node[fnode] (weth) at (5.6,-2.5) {Weak ETH};
  \node[fnode] (sh)   at (8.6,-2.5) {SH};
  \node[fnode,line width=0.9pt,draw=hypothesisink] (w1) at (11.6,-5) {\(\mathrm{FPT}\ne\mathrm{W}[1]\)};
  \draw[fimp] (eth) -- (ueth);
  \draw[fimp] (ueth) -- (seth);
  \draw[fiff] (seth) -- node[flab,above=1pt]{\cite{cef,fg}} (m1);
  \draw[fiff] (weth) -- node[flab,above=1pt]{\cite{sh}} (sh);
  \draw[fimp] (eth) -- node[flab,left=2pt,pos=.55]{\cite{lms}} (kk);
  \draw[fimp] (seth) -- (weth);
  \draw[fimp] (m1) -- node[flab,right=2pt]{\(\mathrm{M}[1]\subseteq\mathrm{W}[1]\)} (w1);
  \draw[fimp] (kk.south) |- node[flab,above=1pt,pos=.75]{\cite{square}} ([yshift=3pt]w1.west);
  \draw[farc,dashed,figgray] ([yshift=-3pt]w1.west) -| node[flab,below=1pt,pos=.25,text=black]{direct encoding fails}
        ([xshift=-12pt]kk.south);
  \draw[farc,dashed,figgray] (w1.north west) -- node[flab,below left=-1pt,text=black,pos=.55]{direct encoding fails} (sh.south east);
\end{tikzpicture}
\caption{Relations between the hypotheses. A solid arrow with one head denotes an implication, and a solid arrow with two heads denotes an equivalence. A dashed arrow marks a direction that the direct encodings discussed in the text do not establish. Strongly uniform ETH implies Weak ETH, since an exponent in \(O(n^{1-\varepsilon})\) is effectively sublinear.}
\label{fig:hypotheses}
\end{figure}
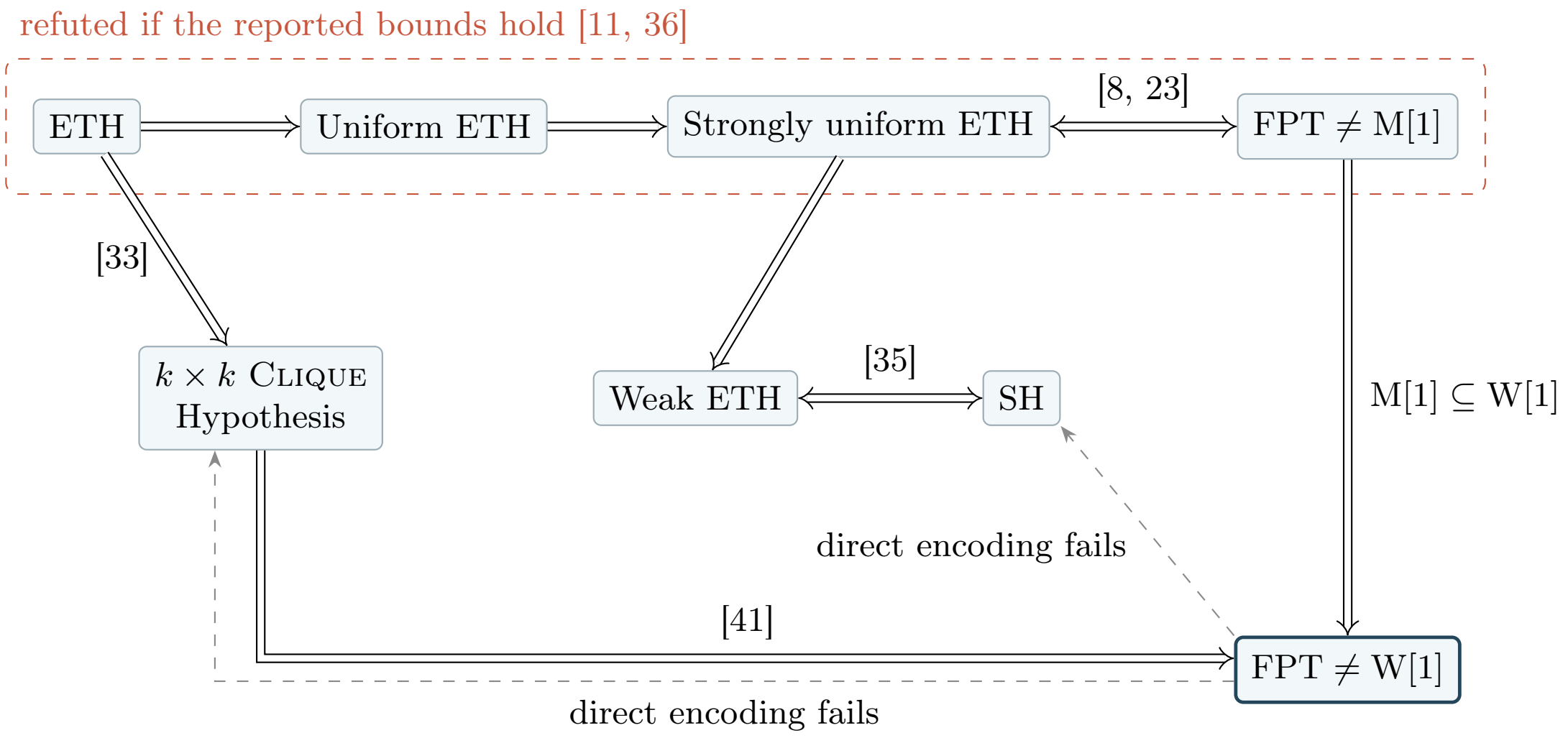

The three hypotheses address different questions. SH concerns fixed
power savings for SAT. The \(k\times k\) \problem{Clique} Hypothesis
preserves a specific exponent useful in reductions from that problem.
The separation \(\mathrm{FPT}\ne\mathrm{W}[1]\) concerns \problem{Clique}, with a requested size independent of the graph order.
There is a known implication. Szeider shows that the \(k\times k\)
\problem{Clique} Hypothesis rules out \(F(k)n^{o(k)}\) time for
\problem{Clique} for any computable \(F\), and thus implies
\(\mathrm{FPT}\ne\mathrm{W}[1]\)~\cite[Proposition~1]{square}.
The converse would require a way to use a fast \(k\times k\)
\problem{Clique} algorithm on \problem{Clique} instances.
Strongly uniform ETH implies Weak ETH. For a fixed power saving
\(h(n)\le c n^{1-\varepsilon}\), a cutoff for \(h(n)\le n/j\)
can be computed by bounding \(c\) from above by an integer and
\(1/\varepsilon\) from above by an integer.
\Cref{fig:hypotheses} summarizes these relations. The dashed arrows refer to
the two reductions discussed below, padding a \problem{Clique}
instance to a square grid and encoding \problem{Clique} as SAT.

For \problem{Clique}, the graph order can grow while the
requested clique size stays fixed. In \(k\times k\) \problem{Clique},
the number of choices per row equals the number of rows. Padding
an ordinary instance to meet this requirement can make the side
length as large as the graph order, so a fast \(k\times k\)
\problem{Clique} algorithm need not give an FPT algorithm for
\problem{Clique} parameterized by the requested size through this encoding.

Likewise, the direct encoding of \problem{Clique} as SAT
uses few variables but can have polynomially many clauses in the
graph order. A fixed power saving in the number of variables plus
clauses can still leave a superpolynomial running time for fixed
clique size. These encodings therefore do not give the reductions
needed to derive either SH or the \(k\times k\) \problem{Clique}
Hypothesis from \(\mathrm{FPT}\ne\mathrm{W}[1]\).

\subsection{Our contribution}\label{sec:contribution}

We study which historical ETH lower bounds can be recovered under
\(\mathrm{FPT}\ne\mathrm{W}[1]\).

\paragraph{Bounds from earlier work.}
Published reductions already recover many of these bounds.
\Cref{tab:prior-benchmarks} compares the resulting lower bounds with
their historical ETH benchmarks. Some
follow directly from hardness for a logarithmically rescaled
parameter. Others use an additional reduction or a more detailed
analysis of a published construction.
\Cref{sec:prior-bound-derivations} gives the derivations.

\begin{table}[t!]
\centering
\small
\setlength{\tabcolsep}{4pt}
\renewcommand{\arraystretch}{1.13}
\begin{tabular}{@{}>{\raggedright\arraybackslash}p{.38\linewidth}>{\raggedright\arraybackslash}p{.28\linewidth}>{\raggedright\arraybackslash}p{.29\linewidth}@{}}
\toprule
Problem and parameter & Historical ETH benchmark & Under \(\mathrm{FPT}\ne\mathrm{W}[1]\) \\
\midrule
\problem{CNF-SAT}, variables \(n\) &
 \(2^{o(n)}\) \newline \cite{ipz} &
 \(2^{o_{\rm eff}(n)}\) \newline \cite[Cor.~3.1]{cj} \\
\problem{Closest String}, radius \(d\) &
 \(2^{o(d\log d)}\) \newline \cite[Thm.~3.1]{lms} &
 \(2^{o_{\rm eff}(d)}\) \newline \cite[Cor.~3.1]{cj} and \cite[Thm.~3.1]{lms} \\
Binary \problem{Distinct Vectors}, column budget \(p\) &
 \(2^{O(2^{\gamma p})}\), \(0<\gamma<1/2\) \newline \cite[Thm.~1]{ps} &
 \(2^{o_{\rm eff}(2^p)}\) \newline \cite[Thm.~4.1]{froese} and \cite[Lemma~14]{lin} \\
\problem{Integer Programming Feasibility}, number of equations \(p\) &
 \(2^{o(p\log p)}\) \newline General right-hand side~\cite[Thm.~1]{kpwconf} &
 \(2^{o_{\rm eff}(p)}\) \newline All-ones right-hand side~\cite[Lemma~14]{lin} \\
\addlinespace[2pt]
\multicolumn{3}{@{}l}{\textit{Treewidth \(t\)}} \\
\problem{Chromatic Number} &
 \(2^{o(t\log t)}\) \newline \cite[Thm.~6.3]{lms-eth} &
 \(2^{o_{\rm eff}(t\log t)}\) \newline (Prop.~\ref{prop:chromatic-treewidth-prior}) \\
\problem{Independent Set} &
 \(2^{o(t)}\) \newline \cite[Thm.~6.2]{lms-eth} &
 \(2^{o_{\rm eff}(t)}\) \newline \cite[Thm.~6]{xnlp}, \cite[Thm.~11]{xalp} \\
\problem{Odd Cycle Transversal} &
 \(2^{o(t)}\) \newline \cite[Thm.~6.2]{lms-eth} &
 \(2^{o_{\rm eff}(t)}\) \newline \cite[Thm.~6]{xnlp} \\
\problem{Dominating Set} &
 \(2^{o(t)}\) \newline \cite[Thm.~6.2]{lms-eth} &
 \(f(b)2^{o_{\rm eff}(t)}\) \newline Connected bipartite graphs
 \newline \cite[Cor.~12]{xalp}, \cite[Lemma~14]{lin} \\
\problem{Vertex Cover} and \problem{Red-Blue Dominating Set} &
 \(2^{o(t)}\) \newline From \cite[Thm.~3.3]{lms-eth} &
 \(2^{o_{\rm eff}(t)}\) \newline \cite[Cor.~12]{xalp} \\
\problem{Feedback Vertex Set} &
 \(2^{o(t)}\) \newline From \cite[Thm.~3.3]{lms-eth} &
 \(2^{o_{\rm eff}(t)}\) \newline \cite[Thm.~6]{xnlp} \\
\(q\)-\problem{Coloring} and \(q\)-\problem{List Coloring}, fixed \(q\ge3\) &
 \(2^{o(t)}\) \newline From \cite[Thm.~3.2]{lms-eth} &
 \(2^{o_{\rm eff}(t)}\) \newline \cite[Lemma~4]{xnlp} \\
\addlinespace[2pt]
\problem{Dominating Set}, maximum prefix cut rank \(p\) of a supplied vertex order &
 \(2^{o(p^2)}\) \newline \cite[Thm.~3.1]{liu-meng-rw} &
 \(2^{o_{\rm eff}(p^2)}\) \newline \cite[Sec.~3.2 and proof of Thm.~C.1]{liu-meng-rw} \\
\bottomrule
\end{tabular}
\caption{Lower bounds derived from earlier work. Each entry gives an
excluded factor in a deterministic running time with a fixed polynomial
factor in the input length. The notation \(o_{\rm eff}\) requires
computable cutoffs. For Dominating Set by treewidth, \(b\) is the
domination budget and \(f\) is an arbitrary computable positive
integer-valued function. The Dominating Set recovery for a supplied
vertex order uses the monopolar construction.}
\label{tab:prior-benchmarks}
\end{table}

\paragraph{New lower bounds.}
We extend these recoveries by proving seven new lower bounds under
the same assumption. \Cref{tab:benchmarks} compares these bounds with
their historical ETH benchmarks. In six cases, the function inside
little-\(o\) matches the historical benchmark, with computable cutoffs.
For \problem{Chromatic Number} by vertex cover number, the two
functions differ by a logarithmic factor. For \problem{Permutation
Hitting}, the historical bound already holds when each input set has
at most one cell per row. Our construction uses sets that can contain
several cells in the same row. Szeider~\cite[Table~1]{square} recovers the
historical bounds for \problem{Line Distortion}, \problem{Cycle Packing},
and \problem{Vertex-Disjoint Paths} under the \(k\times k\)
\problem{Clique} Hypothesis. For these three problems, our reductions
start from \problem{Clique} and obtain effective versions of the
bounds under \(\mathrm{FPT}\ne\mathrm{W}[1]\).

\begin{table}[t!]
\centering
\small
\setlength{\tabcolsep}{4pt}
\renewcommand{\arraystretch}{1.13}
\begin{tabular}{@{}>{\raggedright\arraybackslash}p{.38\linewidth}>{\raggedright\arraybackslash}p{.28\linewidth}>{\raggedright\arraybackslash}p{.29\linewidth}@{}}
\toprule
Problem and parameter & Historical ETH benchmark & Under \(\mathrm{FPT}\ne\mathrm{W}[1]\) \\
\midrule
\problem{Chromatic Number}, vertex cover number \(p\) &
 \(2^{o(p\log p)}\) \newline \cite[Thm.~6.1]{lms} &
 \(2^{O(p)}\) \newline (Thm.~\ref{cover-w1}) \\
\problem{Independent Set}, linear rank-width \(p\) &
 \(2^{o(p^2)}\) \newline \cite[Thm.~1.1]{bkn} &
 \(2^{o_{\rm eff}(p^2)}\) \newline (Thm.~\ref{independent-set-rankwidth-w1}) \\
\problem{Permutation Hitting}, grid order \(D\) &
 \(2^{o(D\log D)}\), at most one cell per row in each set
 \newline \cite[Thm.~C.2]{ccpre} &
 \(2^{o_{\rm eff}(D\log D)}\) \newline (Thm.~\ref{thm:permutation-hitting}) \\
\problem{Vertex-Disjoint Paths}, width \(p\) of a supplied path decomposition &
 \(2^{o(p\log p)}\) \newline \cite[Sec.~5]{lms} &
 \(2^{o_{\rm eff}(p\log p)}\) \newline (Thm.~\ref{vdp-width-w1}) \\
\problem{Cycle Packing}, width \(p\) of a supplied path decomposition &
 \(2^{o(p\log p)}\) \newline \cite[Thm.~1.3]{cc} &
 \(2^{o_{\rm eff}(p\log p)}\) \newline (Thm.~\ref{cycle-packing-degree-six}) \\
\problem{Maximum Cycle Cover}, width \(p\) of a supplied path decomposition &
 \(2^{o(p\log p)}\) \newline \cite[Thm.~1.3]{cc} &
 \(2^{o_{\rm eff}(p\log p)}\) \newline (Thm.~\ref{maximum-cycle-cover-supplied-pathwidth-w1}) \\
\problem{Line Distortion}, distortion \(d\) &
 \(2^{o(d\log d)}\) \newline \cite[Thm.~4.2]{lms} &
 \(2^{o_{\rm eff}(d\log d)}\) \newline (Thm.~\ref{graph-distortion-line-w1}) \\
\bottomrule
\end{tabular}
\caption{Lower bounds proved in this paper and their historical ETH
benchmarks. Each entry gives an excluded factor in a deterministic
running time with a fixed polynomial factor in the input length.}
\label{tab:benchmarks}
\end{table}

Since rank-width is at most linear rank-width, the \problem{Independent Set}
bound also holds by rank-width. Since treewidth is at most the width
of any supplied path decomposition, the bounds for
\problem{Vertex-Disjoint Paths}, \problem{Cycle Packing}, and
\problem{Maximum Cycle Cover} also hold by treewidth.

\begin{figure}[t]
\centering
\begin{tikzpicture}[x=1cm,y=1cm,
   tgt/.style={fnode,inner xsep=6pt,inner ysep=5pt,text width=5.9cm,align=left},
   src/.style={fnode,inner sep=6pt,line width=0.9pt},
   via/.style={fnode,inner sep=6pt,minimum width=3.4cm}]
  \node[tgt,anchor=north west] (cnvc) at (8.0,0)
    {\problem{Chromatic Number} (Thm.~\ref{cover-w1})\\ {\color{black!62}vertex cover number}};
  \node[tgt,below=8pt of cnvc] (isrw)
    {\problem{Independent Set} (Thm.~\ref{independent-set-rankwidth-w1})\\ {\color{black!62}linear rank-width}};
  \node[tgt,below=8pt of isrw] (vdp)
    {\problem{Vertex-Disjoint Paths} (Thm.~\ref{vdp-width-w1})\\ {\color{black!62}width of a supplied path decomposition}};
  \node[tgt,below=8pt of vdp] (cp)
    {\problem{Cycle Packing} (Thm.~\ref{cycle-packing-degree-six})\\ {\color{black!62}width of a supplied path decomposition}};
  \node[tgt,below=8pt of cp] (mcc)
    {\problem{Maximum Cycle Cover} (Thm.~\ref{maximum-cycle-cover-supplied-pathwidth-w1})\\ {\color{black!62}width of a supplied path decomposition}};
  \node[tgt,below=8pt of mcc] (ld)
    {\problem{Line Distortion} (Thm.~\ref{graph-distortion-line-w1})\\ {\color{black!62}distortion}};
  \node[src,draw=figB,anchor=west] (mcl) at (cnvc -| 0,0) {\problem{Multicolored Clique}};
  \node[src,draw=hypothesisink,anchor=west] (cl) at (isrw -| 0,0) {\problem{Clique}};
  \node[via] (cnf) at (isrw -| 4.9,0) {CNF encoding\\(Section~\ref{sec:preliminaries})};
  \coordinate (phy) at ($(vdp.north)!0.5!(ld.south)$);
  \node[via] (ph) at (phy -| 4.9,0)
    {\problem{Permutation Hitting}\\ {\color{black!62}grid order \(D\)}\\(Thm.~\ref{thm:permutation-hitting}, Lemma~\ref{lem:permutation-hitting})};
  \draw[farc,figB,line width=0.9pt] (mcl) -- (cnvc);
  \draw[farc] (cl) -- (cnf);
  \draw[farc] (cnf) -- (isrw);
  \draw[farc] (cnf) -- (ph);
  \foreach \n in {vdp,cp,mcc,ld} \draw[farc] (ph.east) -- (\n.west);
\end{tikzpicture}
\caption{Overview of the reductions. The blue arrow marks the route that starts from \problem{Multicolored Clique}, and the black arrows mark the routes that start from \problem{Clique}. The reductions to \problem{Independent Set} by linear rank-width and to \problem{Permutation Hitting} start from \problem{Clique} and use its CNF encoding. Each target is listed with its parameter.}
\label{fig:reductions}
\end{figure}
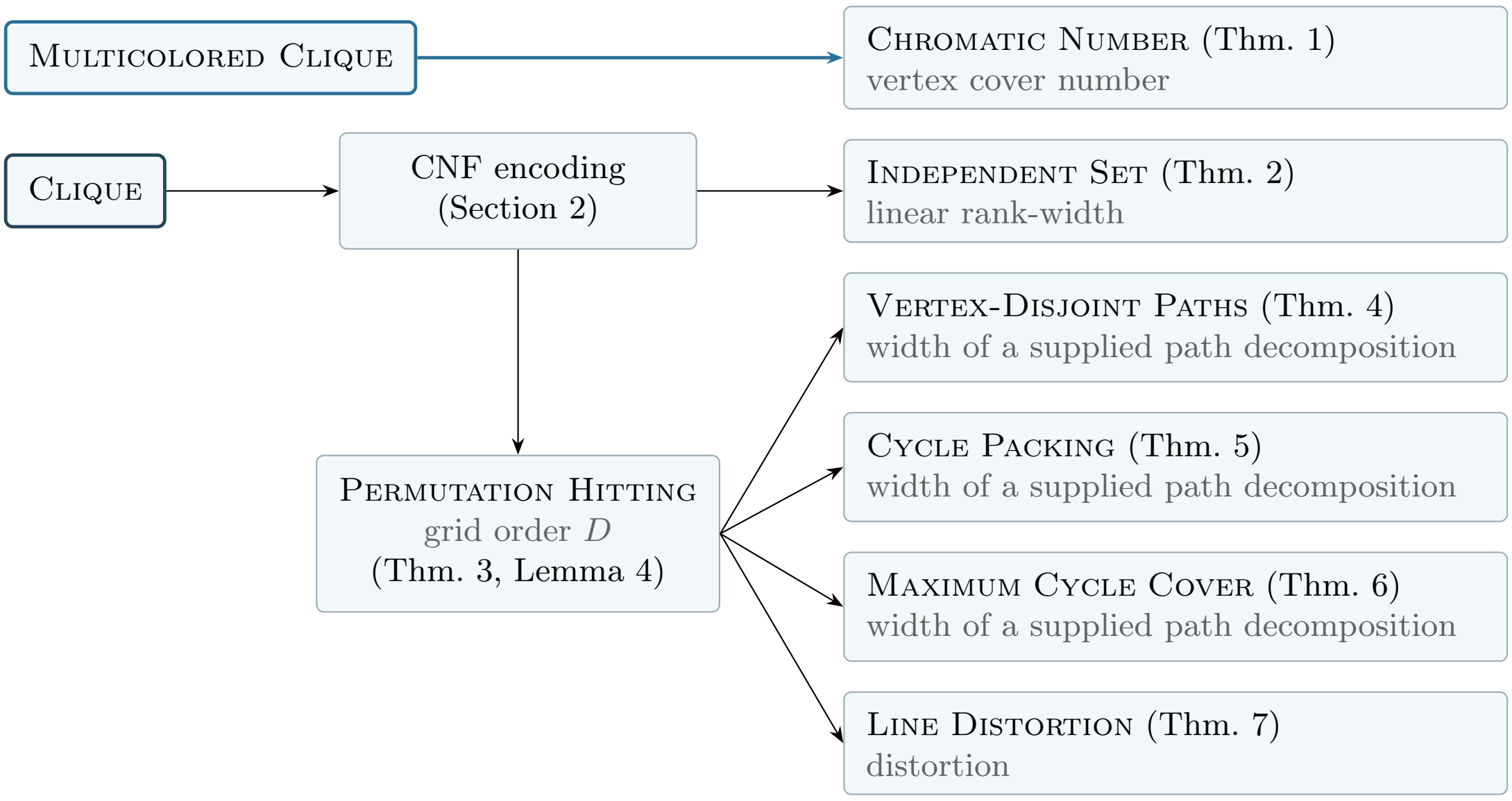

All our reductions start from \problem{Clique} parameterized by the requested
size or \problem{Multicolored Clique} parameterized by the number of
color classes. The \problem{Chromatic Number} proof encodes choices
as colors. The \problem{Independent Set} proof bounds linear rank-width
by ordering the vertices clause by clause and row by row.
The final four proofs use a common reduction to \problem{Permutation
Hitting}, where a permutation must hit every set of cells in a grid.
In \Cref{sec:permutation-hitting}, \Cref{lem:permutation-hitting}
gives the shared reduction, and \Cref{thm:permutation-hitting}
gives a running-time lower bound for \problem{Permutation Hitting}
parameterized by the grid order.
\Cref{fig:reductions} gives an overview of the reductions.

Heggernes and Meister~\cite[Theorem~2.2]{hm} established NP-hardness
for \problem{Line Distortion} on bipartite graphs. We give a
running-time lower bound on the same graph class.

 \section{Preliminaries}\label{sec:preliminaries}

We introduce the source problems for the reductions and the conventions
used to measure parameters and running times.

\subsection{Source problems and conventions}

\paragraph{Parameterized complexity.}
We use standard notions from parameterized complexity and recall the
ones needed here~\cite{cfk,fg}.
A parameterized decision problem belongs to \(\mathrm{FPT}\) if it can
be decided in time \(f(k)N^c\) for a computable function \(f\) and a
constant \(c\), where \(k\) is the parameter and \(N\) is the input
length. An FPT reduction preserves the answer, runs in fixed-parameter
time, and bounds the output parameter by a computable function of the
input parameter. One characterization of \(\mathrm{W}[1]\) is the
class of problems admitting an FPT reduction to \problem{Clique}
parameterized by the requested clique size~\cite{cfk,df}.

\problembox{\problem{Clique}}
  {A simple undirected graph \(G\) and a nonnegative integer \(k\).}
  {Does \(G\) contain a clique of size \(k\)?}
  {The requested clique size \(k\).}

The second source problem prescribes a color class for each choice.

\problembox{\problem{Multicolored Clique}}
  {A simple undirected graph \(G\) whose vertex set is partitioned into
   \(k\) color classes.}
  {Does \(G\) contain a clique with one vertex from each color class?}
  {The number \(k\) of color classes.}

\problem{Clique} is \(\mathrm{W}[1]\)-hard parameterized by the
requested size~\cite{df}. \problem{Multicolored Clique} is
\(\mathrm{W}[1]\)-hard parameterized by the number of color
classes~\cite[Lemma~1]{fhrv}.
Throughout, for a \problem{Clique} or \problem{Multicolored Clique}
instance, let \(n\) denote its number of vertices and \(k\) its
requested clique size or number of color classes, respectively.

We encode clique choices as instances of the following satisfiability problems.

\problembox{\problem{CNF-SAT}}
  {A Boolean formula in conjunctive normal form.}
  {Is there a truth assignment that satisfies the formula?}
  {The number of variables.}

\problembox{\problem{3-SAT}}
  {A Boolean formula in conjunctive normal form with at most three literals per clause.}
  {Is there a truth assignment that satisfies the formula?}
  {The number of variables.}

\paragraph{Encoding clique choices.}
We use the standard binary encoding of \problem{Clique} as a
\problem{CNF-SAT} instance, as described by de Rezende et al.~\cite[Section~2.3]{binary-clique}.
For a \problem{Clique} instance on \(n\ge2\) vertices, use
\(k\) blocks of \(\lceil\log_2 n\rceil\) bits to name the chosen vertices.
Forbid unused names and, for each pair of blocks, forbid names of
equal or nonadjacent vertices. Each forbidden pattern gives one clause
that is false precisely on that pattern. The resulting \problem{CNF-SAT}
instance is a yes-instance exactly when the graph has a \(k\)-clique. It has
\(m=k\lceil\log_2 n\rceil\) variables and \(O(kn+k^2n^2)\) clauses,
each of length at most \(2\lceil\log_2 n\rceil\). For
\problem{Multicolored Clique}, use one block per class and permit only
names from that class.
\Cref{fig:encoding} shows a small example.

\begin{figure}[t]
\centering
\begin{tikzpicture}[x=1cm,y=1cm]
  \node[fv,label={[flab]left:\(\mathtt{00}\)}] (a) at (0,1.4) {};
  \node[fv,label={[flab]right:\(\mathtt{01}\)}] (b) at (1.3,0.7) {};
  \node[fv,label={[flab]left:\(\mathtt{10}\)}] (c) at (0,0) {};
  \draw (a) -- (b) -- (c);
  \draw[dashed,figgray] (a) -- (c);
  \node[flab] at (0.6,-0.7) {\(n=3\), \(k=2\)};
  \begin{scope}[shift={(0.1,-2.6)}]
    \foreach \x/\l in {0/x_1,0.8/x_2,1.9/x_3,2.7/x_4}
      \node[draw,minimum width=0.8cm,minimum height=0.6cm,flab] at (\x,0) {\(\l\)};
    \draw[fbrace] (-0.4,0.42) -- node[flab,above=4pt]{vertex 1} (1.2,0.42);
    \draw[fbrace] (1.5,0.42) -- node[flab,above=4pt]{vertex 2} (3.1,0.42);
  \end{scope}
  \node[anchor=west,flab] at (4.6,-0.6) {%
    \begin{tabular}{@{}lll@{}}
    \toprule
    forbidden pattern & reason & clause\\
    \midrule
    \(x_1x_2=\mathtt{11}\) & unused name & \(\neg x_1\lor\neg x_2\)\\
    \(x_3x_4=\mathtt{11}\) & unused name & \(\neg x_3\lor\neg x_4\)\\
    \(x_1x_2x_3x_4=\mathtt{0101}\) & equal vertices & \(x_1\lor\neg x_2\lor x_3\lor\neg x_4\)\\
    \(x_1x_2x_3x_4=\mathtt{0010}\) & nonadjacent vertices & \(x_1\lor x_2\lor\neg x_3\lor x_4\)\\
    \multicolumn{3}{@{}l}{\(\;\vdots\)}\\
    \bottomrule
    \end{tabular}};
\end{tikzpicture}
\caption{The clique encoding for a path on three vertices and \(k=2\). The first block of two bits names the first chosen vertex and the second block names the second one. The dashed line marks the nonedge. The table lists some of the forbidden patterns together with the clauses that are false precisely on them.}
\label{fig:encoding}
\end{figure}

\paragraph{Grouping variables.}
Enumerating assignments to logarithmic-size groups is a standard
technique. For example, Lokshtanov, Marx, and Saurabh use it for
color assignments in their reduction to \(k\times k\)
\problem{Clique}~\cite[proof of Theorem~2.4]{lms}.
For a formula on \(m\ge16\) variables, we use groups of
\(b=\lfloor(\log_2 m)/2\rfloor\) bits. Pad the last group with
unused variables and enumerate its \(q=2^b\) assignments, as for
every other group. There are \(\lceil m/b\rceil=O(m/\log m)\)
groups, and \(q\le\sqrt m\).

\paragraph{Treewidth and pathwidth.}
A tree decomposition of a graph \(G\) assigns a set of vertices, called
a bag, to each node of a tree. Every vertex appears in a bag, every edge
has both endpoints in a common bag, and the nodes whose bags contain any
given vertex form a connected subtree. A path decomposition is a tree
decomposition whose tree is a path. The width of either decomposition
is its largest bag size minus one. The treewidth \(\operatorname{tw}(G)\)
and pathwidth \(\operatorname{pw}(G)\) are the minimum widths over tree
and path decompositions, respectively~\cite[Section~7.2]{cfk}.

\paragraph{Cut rank and linear rank-width.}
For \(S\subseteq V(G)\), the cut rank of \(S\) is the rank over
\(\mathbb F_2\) of the adjacency matrix with rows indexed by \(S\)
and columns indexed by \(V(G)\setminus S\). The maximum prefix cut rank
of a vertex order is the largest cut rank of any of its prefixes,
including the empty prefix. The linear rank-width of \(G\) is the
minimum of this maximum over all vertex orders. Rank-width is at
most linear rank-width.

\paragraph{Running-time conventions.}
We write \(\log\) for the base-two logarithm and \(\ln\) for the
natural logarithm. In running-time bounds for target problems,
\(N\) denotes the complete binary input length, including any supplied
certificate. For effective little-\(o\) (Section~\ref{sec:transfer}),
one can equivalently request a cutoff for each positive rational
tolerance \(\eta\) by taking \(j=\lceil1/\eta\rceil\). The cutoffs
are computable even if the exponent itself is not. Finite changes to
the comparison function leave the definition unchanged.

\subsection{Transferring running-time bounds}\label{sec:time-transfer-lemma}

We call a deterministic algorithm a
\(2^{o_{\mathrm{eff}}(\varphi(p))}N^{O(1)}\)-time algorithm if it runs in
\(O(2^{h(p)}N^c)\) time for a fixed \(c\ge0\) and a finite nonnegative
\(h=o_{\mathrm{eff}}(\varphi)\). For bounds written using \(p\log p\), we
use the nondecreasing comparison function
\(\Phi(p)=\max\{1,p\log_2(\max\{2,p\})\}\), which agrees with
\(p\log_2 p\) for \(p\ge2\). For \(\varphi(p)=p^2\) or
\(\varphi(p)=\Phi(p)\), we can take the exponent \(h\) to be
nondecreasing by
replacing it with its running maximum.\footnote{Let
\(h=o_{\mathrm{eff}}(\varphi)\) with cutoff \(T\), and let
\(\overline h(p)=\max_{0\le\ell\le p}h(\ell)\). Fix an integer \(c_0\ge0\)
bounding \(h(\ell)\) for \(\ell<T(1)\). Then \(h(\ell)\le c_0+\varphi(\ell)\)
for all \(\ell\). For each \(j\ge1\), terms with \(\ell\ge T(j)\) are at
most \(\varphi(p)/j\), while the other terms are at most
\(c_0+\varphi(T(j))\le c_0+1+T(j)^2\). Both comparison functions
satisfy \(\varphi(p)\ge p\), so \(j(c_0+1+T(j)^2)\) is a computable
cutoff for \(\overline h(p)\le\varphi(p)/j\).}

A disjunctive reduction can answer the source instance directly or
output a family of instances such that the source answer is yes
exactly when at least one output answer is yes. A reduction that
outputs one instance is a special case.

\begin{lemma}\label{lem:time-transfer}
Assume \(\mathrm{FPT}\ne\mathrm{W}[1]\). Let a deterministic
disjunctive reduction from \problem{Clique} by requested size \(k\)
or \problem{Multicolored Clique} by number of classes \(k\) run in
fixed-parameter time. Let \(n\ge2\) be the source graph order and
let \(p\) be the target parameter. The following bounds on \(p\)
must hold for every output instance.
\begin{enumerate}
\item If \(\varphi(p)\le g(k)(1+\log_2 n)\) for a computable
\(g\) and a nonnegative, nondecreasing comparison function \(\varphi\)
for which there is a computable integer-valued function \(B\) satisfying
\(\varphi(p)\le B(p)\) for every \(p\), the target has no \(2^{o_{\mathrm{eff}}(\varphi(p))}N^{O(1)}\)-time
algorithm.
\item If \(p\le a\log_2 n+g(k)\) for a fixed constant \(a\) and a
computable \(g\), the target has no deterministic
\(2^{O(p)}N^{O(1)}\)-time algorithm.
\end{enumerate}
\end{lemma}

\begin{proof}
In the first case, let \(T(j)\) be the computable cutoff for
\(h(p)\le\varphi(p)/j\), and choose a positive integer \(j\ge g(k)\)
computably from \(k\) (an integer computable upper bound suffices).
For \(p\ge T(j)\) this gives \(2^{h(p)}\le 2n\).
Fix a nonnegative integer \(c_0\) that bounds \(h(p)\) for \(p<T(1)\).
Then \(h(p)\le c_0+\varphi(p)\). For \(p<T(j)\) this is bounded
by \(c_0+\varphi(T(j))\le c_0+B(T(j))\).
The last expression is a computable bound depending only on \(k\). In the second case, for any fixed \(c>0\),
\(2^{cp}\le 2^{cg(k)}n^{ca}\). The reduction time bounds both
the number of outputs and their encoding lengths. In either case,
running the target algorithm on all outputs gives an FPT algorithm
for the source problem, a contradiction.
\end{proof}

\section{Deriving bounds from earlier work}\label{sec:prior-bound-derivations}

We derive the bounds in \Cref{tab:prior-benchmarks} using published
reductions and explain the historical comparisons that use an
additional reduction.

\paragraph{CNF-SAT and Closest String.}
Cai and Juedes~\cite[Lemma~3.1 and Corollary~3.1]{cj} gave
a logarithmic reparameterization of a paired-variable variant of
\problem{CNF-SAT}. For \(\mathrm{W}[1]\), their construction encodes a
weight-\(k\) assignment over \(r\) source variables using
\(O(k\log r)\) target variables. The pairing requires exactly one
true variable per pair. Two clauses per pair express this without
adding variables.
Their reduction therefore implies that \problem{CNF-SAT}
on \(n\) variables has no \(2^{o_{\mathrm{eff}}(n)}L^{O(1)}\)-time
algorithm under \(\mathrm{FPT}\ne\mathrm{W}[1]\), where \(L\) is
the formula encoding length.

The \problem{CNF-SAT} bound gives a partial recovery of the ETH lower
bound for \problem{Closest String} by radius. Drop tautological clauses
and add two fresh variables fixed to opposite truth values by unit
clauses. Let \(s\) be the resulting number of variables. Apply the construction of Lokshtanov,
Marx, and Saurabh~\cite[Theorem~3.1]{lms} with one position per
variable and two symbols for truth values. For each clause, write
the satisfying bit at each literal position and form \(s+1\) words
by filling the blank positions with a different additional symbol
in each word. A satisfying assignment agrees with every word.
Conversely, a center of radius \(d=s-1\) agrees with every word
in at least one position. One of the additional symbols is absent
from the center, so agreement with each word filled with that symbol
must occur at a satisfied literal. Replacing the remaining additional
symbols in the center by arbitrary bits preserves these agreements.
The two unit clauses ensure that all \(s+3=d+4\) symbols occur in
the input. Since the construction
takes polynomial time and \(d\) is one more than the original number
of variables, the \problem{CNF-SAT} bound excludes
\(2^{o_{\mathrm{eff}}(d)}N^{O(1)}\) time under
\(\mathrm{FPT}\ne\mathrm{W}[1]\).

\paragraph{Consequences of Lin's Set Cover reduction.}
For binary \problem{Distinct Vectors}, Lin's reduction from
\problem{Multicolored Clique} to \problem{Set Cover}~\cite[Lemma~14]{lin}
produces a universe \(U\) of size \(O(k^3\log n)\) and a budget
\(\binom{k}{2}\), where \(n\) is the graph order and \(k\) is the
number of color classes. View this as a \problem{Hitting Set} instance
by requiring a choice of sets that hits the collection of sets
containing each element of \(U\). Froese's construction from
\problem{Hitting Set}~\cite[Theorem~4.1]{froese} gives a binary matrix
with two rows per element and \(\lceil\log_2 |U|\rceil\) address
columns. For each element, one row has zeros in all set columns,
and the other has a one exactly in the columns for sets containing
that element. Number the elements from zero and use their binary
indices as addresses in both rows. For each address bit, the all-zero address and the address with
just that bit set both occur. Their rows with zero set entries force
that address column to be selected. The remaining columns distinguish the two
rows for each element exactly when the chosen sets cover it.
The resulting column budget is \(p=\binom{k}{2}+\lceil\log_2 |U|\rceil\),
so \(2^p\le g(k)(1+\log_2 n)\) for a computable function \(g\).
The running-time transfer of \Cref{lem:time-transfer} excludes \(2^{o_{\mathrm{eff}}(2^p)}N^{O(1)}\)
time under \(\mathrm{FPT}\ne\mathrm{W}[1]\). In particular, this rules
out \(2^{O(2^{\gamma p})}N^{O(1)}\) time for every fixed
\(0<\gamma<1\). The historical ETH bound of Pilipczuk and
Sorge~\cite[Theorem~1]{ps} rules out the same running time for each
fixed \(0<\gamma<1/2\). Since ETH implies
\(\mathrm{FPT}\ne\mathrm{W}[1]\), this derivation also strengthens the
cited ETH exclusion to every fixed \(0<\gamma<1\).

Lin's construction also gives a partial recovery for
\problem{Integer Programming Feasibility} by number of equations, with
a binary coefficient matrix and an all-ones right-hand side.
Let \(s=\lceil\log_2 n\rceil\) be the number of bits in each vertex
name. In the proof of \cite[Lemma~14]{lin}, the universe has
\(k(k-1)^2s\) elements. At each of its two endpoints, an edge covers
\(k-1\) elements per bit position, so each set has size \(2(k-1)s\).
The total size of \(\binom{k}{2}\) sets is therefore exactly the
universe size. A cover within budget must cover every element exactly
once, and every exact cover meets the budget.

Let \(A\) be the element-set incidence matrix. The equation
\(Ax=\mathbf 1\) for nonnegative integer \(x\) expresses the exact
cover condition. Each column is nonzero, so every feasible coordinate
is at most one. The number \(p\) of equations is \(O(k^3\log n)\),
and the matrix can be written in polynomial time. The running-time
transfer of \Cref{lem:time-transfer} excludes \(2^{o_{\mathrm{eff}}(p)}N^{O(1)}\) time under
\(\mathrm{FPT}\ne\mathrm{W}[1]\).

Lin's construction also gives the bound for \problem{Dominating Set}
by treewidth. Apply the construction
of Bodlaender et al.~\cite[proof of Corollary~12]{xalp} to the
element-set incidence graph. Add a vertex adjacent to every set
vertex and to a private leaf. We can reject an uncovered element
before this step, so the resulting graph is connected and bipartite.
A dominating set can be assumed to include the new vertex by
replacing its leaf if needed. Since the new vertex dominates all
set vertices, each selected element vertex can be replaced by an
incident set vertex. Thus a cover of size at most \(\binom{k}{2}\)
exists exactly when a dominating set of size at most
\(b=\binom{k}{2}+1\) exists. All element vertices and the new
vertex form a vertex cover of size \(|U|+1\), bounding the
treewidth \(t\). Since \(t=O(k^3\log n)\) and \(b\) depends only
on the source parameter \(k\), a computable factor \(f(b)\) in the
running time still gives an FPT algorithm for \problem{Multicolored
Clique} by the running-time calculation in \Cref{lem:time-transfer}.

\paragraph{Treewidth and pathwidth.}
Elberfeld, Stockhusen, and Tantau~\cite{est-xnlp} introduced the class
now called XNLP. Bodlaender, Groenland, Nederlof, and
Swennenhuis~\cite{bgns-xnlp} developed its hardness theory.
Bodlaender et al.~\cite[Lemma~4 and Theorem~6]{xnlp} proved
XNLP-hardness for \problem{Independent Set},
\problem{Dominating Set}, \problem{Odd Cycle Transversal},
\problem{Feedback Vertex Set}, and \(q\)-\problem{Coloring} and
\(q\)-\problem{List Coloring} for each fixed \(q\ge3\), when
parameterized by pathwidth divided by the logarithm of the graph
order. Since XNLP-hardness implies \(\mathrm{W}[1]\)-hardness,
these results give fixed-parameter reductions from \problem{Clique}
parameterized by the requested size \(k\). The output graphs have
pathwidth at most \(g(k)\) times the logarithm of their order, for a
computable function \(g\). The running-time transfer in
\Cref{lem:time-transfer} then rules out an effectively subexponential
dependence on pathwidth for each problem under
\(\mathrm{FPT}\ne\mathrm{W}[1]\). Since treewidth is at most pathwidth,
the exclusions also hold for treewidth. Bodlaender et
al.~\cite[Theorem~11 and Corollary~12]{xalp} directly establish
hardness for \problem{Independent Set}, \problem{Vertex Cover},
\problem{Red-Blue Dominating Set}, and \problem{Dominating Set} with
treewidth divided by the logarithm of the graph order as parameter.

The construction of Bodlaender et al.~\cite[Lemma~4 and Claim~12]{xnlp}
also recovers the historical bound for \problem{Chromatic Number} by
treewidth when the color budget is allowed to grow with the input length.

\begin{proposition}\label{prop:chromatic-treewidth-prior}
Assume \(\mathrm{FPT}\ne\mathrm{W}[1]\). There is no deterministic
\(2^{o_{\mathrm{eff}}(t\log t)}N^{O(1)}\)-time algorithm for
\problem{Chromatic Number} parameterized by the treewidth \(t\).
\end{proposition}

\begin{proof}
Apply the construction of Bodlaender et al.~\cite[Lemma~4]{xnlp}
to a \problem{Chained Positive CNF-SAT} instance of encoding length \(M\)
and parameter \(k\), with \(q\ge3\) colors. The correctness argument
applies for every \(q\ge3\). Each connector has \(O(q)\) vertices per
digit of the encoded choice, so the construction takes time polynomial
in \(M\) and \(q\).

Each choice uses at most \(1+\log_q M\) vertices. As in
\cite[Claim~12]{xnlp}, retain the choice vertices of two consecutive
blocks while processing their clauses. Sweep each clause path in order.
At a clause vertex, retain it and the center of the current connector
while processing the branches one at a time. Each branch uses at most
two further vertices. After all connectors at that clause vertex have
been processed, introduce the next clause vertex to cover the path edge.
The number of additional vertices in each bag is therefore bounded
independently of \(q\). Adding the \(q\)-vertex clique that enforces
the lists to every bag gives a path decomposition of width
\(O(k(1+\log_q M)+q)\), which also bounds the treewidth \(t\).

Choose \(q=\max\{3,\lfloor\sqrt{\log_2 M}\rfloor\}\). For sufficiently
large \(M\), this gives \(t=O((k+1)\log M/\log\log M)\).
The comparison function therefore satisfies
\(\Phi(t)\le g(k)(1+\log M)\) for a computable function \(g\).
\problem{Chained Positive CNF-SAT} is \(\mathrm{W}[1]\)-hard in
\(k\)~\cite[Theorem~9]{xnlp}. A fixed-parameter reduction from
\problem{Clique} on \(n\) vertices with requested size \(k_0\)
bounds \(k\) by a computable function of \(k_0\) and gives
\(M\le f(k_0)n^c\) for a computable function \(f\) and a fixed
constant \(c\). Thus \(\log M\le\log f(k_0)+c\log n\), and
\(\Phi(t)\) is bounded by a computable function of \(k_0\) times
\(1+\log n\). Apply \Cref{lem:time-transfer} with \(\varphi=\Phi\).
\end{proof}

Several of the historical treewidth comparisons use short reductions.
The graph-order bound for \problem{Vertex Cover} implies its treewidth
bound because treewidth is at most the graph order. Subdividing each
edge gives the incidence graph for \problem{Red-Blue Dominating Set}, as
in \cite[Corollary~12]{xalp}. Instead, adding one vertex adjacent to both
endpoints of each edge gives a reduction to \problem{Feedback Vertex Set}.
A vertex cover leaves a forest in the latter graph. Conversely, a
feedback vertex set hits every edge triangle, and replacing any selected
new vertex by an endpoint gives a vertex cover of no larger size.
In both reductions, one can attach a bag for each new vertex at a bag
containing the two endpoints. For fixed \(q\ge3\), joining a
\((q-3)\)-vertex clique to the graph transfers the graph-order bound
for \(3\)-\problem{Coloring} to \(q\)-\problem{Coloring}. Using the
same list of \(q\) colors at every vertex then gives
\(q\)-\problem{List Coloring}. These bounds also apply with treewidth
as the parameter.

\paragraph{Dominating Set with a supplied vertex order.}
Liu and Meng~\cite[Theorem~3.1]{liu-meng-rw} prove the quadratic ETH
lower bound in the maximum prefix cut rank of a supplied vertex order. Their basic construction
produces a monopolar graph, whose vertices partition into an
independent set and a disjoint union of cliques. The construction works
for clauses of arbitrary length, as Liu and Meng note in their
analysis~\cite[proof of Theorem~C.1]{liu-meng-rw}. On a formula with
\(s\) variables, it takes
\(2^{O(\sqrt{s})}\) time times a fixed polynomial in the formula length
and supplies a vertex order of maximum prefix cut rank \(p=O(\sqrt{s})\).
For \problem{Clique} on \(n\) vertices with requested size \(k\),
the CNF encoding gives \(s=O(k\log n)\). Thus the construction
takes fixed-parameter time and \(p^2=O(k\log n)\). The running-time
transfer gives the exclusion in \Cref{tab:prior-benchmarks}. Since
rank-width is at most linear rank-width and linear rank-width is at
most \(p\), the same exclusion holds for both intrinsic parameters.

\paragraph{Bounds on the input-size exponent.}
Class separations also give quantitative exclusions in which the
parameter controls the exponent of the input size. Chen et
al.~\cite[Theorems~5.2 and~5.4]{chkx} use reductions that increase
the parameter by at most a constant factor. Under
\(\mathrm{FPT}\ne\mathrm{W}[1]\) and their computability
conventions for the running-time functions, they exclude
\(f(k)r^{o(k)}N^{O(1)}\) time for \problem{Set Cover},
\problem{Hitting Set}, \problem{Dominating Set}, and weighted
\problem{CNF-SAT}. Here \(k\) is the solution budget or assignment
weight, and \(r\) is the number of available sets, universe elements,
graph vertices, or variables, respectively.

\section{\problem{Chromatic Number} by vertex cover number}

\problembox{\problem{Chromatic Number}}{A simple undirected graph \(G\) and a positive integer \(q\).}{Does \(G\) admit a proper coloring with at most \(q\) colors?}{The vertex cover number \(p\) of \(G\).}

\hypertarget{cover-w1}{}\begin{theorem}\label{cover-w1}
Assume \(\mathrm{FPT}\ne\mathrm{W}[1]\).
There is no deterministic algorithm for \problem{Chromatic Number}
running in \(2^{O(p)}N^{O(1)}\) time.
\end{theorem}

We reduce from \problem{Multicolored Clique}. Write the binary names of the chosen vertices as rows of a matrix.
Each column records one bit from each name. We want the coloring to
decide the order of the columns, so we first guess how many copies
of each column pattern it has to arrange. We do this separately in
blocks, choosing their size so that all count guesses can be
enumerated in fixed-parameter time. For each guess, the coloring
arranges the columns within each block. We adapt the construction of
Lokshtanov, Marx, and Saurabh~\cite[Theorem~6.1]{lms}, which encodes a
permutation as a coloring and uses test vertices to rule out nonedges.
The test vertices are pairwise nonadjacent, so the clique that names
the colors and the vertices that encode the column choices form a
vertex cover. The blockwise guesses let us apply this construction to the binary
vertex names while keeping the vertex cover number at most
\(4\log_2 n+f(k)\), where \(n\) is the number of vertices
and \(f\) depends only on the number \(k\) of color classes.

\begin{proof}
Suppose such an algorithm exists. We use it to decide
\problem{Multicolored Clique} in fixed-parameter time with respect
to the number \(k\) of color classes. Let \(n\) be the total number of
vertices. Handle \(k<2\) and empty color classes directly. Then
\(2\le k\le n\).

\paragraph{Encoding vertex choices.}
Let \(r=\lceil\log_2 n\rceil\), \(b=4^{k+1}\), and
\(L=b\lceil r/b\rceil\). Pad each color class to \(2^L\) vertices with
isolated vertices and assign its vertices distinct \(L\)-bit names.
This preserves the answer since \(k\ge2\). Moreover, \(L<r+b\), so
\(2^L<2^{b+1}n\). Choosing one name from each color class amounts to
choosing a column in \(\{0,1\}^k\) at each of the \(L\) bit positions.

Partition the positions into blocks of \(b\) consecutive positions.
In each block, guess the number of occurrences of each of the \(2^k\)
possible columns. These nonnegative counts must sum to \(b\).
Since \(2^k\log_2(b+1)\le b\), there are at most
\((b+1)^{2^kL/b}\le2^L\) guesses over all blocks, even if we enumerate
all vectors of counts and discard invalid sums. We construct one
instance of \problem{Chromatic Number} per valid guess and accept if any answers
yes. For a fixed guess and a block \(H\), prepare \(b\) labels
indexed by \(1,\ldots,b\), each carrying one \(k\)-bit column, with the
guessed multiplicities. Permuting these labels over the positions
realizes precisely the sequences of columns with those counts.
\Cref{fig:cnvc}(a) illustrates the labels of one block.

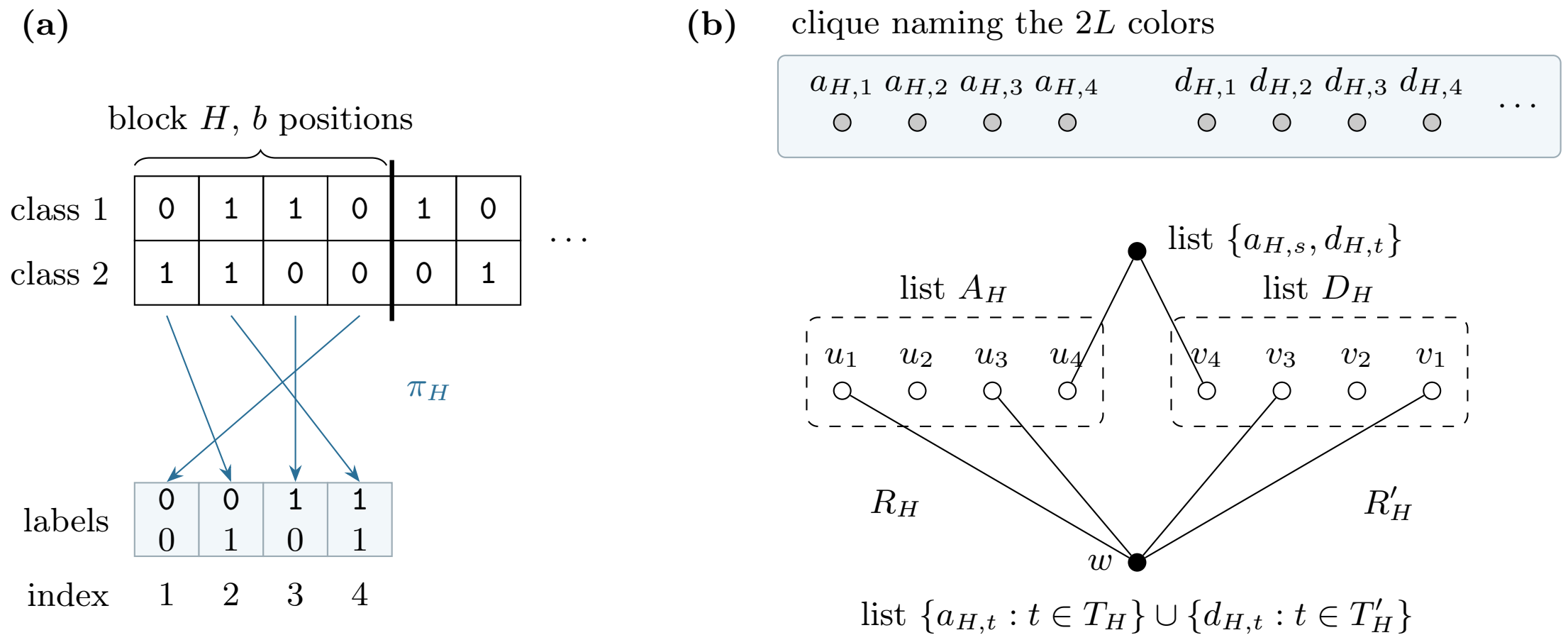
\begin{figure}[t]
\centering
\begin{tikzpicture}[x=1cm,y=1cm]
  \begin{scope}
    \node[flab,anchor=west] at (-1.5,2.3) {\textbf{(a)}};
    \foreach \r/\bits in {1/{0,1,1,0,1,0},0/{1,1,0,0,0,1}}
      \foreach \b [count=\c from 0] in \bits
        \node[draw,minimum size=0.6cm,flab,inner sep=0pt] (m\r\c) at (0.6*\c,0.6*\r) {\ttfamily\b};
    \node[flab] at (3.75,0.3) {\(\cdots\)};
    \node[flab,anchor=east] at (-0.4,0.6) {class 1};
    \node[flab,anchor=east] at (-0.4,0) {class 2};
    \draw[very thick] (0.6*3+0.3,-0.45) -- (0.6*3+0.3,1.05);
    \draw[fbrace] (-0.3,1.0) -- node[flab,above=4pt]{block \(H\), \(b\) positions} (0.6*3+0.25,1.0);
    \foreach \t/\col [count=\c from 0] in {1/{0\\0},2/{0\\1},3/{1\\0},4/{1\\1}}
      \node[draw=hypothesisink!45,fill=hypothesisfill,minimum width=0.6cm,flab,align=center,inner sep=1.5pt] (l\t) at (0.6*\c,-2.3) {\ttfamily\col};
    \foreach \t [count=\c from 0] in {1,...,4} \node[flab] at (0.6*\c,-3.0) {\(\t\)};
    \node[flab,anchor=east] at (-0.4,-2.3) {labels};
    \node[flab,anchor=east] at (-0.4,-3.0) {index};
    \foreach \c/\t in {0/2,1/4,2/3,3/1}
      \draw[farc,figB] (0.6*\c,-0.4) -- (l\t.north);
    \node[flab,text=figB] at (2.45,-1.1) {\(\pi_H\)};
  \end{scope}
  \begin{scope}[shift={(6.1,0.2)}]
    \node[flab,anchor=west] at (-1.4,2.1) {\textbf{(b)}};
    \node[fbox,minimum width=7.3cm,minimum height=0.95cm] (pal) at (3.25,1.35) {};
    \node[flab,anchor=south west] at (pal.north west) {clique naming the \(2L\) colors};
    \foreach \t [count=\c from 0] in {1,...,4} {
      \node[fvg,label={[flab]above:\(a_{H,\t}\)}] (a\t) at (0.2+0.7*\c,1.2) {};
      \node[fvg,label={[flab]above:\(d_{H,\t}\)}] (d\t) at (3.6+0.7*\c,1.2) {};
    }
    \node[flab] at (6.5,1.35) {\(\cdots\)};
    \foreach \j [count=\c from 0] in {1,...,4} {
      \node[fv,label={[flab]above:\(u_{\j}\)}] (x\j) at (0.2+0.7*\c,-1.3) {};
      \node[fv,label={[flab]above:\(v_{\j}\)}] (y\j) at (5.7-0.7*\c,-1.3) {};
    }
    \begin{scope}[on background layer]
      \node[draw,dashed,rounded corners=3pt,fit=(x1)(x4),inner xsep=7pt,inner ysep=12pt,yshift=5pt] (xc) {};
      \node[draw,dashed,rounded corners=3pt,fit=(y1)(y4),inner xsep=7pt,inner ysep=12pt,yshift=5pt] (yc) {};
    \end{scope}
    \node[flab,anchor=south] at (xc.north) {list \(A_H\)};
    \node[flab,anchor=south] at (yc.north) {list \(D_H\)};
    \node[fvb] (t) at (2.95,0.0) {};
    \draw (x4) -- (t) -- (y4);
    \node[flab,anchor=west] at (3.1,0.1) {list \(\{a_{H,s},d_{H,t}\}\)};
    \node[fvb,label={[flab]left:\(w\)}] (z) at (2.95,-2.9) {};
    \draw (z) -- (x1); \draw (z) -- (x3);
    \draw (z) -- (y3); \draw (z) -- (y1);
    \node[flab,anchor=north] at (2.95,-3.1) {list \(\{a_{H,t}:t\in T_H\}\cup\{d_{H,t}:t\in T'_H\}\)};
    \node[flab] at (0.7,-2.35) {\(R_H\)};
    \node[flab] at (5.3,-2.35) {\(R'_H\)};
  \end{scope}
\end{tikzpicture}
\caption{The reduction to \problem{Chromatic Number} parameterized by the vertex cover number, drawn schematically for two color classes. The proof uses blocks of \(b=4^{k+1}\) positions, where \(k\) is the number of color classes. The drawing shortens the block to four positions, which suffices to illustrate the permutation. (a)~The names of the chosen vertices form a bit matrix with one row per color class. In each block \(H\), the permutation \(\pi_H\) maps the positions to the labels that carry the guessed columns. (b)~The vertices of one block. The dashed boxes show the two cliques of position vertices, with allowed color sets \(A_H\) and \(D_H\). The black vertices are tests, with their own allowed color sets. Edges to the clique that names the colors enforce these sets. The upper black vertex is one of the permutation tests for position~\(4\), and \(w\) tests a nonedge.}
\label{fig:cnvc}
\end{figure}

\paragraph{Encoding permutations by colors.}
Our graph contains a clique with \(2L\) vertices, and we set the color
budget to \(q=2L\). Thus its vertices name all available colors. To allow
a vertex a specified set of colors, make it adjacent to the clique
vertices for all other colors. For every block \(H\), reserve two sets
of colors \(A_H=\{a_{H,1},\ldots,a_{H,b}\}\) and
\(D_H=\{d_{H,1},\ldots,d_{H,b}\}\). All these sets are disjoint.
For each position \(j\in H\) introduce vertices \(u_j,v_j\) with
allowed sets \(A_H,D_H\), respectively. Within \(H\), make the
\(u_j\) pairwise adjacent and make the \(v_j\) pairwise adjacent.
Each set of vertices therefore uses a permutation of the \(b\) colors.

We enforce the same permutation in both sets. For every \(j\in H\)
and every ordered pair \(s\ne t\) of indices, give a new vertex the
allowed colors \(\{a_{H,s},d_{H,t}\}\) and join it to \(u_j,v_j\).
Both colors are blocked exactly when \(u_j,v_j\) choose indices
\(s,t\), respectively. Once all these vertices can be colored, there
is a single permutation \(\pi_H\) from the positions of \(H\) to the
label indices, realized by both sets. Conversely, any such permutation
allows every one of these vertices to be colored.
See \Cref{fig:cnvc}(b) for an illustration.

\paragraph{Forbidding pairs of nonadjacent vertices.}
For each nonedge between a vertex \(u\) in color class \(i\) and a
vertex \(v\) in color class \(i'>i\), we want to forbid the
simultaneous choice of \(u\) and \(v\). For each block \(H\), let
\(R_H\) be the positions in \(H\) where the name of \(u\) has a one, and
\(T_H\) the label indices whose column has a one in coordinate \(i\).
Define \(R'_H,T'_H\) in the same way for \(v,i'\).
The encoded names agree with these two names on the whole block
exactly when
\[
   \pi_H(R_H)=T_H \qquad\text{and}\qquad \pi_H(R'_H)=T'_H .
\]
If either equality of cardinalities fails in any block, the forbidden
pair cannot be selected with this guess, and we skip the nonedge.

Otherwise introduce one vertex \(w\) for this nonedge. Over all
blocks \(H\), its allowed colors are the \(a_{H,t}\) for \(t\in T_H\)
and the \(d_{H,t}\) for \(t\in T'_H\). Its neighbors outside the
clique that names the colors are the \(u_j\) for \(j\in R_H\) and the \(v_j\) for
\(j\in R'_H\) over all blocks. Then \(w\) has no available color
exactly when \(T_H\subseteq\pi_H(R_H)\) and
\(T'_H\subseteq\pi_H(R'_H)\) in every block. Since the cardinalities
match, these inclusions are exactly the two required equalities.
Thus \(w\) has no available color exactly when both forbidden
vertices are selected. This also covers empty sets, since the
corresponding inclusion then holds for every permutation.

All vertices added to test permutations or nonedges are mutually
nonadjacent, so their colors can be chosen independently.
\paragraph{Completeness.}
Suppose the \problem{Multicolored Clique} instance is a yes-instance.
Write the binary names of the clique vertices as rows, in color-class
order. In each block, count how often each column pattern occurs.
These counts sum to the block size, so the reduction enumerates this
guess. Match the positions carrying each pattern to the labels
carrying that same pattern. This gives a permutation in each block.
Use it for both sets of color choices, so all permutation tests can
be colored. Since the chosen vertices form a clique, every nonedge
test can also be colored.

\paragraph{Soundness.}
Suppose one of the produced graphs has a proper coloring with the
prescribed budget. The coloring determines one vertex name per
color class, and the nonedge tests ensure that the selected vertices
are pairwise adjacent. Since \(k\ge2\), none of the selected vertices
is an isolated vertex added by padding. The selected vertices form
a multicolored clique.

\paragraph{Parameter bound.}
The clique that names the colors and the vertices \(u_j,v_j\) form
a vertex cover of size \(4L\), since the test vertices are pairwise
nonadjacent. As \(L<\lceil\log_2 n\rceil+b\), the vertex cover
number of each output graph satisfies
\[
 p\le 4L\le 4\log_2 n+4(b+1).
\]
The coefficient of \(\log_2 n\) is absolute, and \(b\) depends only
on the number \(k\) of color classes.

\paragraph{Construction time.}
There are at most \(2^L\) guesses. Each output graph has at most
\(Lb^2\) permutation tests and \(\binom{k}{2}2^{2L}\) nonedge tests.
Each test vertex has at most \(4L\) neighbors. Since
\(2^L<2^{b+1}n\), generating all output instances, including their
explicit edge lists, takes \(f(k)n^{O(1)}\) time for a computable
function \(f\). The second case of \Cref{lem:time-transfer} now gives
the claimed lower bound.
\end{proof}

\section{\problem{Independent Set} by linear rank-width}

\problembox{\problem{Independent Set}}{A simple undirected graph \(G\) and a nonnegative integer \(r\).}{Does \(G\) contain an independent set of size at least \(r\)?}{The linear rank-width \(p\) of \(G\).}

\hypertarget{independent-set-rankwidth-w1}{}
\begin{theorem}\label{independent-set-rankwidth-w1}
Assume \(\mathrm{FPT}\ne\mathrm{W}[1]\).
There is no deterministic \(2^{o_{\mathrm{eff}}(p^2)}N^{O(1)}\)-time algorithm
for \problem{Independent Set} parameterized by the linear rank-width \(p\).
\end{theorem}

We reduce from \problem{Clique} using its CNF encoding.
Bergougnoux, Korhonen and Nederlof~\cite[Theorem 1.1]{bkn} establish
the quadratic ETH lower bound parameterized by linear rank-width.
Since rank-width is at most linear rank-width, their result also
rules out the corresponding algorithm parameterized by rank-width
when no decomposition is supplied. We order the literal vertices by
row to extend their construction from 3-CNF formulas to clauses of
arbitrary length. This lets us use the CNF encoding of \problem{Clique}
directly, with no extra variables for splitting clauses.

We arrange the formula variables in a \(d\times d\) array, where \(d\) is
the square root of the number of variables rounded up. Each clause gets its
own copy of the row assignments. A shared set of vertices forces these
copies to agree, and a literal vertex can be selected only if the common
assignment makes that literal true.

The vertex order lists all vertices of the shared set before listing
the clause vertices, clause by clause and row by row. The edges from
the shared set to the assignment vertices have low cut rank because
they are defined by dot products of short vectors. The edges that
test literals can cross a cut only in the current row of the current
clause. This keeps the linear rank-width at \(O(d)\).

\subsection{Encoding assignments in an independent set}

We first construct a weighted graph in which an independent set reaches
the required weight exactly when the formula is satisfiable. Each clause
gets its own copy of the assignment choices, and a shared independent set
rewards choices that agree across clauses.

Take a CNF formula and remove duplicate literals and tautological clauses.
Reject if there is an empty clause, and accept if no clauses remain.
Normalize the remaining variable names. Let \(m_{\mathrm{var}}\) be
the number of occurring variables and \(m\) the number of remaining
clauses. With \(d=\lceil\sqrt{m_{\mathrm{var}}}\rceil\), pad with unused
variables to make \(d\) rows of \(d\) variables each. Let \(B\) be an independent set with a
vertex \(u_z\) for each vector \(z\in\mathbb F_2^{2d}\). For each clause \(i\) and
row \(j\) make a clique \(A_{i,j}\) with one vertex per assignment
\(a\in\mathbb F_2^d\). Its vector is \((e_j,a)\), with \(e_j\) the
unit vector at \(j\). Join the assignment vertex to \(u_z\) exactly
when \(\langle(e_j,a),z\rangle=1\). Each clause has its own copy of
the assignment. The dot-product edges make the vertices of \(B\) that
can be selected alongside that copy depend on the assignment it chooses.
If two clauses choose different assignments, fewer vertices of \(B\)
are available to both.

For each clause \(i\), make a clique \(T_i\) with one vertex per literal. Join each literal vertex
to the row-assignment vertices for its clause and row whose assignments
make the literal false. There are no other edges.

\begin{figure}[t]
\centering
\begin{tikzpicture}[x=1cm,y=1cm,
   blk/.style={fbox,minimum width=1.5cm,minimum height=0.8cm,flab,align=center},
   lit/.style={draw,rounded corners=2pt,minimum width=5.4cm,minimum height=0.7cm,flab,align=center}]
  \node[blk,minimum width=12.4cm] (B) at (6.2,0) {\(B\): independent set, one vertex \(u_z\) for each vector \(z\in\mathbb F_2^{2d}\)};
  \foreach \i/\x in {1/0,2/7.0} {
    \foreach \j/\l/\dx in {1/1/0,2/2/1.95,3/d/3.9}
      \node[blk] (A\i\j) at (\x+\dx+0.75,2.4) {\(A_{\i,\l}\)};
    \node[flab] at (\x+3.68,2.4) {\(\cdots\)};
    \node[lit] (T\i) at (\x+2.7,4.6) {\(T_{\i}\): clique of the literals of clause \(\i\)};
    \foreach \j in {1,2,3} {
      \draw[hiB] (A\i\j.south) -- (A\i\j.south |- B.north);
      \draw[hiA] (A\i\j.north) -- (A\i\j.north |- T\i.south);
    }
  }
  \node[flab] at (12.9,2.4) {\(\cdots\)};
  \draw[hiB] (0.2,-1.05) -- (1.0,-1.05);
  \node[flab,anchor=west] at (1.1,-1.05) {the vertex of \(A_{i,j}\) with vector \((e_j,a)\) is adjacent to \(u_z\) if \(\langle(e_j,a),z\rangle=1\)};
  \draw[hiA] (0.2,-1.6) -- (1.0,-1.6);
  \node[flab,anchor=west] at (1.1,-1.6) {a literal of row \(j\) is adjacent to the assignments \(a\) in \(A_{i,j}\) that falsify it};
\end{tikzpicture}
\caption{The structure of the graph encoding assignments. Each box \(A_{i,j}\) is a clique with one vertex for each assignment \(a\in\mathbb F_2^d\) to the variables of row \(j\). The blue edges to \(B\) enforce a common assignment when the independent set reaches the weight threshold. The red edges test clause satisfaction. There are no edges other than these two types and the ones inside the cliques.}
\label{fig:rw-layout}
\end{figure}
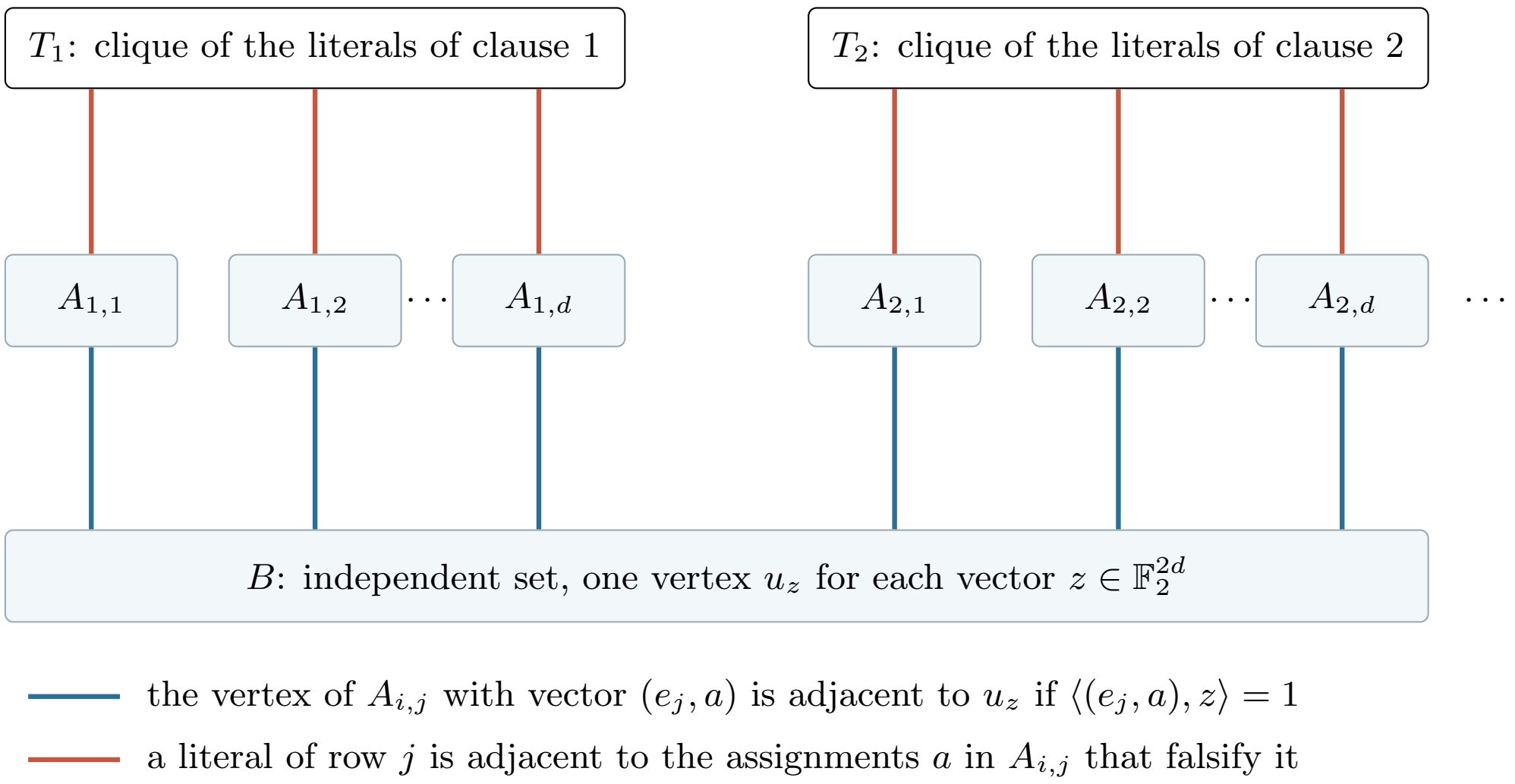

We use weights so that an optimal independent set must choose one
assignment vertex for each row of each clause. In the next subsection,
we replace the weights by copies of vertices to obtain an unweighted
instance. Give each assignment vertex weight \(M=2^{2d}\) and each
other vertex weight one. The target weight is \(W=Mdm+2^d+m\).
The three terms count one assignment vertex of weight \(M\) for each
row of each clause, the \(2^d\) vertices of \(B\) available for a
common assignment, and one true literal for each clause.
\Cref{fig:rw-layout} shows the two kinds of adjacency. The edges between the
cliques \(A_{i,j}\) and \(B\) force the assignments chosen for the different
clauses to agree, and the edges between \(T_i\) and the cliques \(A_{i,j}\)
test clause \(i\).

The chosen vectors of one assignment matrix are linearly independent
because their first blocks are the distinct standard basis vectors
\(e_1,\ldots,e_d\). The indices of their common nonneighbors in \(B\)
form the
orthogonal complement of their span, so there are \(2^d\) common
nonneighbors. These indices determine the assignment matrix. In the
orthogonal complement of their subspace, the unique vector with first
block \(e_j\) encodes row \(j\).

The weight threshold distinguishes satisfying from unsatisfying formulas.

\begin{lemma}\label{lem:rw-weighted-encoding}
The formula is satisfiable if and only if the constructed weighted
graph has an independent set of weight at least \(W\).
\end{lemma}

\begin{proof}
The independent set records a common assignment across the clause layers.

\paragraph{Completeness.}
Fix a satisfying assignment. For each clause and each row, select the
vertex representing that row's assignment. These \(dm\) vertices lie in distinct
assignment cliques and have no edges between them. Every clause
uses the same assignment, so the same \(2^d\) vertices of \(B\)
are nonadjacent to all the selected assignment vertices. Select all
of them. Finally, select one vertex of \(T_i\) for a true literal of
each clause \(i\). Its edges to assignment vertices go only to choices
that make it false. It has no edges to \(B\) or to literal vertices
of other clauses. The selected vertices therefore form an independent
set of weight \(Mdm+2^d+m=W\).

\paragraph{Soundness.}
Take a maximum-weight independent set of weight at least \(W\).
It contains one vertex from every assignment clique. If some assignment
clique contributed no vertex, we could insert one of weight \(M\). Its
vector has a nonzero first block, so it has exactly \(M/2\) neighbors
in \(B\). It also has at most one selected neighbor in the relevant
literal clique \(T_i\). Removing these neighbors costs at most \(M/2+1\)
in weight, strictly less than \(M\).
The selected assignment vertices contribute weight \(Mdm\). For each
clause \(i\), let \(B_i\) be the set of vertices in \(B\) nonadjacent
to the \(d\) selected assignment vertices for that clause. Each \(B_i\)
has size \(2^d\). Any selected vertex of \(B\) lies in every \(B_i\),
so at most \(2^d\) vertices of \(B\) can be selected. At most one literal
vertex can be selected per clause. Reaching \(W\) therefore requires
selecting \(2^d\) vertices of \(B\) and one literal vertex per clause.
Since the selected vertices of \(B\) lie in every \(B_i\), all the
\(B_i\) are equal. Each \(B_i\) determines the assignment in its
clause, so all clauses use the same assignment. The selected literal
in each clause is true under that assignment, as it is nonadjacent to
the selected assignment vertex for its row.
\end{proof}

\subsection{Bounding linear rank-width}

We first order the vertices of the weighted graph to bound the cut rank
of every prefix. We then remove the weights by replacing each vertex with
an independent set of copies and keeping those copies consecutive in the
order.

\begin{lemma}\label{lem:is-rank-width-order}
The weighted graph from the preceding subsection can be converted to an
unweighted graph \(G\) with the same optimum and linear rank-width
at most \(3d+4\). The construction takes \(2^{O(d)}\) time times a
fixed polynomial in the formula length.
\end{lemma}

\begin{proof}
First consider the underlying graph of the weighted construction.
List \(B\) first. Then, for each clause \(i\) and each row \(j\), list
\(A_{i,j}\) followed by the vertices of \(T_i\) for literals in row \(j\).

At any prefix cut, the edges between \(B\) and the assignment vertices
contribute rank at most \(2d\), since their entries are dot products of
\(2d\)-dimensional vectors and the vertices of \(B\) come first.
A split assignment clique and a split literal clique each contribute
rank at most one. At most one clique of each type is split. The edges
between assignment and literal vertices cross only within the current
row of the current clause. For that row, the column of a literal
vertex is a coordinate column of the row assignments or its
complement. These columns have rank at most \(d+1\), as they lie in
the span of the \(d\) coordinate columns and the all-ones column.
Thus every prefix cut has rank at most \(3d+3\).

Following Bergougnoux, Korhonen, and Nederlof~\cite[Lemma~4.12]{bkn},
replace each vertex of weight \(\omega\) by \(\omega\) independent
false twins and each edge by the complete join of the two corresponding
classes. An independent set can be enlarged to take every twin in each
class it meets, so the new maximum independent-set size is the old
maximum weight. Keep each class consecutive in the order. A cut
between classes only repeats rows and columns. For a cut inside a
class, compare with the cut just after that class. The twins left on
the prefix side repeat the same row, and the twins now on the suffix
side add at most one column type. The cut rank is therefore at most
\(3d+4\). The largest weight is \(2^{2d}\), so the expanded graph can
be constructed within the claimed time bound.
\end{proof}

\subsection{Putting the reduction together}

We now apply the construction to the CNF encoding of \problem{Clique}.
The bound on linear rank-width is linear in the side length of the
variable array, so its square is proportional to the number of variables
in the encoding.

\begin{proof}[Proof of \Cref{independent-set-rankwidth-w1}]
Handle \(n<2\) and \(k>n\) directly. On a \problem{Clique} instance
with \(n\ge2\) vertices and requested clique size \(k\), use the CNF
encoding from Section~\ref{sec:preliminaries}. The formula has
\(m_{\mathrm{var}}\le k\lceil\log_2 n\rceil\) variables, and its length
and construction time are polynomial in \(n\).

By \Cref{lem:rw-weighted-encoding}, the formula is satisfiable
exactly when the weighted graph has an independent set of weight
at least \(W\). By \Cref{lem:is-rank-width-order}, this holds
exactly when \(G\) has an independent set of size at least \(W\).
We output \(G\) and threshold \(W\).

\paragraph{Parameter bound.}
By \Cref{lem:is-rank-width-order}, the linear rank-width \(p\) is
\(O(d)\). Since \(d=\lceil\sqrt{m_{\mathrm{var}}}\rceil\) and
\(m_{\mathrm{var}}\le k\lceil\log_2 n\rceil\), we have
\(p^2=O(k\log_2 n+1)\).

\paragraph{Construction time.}
The construction takes \(2^{O(d)}\) time times a fixed polynomial
in the formula length. The inequality
\(2\sqrt{k\lceil\log_2 n\rceil}\le k+\lceil\log_2 n\rceil\)
shows that this is fixed-parameter time.
Apply \Cref{lem:time-transfer} with \(\varphi(p)=p^2\).
\end{proof}

\section{\problem{Permutation Hitting} by grid order}\label{sec:permutation-hitting}

\problembox{\problem{Permutation Hitting}}{A positive integer \(D\) and an explicit family of subsets of \([D]\times[D]\).}{Is there a permutation \(\pi\) of \([D]\) such that the cells \((i,\pi(i))\), for \(i\in[D]\), hit every set in the family?}{The grid order \(D\).}

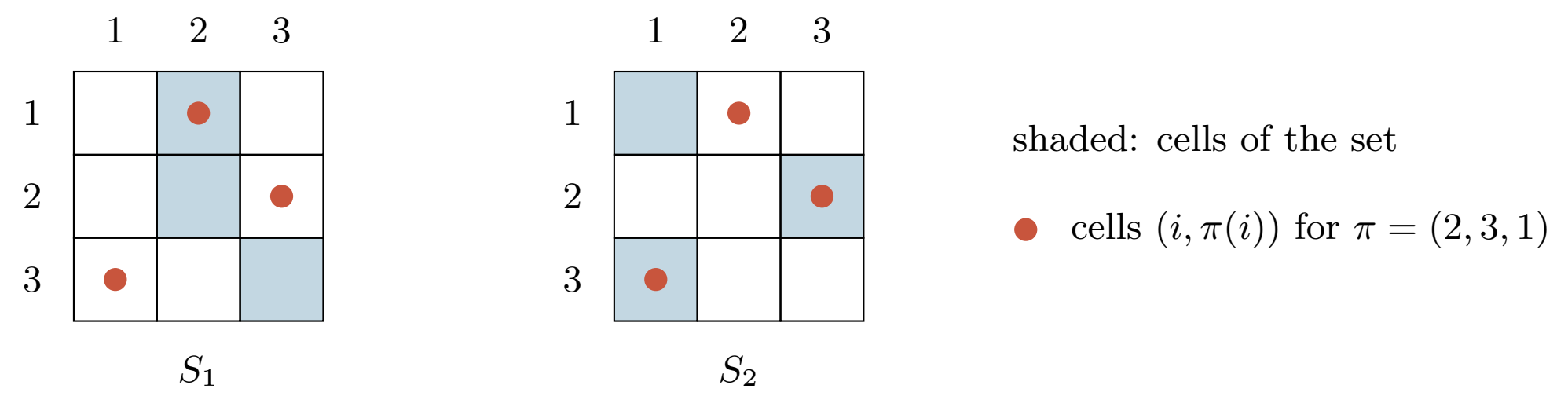
\begin{figure}[t]
\centering
\begin{tikzpicture}[x=0.7cm,y=0.7cm]
  \foreach \g/\ox/\cells in {1/0/{1/2,2/2,3/3},2/6.5/{1/1,2/3,3/1}} {
    \begin{scope}[shift={(\ox,0)}]
      \foreach \i/\j in \cells
        \fill[figB!28] (\j-0.5,-\i+0.5) rectangle (\j+0.5,-\i-0.5);
      \foreach \i in {1,2,3} \foreach \j in {1,2,3}
        \draw (\j-0.5,-\i+0.5) rectangle (\j+0.5,-\i-0.5);
      \foreach \i/\j in {1/2,2/3,3/1}
        \node[circle,fill=figA,inner sep=0pt,minimum size=5.5pt] at (\j,-\i) {};
      \foreach \k in {1,2,3} {
        \node[flab] at (\k,0) {\(\k\)};
        \node[flab] at (0,-\k) {\(\k\)};
      }
      \node[flab] at (2,-4.1) {\(S_{\g}\)};
    \end{scope}
  }
  \node[flab,anchor=west,align=left] at (11.6,-1.3) {shaded: cells of the set};
  \node[circle,fill=figA,inner sep=0pt,minimum size=5.5pt] at (11.95,-2.4) {};
  \node[flab,anchor=west,align=left] at (12.3,-2.4) {cells \((i,\pi(i))\) for \(\pi=(2,3,1)\)};
\end{tikzpicture}
\caption{A \problem{Permutation Hitting} instance of order \(D=3\) with the two sets \(S_1=\{(1,2),(2,2),(3,3)\}\) and \(S_2=\{(1,1),(2,3),(3,1)\}\). The permutation \(\pi\) with \(\pi(1)=2\), \(\pi(2)=3\), and \(\pi(3)=1\) hits both sets.}
\label{fig:ph-instance}
\end{figure}

\Cref{fig:ph-instance} illustrates how a permutation can hit every
set in an instance of grid order three.

\begin{theorem}\label{thm:permutation-hitting}
Assume \(\mathrm{FPT}\ne\mathrm{W}[1]\).
There is no deterministic \(2^{o_{\mathrm{eff}}(D\log D)}N^{O(1)}\)-time
algorithm for \problem{Permutation Hitting}, where \(D\) is the grid order.
\end{theorem}

We reduce from \problem{Clique} by encoding its CNF formula as a
family of grid instances.
Cygan et al.~\cite[Theorem~C.2]{ccpre} observed that the proof of
Lokshtanov, Marx, and Saurabh~\cite[Theorem~2.12]{lms} also gives
an ETH lower bound for \problem{Permutation Hitting}, ruling out
\(2^{o(D\log D)}N^{O(1)}\) time even when each input set contains
at most one cell per row. The sets produced by our construction can
contain several cells in a row. This difference raises the question
of whether the historical input restriction can be preserved under
the new hypothesis. Does \(\mathrm{FPT}\ne\mathrm{W}[1]\) also rule out
\(2^{o_{\mathrm{eff}}(D\log D)}N^{O(1)}\) time when each input set
contains at most one cell per row?

We encode the \problem{Clique} instance as a formula with \(m\)
variables. We group these variables so that a choice of one cell in each
row specifies an assignment. A cell represents an assignment to that
row's group, and each clause gives a set of cells whose assignments
make one of its literals true. Thus choosing one cell per row and
hitting all the sets is equivalent to satisfying the formula. Using
groups of \(\lfloor(\log_2 m)/2\rfloor\) variables keeps both the
number of groups and the number of assignments per group within
\(O(m/\log m)\).

Different rows may choose the same column, so a satisfying selection
need not be a permutation. We use the coloring construction of
Lokshtanov, Marx, and Saurabh~\cite[Lemma~2.8]{lms} and use the colors
as new columns. For any satisfying selection, some
coloring gives the chosen cells distinct colors and makes each chosen
color unique in its row. We retain only cells whose color is unique in
their row. This lets us recover the original cell from its new column,
so any permutation that hits the recolored sets gives a satisfying
selection in the original grid.

The four later reductions start from these grid instances and bound
their target parameters in terms of the grid order.

\subsection{Reduction from \problem{Clique}}

We first encode the formula as a choice of one cell per row, then use
colorings to turn satisfying choices into permutations. The following
lemma records the grid order and construction time for this reduction
and its later applications.

\begin{lemma}\label{lem:permutation-hitting}
Given a \problem{Clique} instance on \(n\ge2\) vertices with requested
size \(k\), let \(m=k\lceil\log_2 n\rceil\). A deterministic disjunctive reduction
either decides the instance in polynomial time or, for \(m\ge16\), outputs a family of
\problem{Permutation Hitting} instances of order \(O(m/\log m)\).
In the latter case, the original instance is a yes-instance if and
only if at least one output instance is a yes-instance. Each output
instance has length polynomial in \(n\) and \(m\). The total construction time is
\(f(k)n^{O(1)}\) for a computable \(f\).
\end{lemma}

\begin{proof}
Decide trivial \problem{Clique} inputs directly. Otherwise apply the
Boolean encoding from Section~\ref{sec:preliminaries}. The encoding
gives a CNF formula with \(m=k\lceil\log_2 n\rceil\) variables and
\(O(kn+k^2n^2)\) clauses, each of length at most \(2\lceil\log_2 n\rceil\).
The encoding length and construction time are polynomial in \(n\)
and \(m\), with fixed exponents.

If \(m<16\), test all assignments. Reject if the formula contains
an empty clause, and accept if it has no clauses. Otherwise, apply
the grouping of \Cref{sec:preliminaries}
with \(s\) groups of \(b\) bits each and \(q=2^b\) assignments
per group. Let \(d=\max\{s,q\}\). In row \(i\) of a \(d\times d\) table, the first \(q\) columns encode
assignments to group \(i\), for each \(i\le s\). Require a row selection to
hit the set of these \(q\) cells for each real row. For each clause, require
it to hit the set of all real cells whose group assignment satisfies a
literal of the clause. Thus a mapping selecting one column per row hits all
sets exactly when its real rows encode a satisfying assignment. Dummy
rows have no domain constraint. The set family is explicit and has polynomial
size in the formula length and \(m\).
\Cref{fig:ph-table} shows a toy example.

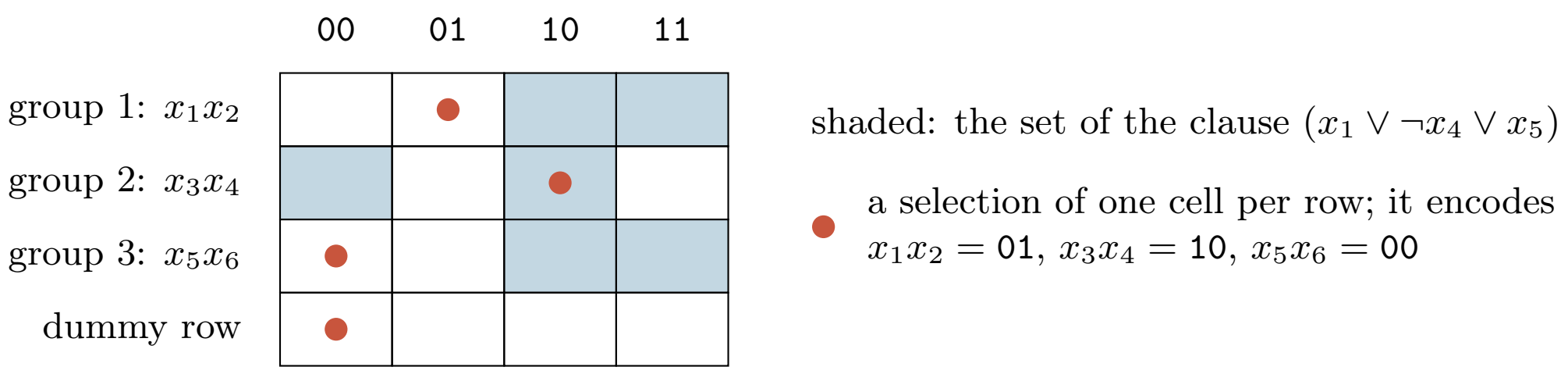
\begin{figure}[t]
\centering
\begin{tikzpicture}[x=0.95cm,y=0.62cm]
  \foreach \i/\j in {1/3,1/4,2/1,2/3,3/3,3/4}
    \fill[figB!28] (\j-0.5,-\i+0.5) rectangle (\j+0.5,-\i-0.5);
  \foreach \i in {1,2,3,4} \foreach \j in {1,2,3,4}
    \draw (\j-0.5,-\i+0.5) rectangle (\j+0.5,-\i-0.5);
  \foreach \j/\a in {1/00,2/01,3/10,4/11} \node[flab] at (\j,0.1) {\ttfamily\a};
  \foreach \i/\l in {1/{group 1: \(x_1x_2\)},2/{group 2: \(x_3x_4\)},3/{group 3: \(x_5x_6\)},4/{dummy row}}
    \node[flab,anchor=east] at (0.3,-\i) {\l};
  \foreach \i/\j in {1/2,2/3,3/1,4/1}
    \node[circle,fill=figA,inner sep=0pt,minimum size=5.5pt] at (\j,-\i) {};
  \node[flab,anchor=west,align=left] at (5.1,-1.2) {shaded: the set of the clause \((x_1\lor\neg x_4\lor x_5)\)};
  \node[circle,fill=figA,inner sep=0pt,minimum size=5.5pt] at (5.35,-2.6) {};
  \node[flab,anchor=west,align=left] at (5.6,-2.6) {a selection of one cell per row; it encodes\\
     \(x_1x_2=\mathtt{01}\), \(x_3x_4=\mathtt{10}\), \(x_5x_6=\mathtt{00}\)};
\end{tikzpicture}
\caption{A toy example of the table and of a clause set, with \(b=2\), \(q=4\), \(s=3\), and \(d=4\). These values illustrate the encoding. They do not occur together under the choice of \(b\) and \(s\) in the proof. The first \(q\) cells of a real row correspond to the assignments to its group. A selection of one cell per row hits the set of a clause exactly when the encoded assignment satisfies the clause. Here the selection hits the shaded set in the row of group~\(2\).}
\label{fig:ph-table}
\end{figure}

Use the coloring construction of Lokshtanov, Marx, and
Saurabh~\cite[Lemma~2.8]{lms}. Its maps have the form
\(f\colon[d]\times[d]\to[d+1]\). For any mapping selecting one cell
per row, some \(f\) gives the selected cells distinct colors and gives each
selected cell a color absent from the other cells of its row. Indeed, the
graph formed by a clique on the selected cells and the stars joining each
selection to its row is the cactus-grid graph of that lemma.

To enumerate the maps, precompute a perfect hash family from \([d]\)
to \([r]\) for each \(1\le r\le d\). By Alon and Gutner~\cite{ag-hash},
each family has size \(2^{O(r\log\log(\max\{r,4\}))}\log d\) and can
be constructed deterministically in
\(2^{O(r\log\log(\max\{r,4\}))}d\log d\) time. Enumerate the occupied
columns and their
positive occupancies \(r_1,\ldots,r_a\), which sum to \(d\).
There are \(2^{O(d)}\) such choices. For each choice, enumerate
the product of the corresponding hash families, one function per
occupied column. As \(\sum_i r_i=d\) and \(a\le d\), the product
has size \(2^{O(d\log\log(\max\{d,4\}))}\). Each tuple specifies a
map using disjoint color ranges for the occupied columns and one
extra color for the other columns. Writing each map takes
polynomial time in \(d\), so the full enumeration takes
\(2^{O(d\log\log(\max\{d,4\}))}d^{O(1)}\) time.

For each \(f\), retain the cells whose colors are unique within their rows, and
replace the column of every retained cell by its color. Apply this operation to
every set. For each of the first \(d\) rows, add the set of images of all
retained cells in that row.
Add the unconstrained row \(d+1\) to make a square of order \(D=d+1\).
Discard any output instance containing an empty set, since no permutation can hit it.

\paragraph{Completeness.}
A clique gives a choice of one cell in each of the first \(d\) rows that hits every
original set. Choose a coloring \(f\) that retains all the chosen
cells and gives them distinct colors. If the chosen cell in row
\(i\) is \((i,j)\), assign color \(f(i,j)\) to row \(i\). Assign the
one unused color to row \(d+1\) to obtain a permutation.
For each original set, a chosen cell that hits it is retained, so
its image hits the transformed set. The permutation also hits the
added set for each of the first \(d\) rows, since its chosen color comes from
a retained cell in that row.

\paragraph{Soundness.}
Take a permutation that hits every set in an output instance. The added set
for each of the first \(d\) rows forces its chosen color to come from a retained
cell in that row. There is exactly one such cell because the
construction retains only cells whose colors are unique in their
rows. Recover these cells and ignore row \(d+1\). Whenever
the permutation hits a transformed set, the recovered cell belongs
to the original set. The recovered choices therefore hit every
original set and encode a satisfying assignment, which gives a
clique of the requested size.

\paragraph{Parameter bound.}
The initial grid has order \(d=\max\{s,q\}\), where
\(s=\lceil m/b\rceil=O(m/\log m)\) is the number of groups
and \(q=2^b\le\sqrt m\) is the number of assignments per group.
The extra row and the \(d+1\) colors give output grids of order
\(D=d+1=O(m/\log m)\).

\paragraph{Construction time.}
The number of queries and the time
to enumerate the family are \(2^{o(m)}\) up to a fixed polynomial factor in
the explicit formula length, with a computable sublinear exponent bound
\(O(m\log\log m/\log m)\). Its factor \(2^{O(m\log\log m/\log m)}\)
is at most \(2^{m/k}\le 2n\) beyond a computable cutoff in \(m\)
depending on \(k\). Below the cutoff it is bounded by a computable
factor depending only on \(k\). The polynomial in the formula
length remains a fixed polynomial in \(n\) and \(m\), proving the
claimed construction time.
\end{proof}

\subsection{Putting the reduction together}

We now use the grid-order bound to check that \(D\log D\) is at most a
constant times the number of variables in the \problem{Clique} encoding.
This gives the parameter bound needed for the running-time transfer.

\begin{proof}[Proof of \Cref{thm:permutation-hitting}]
Apply the disjunctive reduction of \Cref{lem:permutation-hitting}.
For each output instance, \(D=O(m/\log m)\), where
\(m=k\lceil\log_2 n\rceil\). Thus \(\Phi(D)=O(k\log n)\).
The reduction runs in fixed-parameter time, so
\Cref{lem:time-transfer} applies with \(\varphi=\Phi\).
\end{proof}

\section{\problem{Vertex-Disjoint Paths} by the width of a supplied path decomposition}

\problembox{\problem{Vertex-Disjoint Paths}}%
{A simple undirected graph \(G\), a path decomposition of \(G\),
and an explicit list of demand pairs of vertices.}%
{Is there one path for each demand pair, joining its endpoints, with all
paths pairwise vertex-disjoint, including at their endpoints?}%
{The width \(p\) of the supplied path decomposition.}

\begin{theorem}\label{vdp-width-w1}\hypertarget{vdp-width-w1}{}
Assume \(\mathrm{FPT}\ne\mathrm{W}[1]\).
There is no deterministic \(2^{o_{\mathrm{eff}}(p\log p)}N^{O(1)}\)-time algorithm
for \problem{Vertex-Disjoint Paths} parameterized by the width \(p\)
of a supplied path decomposition. This holds on connected bipartite
graphs of maximum degree at most three with distinct leaf terminals.
\end{theorem}

We reduce from \problem{Clique} via \problem{Permutation Hitting}.
We use the directed construction of Lokshtanov, Marx, and
Saurabh~\cite[Section~5]{lms}, with one gadget for each set in the
\problem{Permutation Hitting} instance. In each row, a path chooses a column by passing through a
vertex shared by all paths that could choose that column. Disjointness
forces the rows to choose distinct columns, so their choices define a
permutation. The row path and an auxiliary path leave only the chosen
cell available to a test path. The test path can be routed exactly when
some chosen cell belongs to the set.

We identify the output vertices of one gadget with the input vertices
of the next. Only the outputs at the chosen cells are free, so the next
gadget must use the same permutation. Processing the gadgets row by row
gives a path decomposition of width \(O(D)\), where \(D\) is the grid
order. We adapt their use of auxiliary demands to simulate direction
in a graph of maximum degree three with distinct leaf terminals.
Subdividing each edge once makes the graph bipartite and increases the width of the supplied path decomposition by at most one.

\subsection{Testing the same permutation in every set}

We first describe the directed gadget of Lokshtanov, Marx, and
Saurabh~\cite[Section 5]{lms} for testing whether a permutation hits one
set of a \problem{Permutation Hitting} instance from
\Cref{lem:permutation-hitting}. We then join the gadgets so that every
set is tested against the same permutation, and construct a path
decomposition for the resulting graph.

For a set \(S\subseteq [D]\times[D]\), make vertices \(a_i,b_j,u_{i,j}\) and
arcs \(a_i\to u_{i,j}\to b_j\). For each row \(i\), make a directed path
\[
 P_i=(v_{i,0},d_{i,1},w_{i,1},v_{i,1},\dots,d_{i,D},w_{i,D},v_{i,D}).
\]
For each cell \((i,j)\), add vertices \(q_{i,j},\ell_{i,j},r_{i,j}\)
and arcs
\[
 b_j\to q_{i,j}\to v_{i,j},\qquad
 \ell_{i,j}\to q_{i,j}\to r_{i,j},\qquad
 \ell_{i,j}\to v_{i,0},\qquad v_{i,j-1}\to r_{i,j}.
\]
Make a source \(s\) and a sink \(t\), and add \(s\to d_{i,j}\to t\)
for \((i,j)\in S\). The demands are \((a_i,v_{i,D})\) for each row,
\((\ell_{i,j},r_{i,j})\) for each cell, and \((s,t)\). All demand endpoints
are distinct. Each source has indegree zero and each sink has outdegree zero.

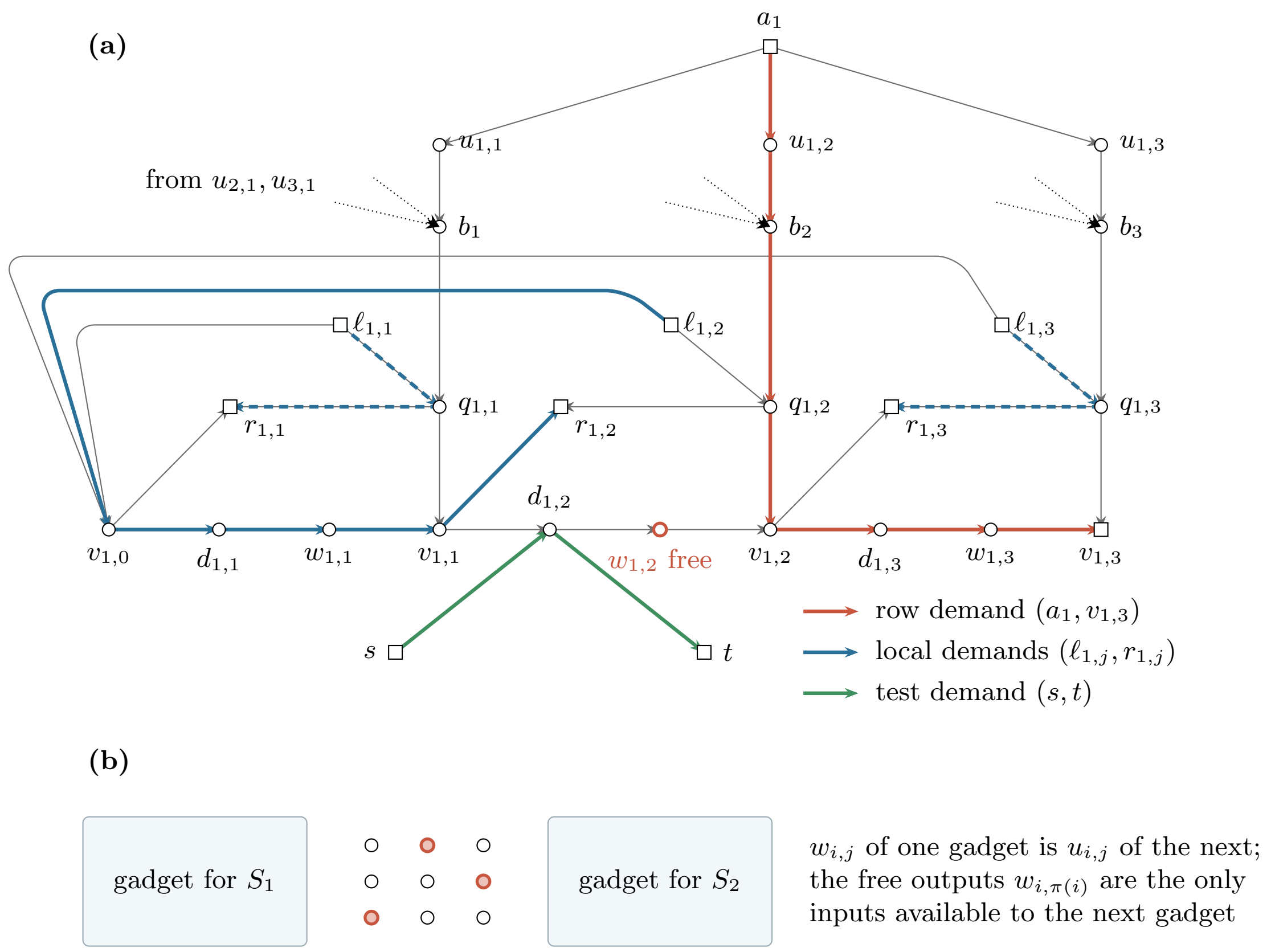
\begin{figure}[t]
\centering
\begin{tikzpicture}[x=1.28cm,y=0.95cm,>={Stealth[length=4.5pt]},
    pv/.style={fv,minimum size=4.2pt}]
  \node[flab,anchor=west] at (-0.3,5.9) {\textbf{(a)}};
  \foreach \n/\x in {c0/0,d1/1,w1/2,c1/3,d2/4,w2/5,c2/6,d3/7,w3/8,c3/9}
    \coordinate (\n) at (\x,0);
  \foreach \j in {1,2,3} {
    \coordinate (b\j) at (3*\j,3.7);
    \coordinate (f\j) at (3*\j,1.5);
    \coordinate (l\j) at (3*\j-0.9,2.5);
    \coordinate (r\j) at (3*\j-1.9,1.5);
    \coordinate (u\j) at (3*\j,4.7);
  }
  \coordinate (a) at (6,5.9);
  \coordinate (s) at (2.6,-1.5);
  \coordinate (t) at (5.4,-1.5);
  \begin{scope}[figgray!80!black]
    \draw[->] (c0)--(d1); \draw[->] (d1)--(w1); \draw[->] (w1)--(c1); \draw[->] (c1)--(d2);
    \draw[->] (d2)--(w2); \draw[->] (w2)--(c2); \draw[->] (c2)--(d3); \draw[->] (d3)--(w3); \draw[->] (w3)--(c3);
    \foreach \j in {1,2,3} {
      \draw[->] (b\j)--(f\j); \draw[->] (f\j)--(c\j);
      \draw[->] (l\j)--(f\j); \draw[->] (f\j)--(r\j);
      \draw[->] (u\j)--(b\j);
    }
    \draw[->] (c0)--(r1); \draw[->] (c1)--(r2); \draw[->] (c2)--(r3);
    \draw[->][rounded corners=7pt] (l1) -- (-0.32,2.5) -- (c0);
    \draw[->][rounded corners=7pt] (l2) -- ++(-0.4,0.42) -- (-0.64,2.92) -- (c0);
    \draw[->][rounded corners=7pt] (l3) -- ++(-0.4,0.84) -- (-0.96,3.34) -- (c0);
    \draw[->] (a) -- (u1);
    \draw[->] (a) -- (u2);
    \draw[->] (a) -- (u3);
    \draw[->] (s)--(d2); \draw[->] (d2)--(t);
  \end{scope}
  \draw[hiA,->] (a) -- (u2);
  \draw[hiA,->] (u2)--(b2); \draw[hiA,->] (b2)--(f2); \draw[hiA,->] (f2)--(c2);
  \draw[hiA,->] (c2)--(d3); \draw[hiA,->] (d3)--(w3); \draw[hiA,->] (w3)--(c3);
  \draw[hiB,->][rounded corners=7pt] (l2) -- ++(-0.4,0.42) -- (-0.64,2.92) -- (c0);
  \draw[hiB,->] (c0)--(d1); \draw[hiB,->] (d1)--(w1); \draw[hiB,->] (w1)--(c1); \draw[hiB,->] (c1)--(r2);
  \foreach \j in {1,3} { \draw[hiB,->,densely dashed] (l\j)--(f\j); \draw[hiB,->,densely dashed] (f\j)--(r\j); }
  \draw[hiC,->] (s)--(d2); \draw[hiC,->] (d2)--(t);
  \foreach \n/\l/\p in {d1/{d_{1,1}}/below,w1/{w_{1,1}}/below,c1/{v_{1,1}}/below,d2/{d_{1,2}}/above,
                        c2/{v_{1,2}}/below,d3/{d_{1,3}}/below,w3/{w_{1,3}}/below}
    \node[pv,label={[flab]\p:\(\l\)}] at (\n) {};
  \node[pv,label={[flab]below:\(v_{1,0}\)}] at (c0) {};
  \node[pv,line width=1pt,draw=figA,label={[flab,text=figA]below:\(w_{1,2}\) free}] at (w2) {};
  \node[fterm,label={[flab]below:\(v_{1,3}\)}] at (c3) {};
  \foreach \j in {1,2,3} {
    \node[pv,label={[flab]right:\(b_{\j}\)}] at (b\j) {};
    \node[pv,label={[flab]right:\(q_{1,\j}\)}] at (f\j) {};
    \node[fterm,label={[flab,inner sep=1.5pt]right:\(\ell_{1,\j}\)}] at (l\j) {};
    \node[fterm,label={[flab,anchor=north west,inner sep=2pt]below right:\(r_{1,\j}\)}] at (r\j) {};
    \node[pv,label={[flab]right:\(u_{1,\j}\)}] at (u\j) {};
    \draw[densely dotted,->] (3*\j-0.6,4.3) -- (b\j);
    \draw[densely dotted,->] (3*\j-0.95,4.0) -- (b\j);
  }
  \node[fterm,label={[flab]above:\(a_1\)}] at (a) {};
  \node[flab,anchor=east] at (2.0,4.25) {from \(u_{2,1},u_{3,1}\)};
  \node[fterm,label={[flab]left:\(s\)}] at (s) {};
  \node[fterm,label={[flab]right:\(t\)}] at (t) {};
  \begin{scope}[shift={(6.3,-1.25)}]
    \draw[hiA,->] (0,0.25) -- (0.5,0.25); \node[flab,anchor=west] at (0.55,0.25) {row demand \((a_1,v_{1,3})\)};
    \draw[hiB,->] (0,-0.25) -- (0.5,-0.25); \node[flab,anchor=west] at (0.55,-0.25) {local demands \((\ell_{1,j},r_{1,j})\)};
    \draw[hiC,->] (0,-0.75) -- (0.5,-0.75); \node[flab,anchor=west] at (0.55,-0.75) {test demand \((s,t)\)};
  \end{scope}
  \begin{scope}[shift={(0,-4.3)},x=1cm,y=0.42cm]
    \node[flab,anchor=west] at (-0.38,3.3) {\textbf{(b)}};
    \node[fnode,minimum width=2.6cm,minimum height=1.5cm] (g1) at (1.0,0) {gadget for \(S_1\)};
    \node[fnode,minimum width=2.6cm,minimum height=1.5cm] (g2) at (6.4,0) {gadget for \(S_2\)};
    \foreach \i in {1,2,3} \foreach \j in {1,2,3}
      \node[pv] (x\i\j) at (2.4+0.65*\j,2-\i*1) {};
    \foreach \i/\j in {1/2,2/3,3/1}
      \node[pv,line width=1pt,draw=figA,fill=figA!35] at (x\i\j) {};
    \node[flab,anchor=west,align=left] at (8.0,0) {\(w_{i,j}\) of one gadget is \(u_{i,j}\) of the next;\\
        the free outputs \(w_{i,\pi(i)}\) are the only\\ inputs available to the next gadget};
  \end{scope}
\end{tikzpicture}
\caption{The directed construction for \problem{Vertex-Disjoint Paths} on the instance of \Cref{fig:ph-instance}. (a)~Row~\(1\) of the gadget for \(S_1\). Squares are demand endpoints. The vertices \(b_j\), \(s\), and \(t\) are shared by all rows. The thick arcs show the routes for \(\pi(1)=2\): the row path enters through \(u_{1,2}\) and \(b_2\) and uses the suffix of the row, the local demand of the cell \((1,2)\) uses the prefix, and the test path can pass through \(d_{1,2}\) since \((1,2)\in S_1\). The remaining local demands are routed directly (dashed). (b)~Consecutive gadgets share the vertices \(w_{i,j}=u_{i,j}\).}
\label{fig:vdp}
\end{figure}

A row path choosing \(u_{i,j}\) must follow \(b_j,q_{i,j}\) into its own
row and then traverse the suffix from \(v_{i,j}\) to \(v_{i,D}\). Any other
route from \(b_j\) would enter a different row or end at another demand's
terminal. Distinct row paths use distinct vertices \(b_j\). Their choices
therefore define a permutation \(\pi\).
The local demand for \((i,\pi(i))\) must follow the prefix through
\(v_{i,\pi(i)-1}\) since its \(q\)-vertex is occupied. The prefix and suffix
together occupy all \(d_{i,j}\) and \(w_{i,j}\) except those at
\(j=\pi(i)\). The test path must start through an unoccupied \(d_{i,j}\),
so a hit exists. Conversely, route the rows and the special local demands
as above, and route all other local demands directly through their own
\(q\)-vertices. The test path uses \(s\to d_{i,\pi(i)}\to t\) for a hit.
These routes are disjoint and leave exactly \(w_{i,\pi(i)}\) free among
the output vertices of each row.
\Cref{fig:vdp}(a) shows one row of the gadget together with these routes.

For the full set family, make one such graph per set. Identify each
\(w_{i,j}\) of a graph with \(u_{i,j}\) of the next graph. Retain all demands.
See \Cref{fig:vdp}(b).
Identifying consecutive gadgets forces the same permutation to be
used for every set.

\begin{lemma}\label{lem:vdp-permutation}
The constructed directed graph admits vertex-disjoint paths for all demands
if and only if some permutation hits every set of the
\problem{Permutation Hitting} instance.
\end{lemma}

\begin{proof}\leavevmode
\paragraph{Completeness.}
Let \(\pi\) be a permutation that hits every set, and use the local
routes for \(\pi\) in every gadget. These routes are vertex-disjoint
within each gadget. Consecutive gadgets share only the vertices
obtained by identifying each output \(w_{i,j}\) of the first gadget
with the corresponding input \(u_{i,j}\) of the second. In row \(i\),
the routes of the second gadget use only the input \(u_{i,\pi(i)}\).
The corresponding output \(w_{i,\pi(i)}\) is unused by the routes of
the first gadget. Thus no shared vertex is used by routes from both
gadgets, and the routes together satisfy all demands.

\paragraph{Soundness.}
Suppose the directed graph admits vertex-disjoint paths for all demands.
Number the gadgets in their order along the construction. Each arc
belongs to one gadget, and a path can switch between gadgets only at a
shared vertex. We first show that a path for a demand of gadget \(g\)
cannot use an arc of a later gadget. If it did, it would eventually have
to take an arc of gadget \(h-1\) immediately after an arc of gadget
\(h\), for some \(h>g\), to reach its terminal in gadget \(g\).
The switch would occur at an input of gadget \(h\). Every arc into
such an input within gadget \(h\) comes from a row source of that
gadget. A path starting in gadget \(g\) cannot reach this source,
which has indegree zero in the full graph.

We now show by induction on \(g\) that no path for a demand of
gadget \(g\) uses an arc of an earlier gadget. For \(g=1\) there is
no earlier gadget. Suppose all paths for demands of gadget \(g-1\)
stay in that gadget. By the single-gadget argument, in every row its
row path and the local path for the chosen cell occupy all outputs
except the one for the chosen column. If a path for a demand of
gadget \(g\) first takes an arc of gadget \(g-1\), it must switch at
that sole available output in some row. Within gadget \(g-1\), it
can only follow the suffix of that row or reach a demand sink.
The suffix reaches no input of gadget \(g-1\), and its only
possible switches back to gadget \(g\) are at outputs of the same
row. Every other such output is occupied by the paths for demands
of gadget \(g-1\). Returning at the original output would repeat
a vertex, so the path cannot reach its terminal in gadget \(g\)
after making this switch. This completes the induction.

The paths for each gadget therefore determine a permutation
hitting its set. In each row, the input used by the row path of
the next gadget must be the sole available output of the
preceding gadget. Hence the permutations agree in every row
and hit every set.
\end{proof}

We construct a path decomposition of the underlying undirected graph.
Let \(M\) be the number of gadgets and let \(K_g\) consist of the
\(a_i\) and \(b_j\) vertices of gadget \(g\). Set \(K_{M+1}=\emptyset\).
Use the following bags, in order.
\begin{enumerate}
\item Start with \(K_1\), then use one bag \(K_1\cup\{u_{i,j}\}\)
for each input of the first gadget. These bags cover its input arcs.
\item For gadget \(g\), use the base bag \(K_g\cup K_{g+1}\cup\{s,t\}\),
where \(s,t\) are its test terminals. For each row \(i\), add all
vertices of \(P_i\) to the base bag. Use that row bag followed by
one copy per column \(j\) with \(q_{i,j},\ell_{i,j},r_{i,j}\) added.
This covers the row path, the test arcs, and the arcs incident with
the three added vertices. When \(g<M\), the row bag also covers
the two input arcs of gadget \(g+1\) at each shared vertex
\(w_{i,j}=u_{i,j}\), because it contains \(K_{g+1}\).
\item After the last row of gadget \(g\), use the bag \(K_{g+1}\)
before the base bag of gadget \(g+1\), if that gadget exists.
\end{enumerate}
Each shared input is introduced in the preceding gadget's row block
and needs no bag in its own gadget block. The vertices of \(K_g\)
appear throughout their gadget block and the preceding block, with
the intervening transition bag. All other vertices occur within the
initial input bags, one gadget block, or one row block, as specified
above. Thus each vertex appears in a consecutive interval of bags.
The largest bags have at most \((4D+2)+(3D+1)+3=7D+6\) vertices.
The graph and the bag sequence have polynomial size in the table
and the set list.

\subsection{Simulating direction on a subcubic graph}\label{sec:vdp-direction}

We convert the directed instance to an equivalent undirected instance
with maximum degree three and distinct leaf terminals. We first bound
the indegree and outdegree of the directed graph, then use auxiliary
demands to enforce the traversal directions.

Let \(p_0\) be the width of the directed instance's path decomposition.
We first modify the digraph so every vertex has indegree and outdegree at most
two. Assign each arc to a bag containing both endpoints and order the arcs by
bag and a fixed tie order. Replace each vertex by a center with a directed
chain of incoming ports leading to it and a directed chain of outgoing ports
leading away from it. Order ports in each chain by their assigned arcs.
Every original arc now connects its outgoing port at its tail to its incoming
port at its head. Place the demands on the corresponding centers. A traversal
from an incoming incidence to an outgoing one uses the center, so directed
routes expand and contract to give equivalent solutions. Any route that
starts or ends at a demanded center also uses that center. A source center
has indegree zero and at most one outgoing arc. A sink center has outdegree
zero and at most one incoming arc.

To construct a path decomposition, keep every center throughout its original bag interval.
Process the assigned arcs at their bags. For each live center and each port
chain retain the last introduced port until the next port of that chain
arrives, or until the final incident chain edge is covered. Introduce the
two fresh ports for an arc together to cover its replacement edge and the
chain edges to retained predecessors. The first outgoing port and the last
incoming port meet the center. Each bag contains at most three retained
vertices per old live vertex and two new ports. Every port occurrence is
consecutive. This gives a polynomial-length path decomposition of width at most
\(3(p_0+1)+1\).

\begin{figure}[t]
\centering
\begin{tikzpicture}[x=1cm,y=1cm,>={Stealth[length=4.5pt]}]
  \node[flab,anchor=west] at (-0.6,2.3) {\textbf{(a)}};
  \node[fv,label={[flab]left:\(u_1\)}] (u1) at (0,1.2) {};
  \node[fv,label={[flab]left:\(u_2\)}] (u2) at (0,-1.2) {};
  \node[fv,label={[flab]above:\(v\)}] (v) at (2.0,0) {};
  \draw[->] (u1) -- (v); \draw[->] (u2) -- (v); \draw[->] (v) -- ++(1.1,0);
  \begin{scope}[shift={(5.6,0)}]
    \node[flab,anchor=west] at (-1.2,2.3) {\textbf{(b)}};
    \foreach \i/\y/\pos in {1/1.4/above,2/-1.4/below} {
      \node[fv,label={[flab]left:\(u_{\i}^-\)}] (um\i) at (-0.4,\y) {};
      \node[fv,label={[flab]\pos:\(u_{\i}^+\)}] (up\i) at (0.8,\y) {};
      \node[fv,label={[flab]\pos:\(p_{\i}\)}] (p\i) at (2.6,\y) {};
      \node[fv,label={[flab]\pos:\(q_{\i}\)}] (q\i) at (4.4,\y) {};
      \draw (um\i) -- (up\i) -- (p\i) -- (q\i);
    }
    \node[fv,label={[flab]below:\(v^-\)}] (vm) at (6.2,0) {};
    \node[fv,label={[flab]below:\(v^+\)}] (vp) at (7.4,0) {};
    \draw (q1) -- (vm) -- (vp); \draw (q2) -- (vm);
    \draw (vp) -- ++(0.8,0);
    \node[fvb,label={[flab]right:\(a\)}] (a) at (2.6,0) {};
    \node[fvb,label={[flab]left:\(b\)}] (b) at (4.4,0) {};
    \node[fterm,label={[flab]left:\(a'\)}] (ap) at (1.7,0) {};
    \node[fterm,label={[flab]right:\(b'\)}] (bp) at (5.3,0) {};
    \draw (ap) -- (a); \draw (b) -- (bp);
    \draw (a) -- (p1); \draw (a) -- (p2); \draw (b) -- (q1); \draw (b) -- (q2);
    \begin{scope}[on background layer]
      \draw[figA!45,line width=4.5pt,line cap=round,line join=round] (um1) -- (up1) -- (p1) -- (q1) -- (vm) -- (vp) -- ++(0.8,0);
      \draw[figB!45,line width=4.5pt,line cap=round,line join=round] (ap) -- (a) -- (p2) -- (q2) -- (b) -- (bp);
    \end{scope}
  \end{scope}
\end{tikzpicture}
\caption{Simulating direction. (a)~A vertex \(v\) with two incoming arcs. (b)~The undirected replacement. Squares are the endpoints of the auxiliary demand \((a',b')\). If a main route (red) enters \(v^-\) along the arc path from \(u_1\), the auxiliary route (blue) uses the other arc path.}
\label{fig:direction}
\end{figure}
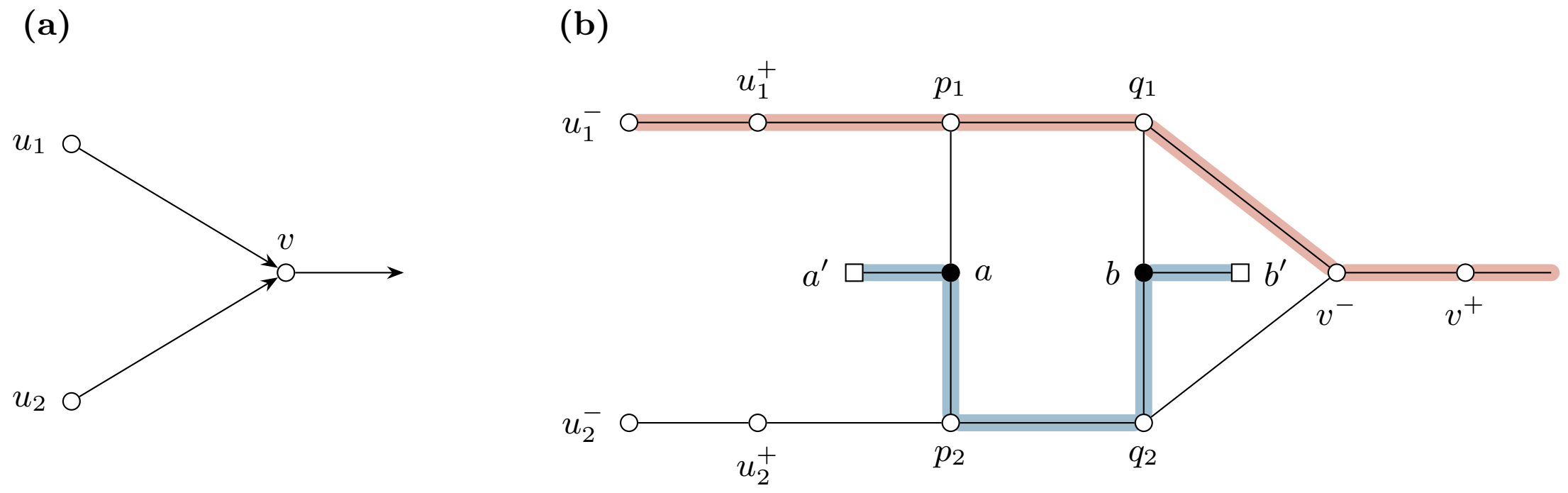

For each split digraph vertex \(v\), make adjacent vertices \(v^-\) and
\(v^+\). Replace an arc \(u\to v\) by an undirected path
\(u^+,p_{uv},q_{uv},v^-\) with private internal vertices. A main demand
from a source \(s\) to a sink \(t\) now joins \(s^-\) to \(t^+\), both
leaves. If \(v\) has two incoming arcs with private pairs \(p_i,q_i\) for
\(i=1,2\), add \(a\) adjacent to both \(p_i\) and \(b\) adjacent to both
\(q_i\). Attach a private leaf \(a'\) to \(a\) and \(b'\) to \(b\), and add
the auxiliary demand \((a',b')\). All vertices have degree at most three,
and all demand endpoints are distinct leaves.
\Cref{fig:direction} illustrates the replacement.

\begin{lemma}\label{lem:vdp-direction}
The constructed undirected graph admits vertex-disjoint paths for all
main and auxiliary demands if and only if the original directed graph
admits vertex-disjoint paths for all its demands.
\end{lemma}

\begin{proof}
The port-chain replacement preserves feasibility, so it suffices to
compare the split directed graph with the undirected graph.

\paragraph{Completeness.}
A directed routing extends to main routes using the natural arc paths.
At most one incoming arc path at each \(v\) is in use, so the auxiliary
route can use \(a',a,p_i,q_i,b,b'\) on a free arc path.

\paragraph{Soundness.}
In an undirected routing for all demands, each auxiliary route uses \(a,b\) and at least one of the two \(p_i\). A main
route using any private pair must traverse the complete arc path because
the only branches lead to the occupied \(a,b\), and neither private vertex
is a terminal. At most one incoming arc path at each \(v\) is therefore
available to the main routes. A main route starts by crossing from its
source leaf to the source's positive copy. After an outgoing arc path
it reaches another negative copy and must cross to the positive copy,
since no other incoming arc path is available. It continues in this way
until its own sink leaf. The projected directed routes are disjoint since
each visited vertex uses both copies.
\end{proof}

Double the path decomposition's bags and insert a local bag adding two
private vertices for each arc at a bag containing its endpoints. At an
indegree-two head, extend the earlier private pair through the local bag
of the later pair. The extension stays inside the interval of the negative
copy of the common head, which is present in both local bags. Each head
requires at most two extended vertices per bag. Where the two pairs meet,
insert copies adding \(a,a'\) and \(b,b'\) in separate bags. This covers
all auxiliary edges and preserves contiguity. The resulting path decomposition has width \(O(p_0+1)\)
and the listed bag size is polynomial in the preceding input.

\subsection{Bipartiteness and connectivity}

We check that the undirected graph is connected, then make it bipartite
by subdividing every edge once. We also show how to extend the path
decomposition to the subdivision vertices.

The underlying graph of each original directed gadget is connected.
The \(a_i,u_{i,j},b_j\) vertices connect the rows and columns, every
row path and local demand attaches to them, and \(s,t\) attach through
a cell of the nonempty set. Consecutive gadgets share vertices.
The direction simulation uses connected replacements.

Subdivide each edge of the resulting undirected graph once. All
original vertices can receive one color and all subdivision vertices
the other. Expanding or contracting the subdivided edges preserves
feasibility, and the graph stays connected and subcubic with distinct
leaf terminals. For each subdivision vertex, insert a copy of a bag
containing the endpoints of the original edge and add the new vertex
to the copy. This increases the width of the path decomposition by at
most one and takes polynomial time.

\subsection{Putting the reduction together}

We combine the directed construction, the direction simulation, and
the subdivision step. The width of the supplied path decomposition
stays linear in the grid order throughout.

\begin{proof}[Proof of \Cref{vdp-width-w1}]
By \Cref{lem:vdp-permutation}, the directed demands are feasible
exactly when the \problem{Permutation Hitting} instance is a
yes-instance. By \Cref{lem:vdp-direction}, the undirected simulation
preserves feasibility, and so does subdivision. We output the
resulting connected bipartite subcubic graph with its distinct leaf
terminals, demands, and path decomposition.

\paragraph{Parameter bound.}
The directed graph has a path decomposition of width \(O(D)\),
where \(D\) is the grid order. Direction simulation and subdivision
give a supplied path decomposition of width \(p=O(D)\). By
\Cref{lem:permutation-hitting}, \(D=O(m/\log m)\), where
\(m=k\lceil\log_2 n\rceil\), so \(\Phi(p)=O(m)\).

\paragraph{Construction time.}
The graph transformations take polynomial time per output of
\Cref{lem:permutation-hitting}, so the total construction takes
fixed-parameter time. Apply \Cref{lem:time-transfer} with
\(\varphi=\Phi\).
\end{proof}

\section{\problem{Cycle Packing} and \problem{Maximum Cycle Cover} by the width of a supplied path decomposition}

We first prove the lower bound for \problem{Cycle Packing}. In
\Cref{sec:maximum-cycle-cover}, we adapt the same construction to
obtain the lower bound for \problem{Maximum Cycle Cover}.

\problembox{\problem{Cycle Packing}}%
{A simple undirected graph \(G\), a path decomposition of \(G\),
and a nonnegative integer \(r\).}%
{Does \(G\) contain at least \(r\) pairwise vertex-disjoint cycles?}%
{The width \(p\) of the supplied path decomposition.}

\hypertarget{cycle-packing-degree-six}{}
\begin{theorem}\label{cycle-packing-degree-six}
Assume \(\mathrm{FPT}\ne\mathrm{W}[1]\).
There is no deterministic \(2^{o_{\mathrm{eff}}(p\log p)}N^{O(1)}\)-time
algorithm for \problem{Cycle Packing} parameterized by the width \(p\)
of a supplied path decomposition. This holds on connected bipartite
graphs.
\end{theorem}

We reduce from \problem{Clique} via \problem{Permutation Hitting}.
We adapt the construction of Cygan et al.~\cite[Appendix C.2 of the
full version]{ccpre} to allow several cells of a set in the same row.
We process the cells of each set one at a time and use sparse capacity
gadgets to construct a path decomposition whose width is linear in the
grid order. Subdividing each edge once gives a simple bipartite graph.

For each \problem{Permutation Hitting} instance from
\Cref{lem:permutation-hitting}, let \(D\) be its grid order and
let~\(m\) be the number of variables in the Boolean encoding.
The graph has one rail per row and one per column. At the start,
cycles pair row rails with column rails, defining a permutation.
Each set gets a stage with a triangle for every cell it contains.
A capacity gadget prevents at least one of these triangles from
being packed as a cycle in the host, forcing a crossing between
its row and column rails.

The packing threshold forces each host cycle to use exactly two
vertices from a designated set. Before a stage, each path carrying a
row-column pair already contains one such vertex. A crossing between
different pairs would bring together two of those vertices and add a
third. The crossing must therefore occur at a cell selected by the
permutation. A second crossing carries the same pairing forward, so
the permutation hits every set.

\subsection{From \problem{Permutation Hitting} to \problem{Cycle Packing}}\label{sec:cp-host}
Consider a \problem{Permutation Hitting} instance with sets
\(S_1,\ldots,S_M\) of cells in a \(D\times D\) grid. A single set \(S_s\)
may contain several cells in the same row, for example both \((i,j)\)
and \((i,j')\) with \(j\ne j'\). The construction below handles such
sets by processing their cells one at a time.

We reject the instance directly if it contains an empty set and accept
it directly if it has no sets. For every other instance, we construct
a simple graph, a threshold \(r\), and a path decomposition of width
\(O(D)\). The graph has \(r\) vertex-disjoint cycles if and only if some
permutation hits every input set. We first build a multigraph, in which
two parallel edges count as a cycle, and then subdivide each edge once
to obtain the simple graph.

\subsubsection{Capacity gadgets}

The capacity gadgets limit how many designated host vertices a
packing can use while attaining the required number of cycles.
Let \(X=\{v_1,\ldots,v_d\}\) be a set of such vertices, called ports.
Introduce \(d-1\) private vertices \(u_1,\ldots,u_{d-1}\), and add
two parallel edges from \(u_i\) to each of \(v_i\) and \(v_{i+1}\).
We call the resulting doubled alternating path a sparse selector.
For \(d=1\), it has no private vertices.

For every \(A\subseteq X\), the selector has a matching of size
\(\min\{|A|,d-1\}\) between \(A\) and the private vertices.
If \(A=X\), match \(u_i\) to \(v_i\). Otherwise, let \(v_j\) be
a port outside \(A\). Match \(u_i\) to \(v_i\) for \(i<j\) and to
\(v_{i+1}\) for \(i\ge j\). This matches every port except \(v_j\),
so every port in \(A\) is matched. Keep only the edges incident with
ports in \(A\). These edges form a matching of size \(|A|\).

Each matching edge gives a two-cycle using its parallel edges.
Thus if host cycles use at most one port, the remaining ports
can support \(d-1\) disjoint two-cycles. This is the largest
possible contribution from the gadget, because every cycle
through it uses a private vertex.

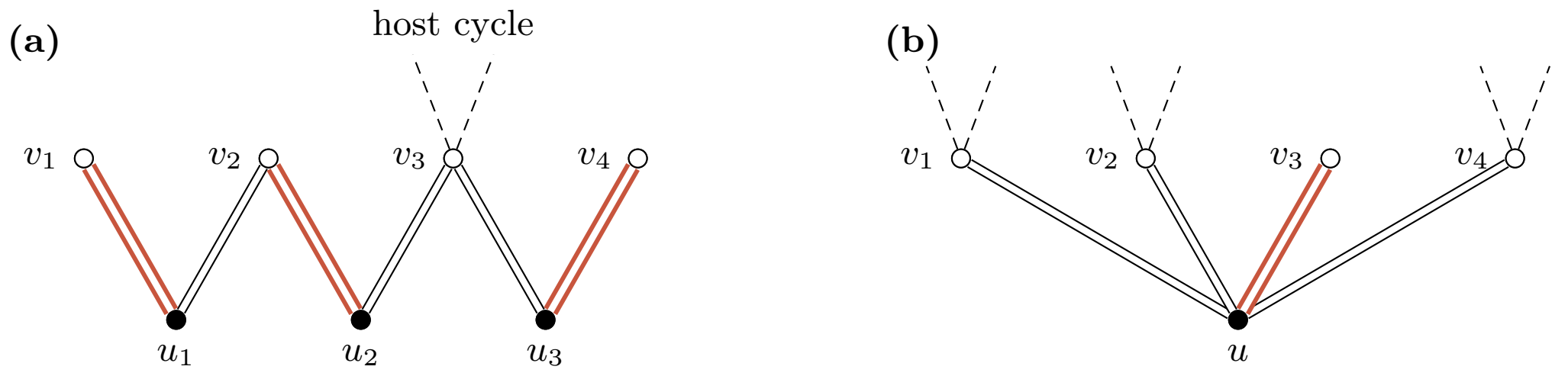
\begin{figure}[t]
\centering
\begin{tikzpicture}[x=1cm,y=1cm,
   dbl/.style={double,double distance=1.6pt}]
  \node[flab,anchor=west] at (-0.8,2.4) {\textbf{(a)}};
  \foreach \i in {1,...,4} \node[fv,label={[flab]left:\(v_{\i}\)}] (v\i) at (1.6*\i-1.6,1.4) {};
  \foreach \i in {1,...,3} \node[fvb,label={[flab]below:\(u_{\i}\)}] (u\i) at (1.6*\i-0.8,0) {};
  \foreach \i/\j in {1/2,2/3,3/3} \draw[dbl] (u\i) -- (v\j);
  \foreach \i/\j in {1/1,2/2,3/4} \draw[dbl,figA,line width=1.1pt] (u\i) -- (v\j);
  \foreach \i in {1,...,4} \node[fv] at (v\i) {};
  \foreach \i in {1,...,3} \node[fvb] at (u\i) {};
  \draw[densely dashed] (v3) -- ++(-0.35,0.9); \draw[densely dashed] (v3) -- ++(0.35,0.9);
  \node[flab] at (3.2,2.55) {host cycle};
  \begin{scope}[shift={(7.6,0)}]
    \node[flab,anchor=west] at (-0.8,2.4) {\textbf{(b)}};
    \foreach \i in {1,...,4} \node[fv,label={[flab]left:\(v_{\i}\)}] (w\i) at (1.6*\i-1.6,1.4) {};
    \node[fvb,label={[flab]below:\(u\)}] (c) at (2.4,0) {};
    \foreach \i in {1,2,4} \draw[dbl] (c) -- (w\i);
    \draw[dbl,figA,line width=1.1pt] (c) -- (w3);
    \foreach \i in {1,...,4} \node[fv] at (w\i) {};
    \node[fvb] at (c) {};
    \foreach \i in {1,2,4} { \draw[densely dashed] (w\i) -- ++(-0.3,0.8); \draw[densely dashed] (w\i) -- ++(0.3,0.8); }
  \end{scope}
\end{tikzpicture}
\caption{The capacity gadgets on \(d=4\) ports. Double lines are pairs of parallel edges, and dashed lines indicate host cycles. (a)~A sparse selector. If host cycles use only the port \(v_3\), the red matching gives \(d-1\) two-cycles. (b)~The gadget with one private vertex. Its two-cycle (red) needs one free port, so host cycles can use at most \(d-1\) ports.}
\label{fig:capacity}
\end{figure}

We also use a gadget that allows host cycles to occupy at most
\(d-1\) ports when it makes its full contribution. It has one
private vertex with two parallel edges to each port in \(X\).
Its contribution is one two-cycle, which requires at least one
port to be free. For \(d=1\), its host capacity is zero.
\Cref{fig:capacity} shows both gadgets.

The following lemma lets us count the private two-cycles separately
from the cycles that encode a permutation.

\begin{lemma}\label{lem:cover-normalization}
Suppose the port sets of the capacity gadgets are pairwise disjoint
and private vertices have no neighbors outside their gadget's
private vertices and ports. There is a maximum cycle packing in
which every private vertex lies on a two-cycle and the cycles
avoiding private vertices respect every gadget's capacity.
\end{lemma}

\begin{proof}
Start with a maximum packing and process the gadgets one at a time.
Each step keeps the size of the packing and leaves the private
two-cycles already installed intact, because the port sets are
pairwise disjoint.

\paragraph{Sparse selector.}
Keep the cycles avoiding the current gadget's private vertices.
Let \(\nu\) be the number of removed cycles and \(A\) the set of ports
unused by the kept cycles. Each removed cycle uses a private vertex
and a port, so \(\nu\le\min\{d-1,|A|\}\). The matching property supplies
at least \(\nu\) two-cycles. Write \(q\) for the number of these
two-cycles installed and \(a\) for the number of ports used by the
kept cycles. Then \(q\le d-a\).
If \(a\le1\), the matching property gives \(d-1\) two-cycles on
the available ports. If \(a\ge2\), delete kept cycles until at most
one port is used. This costs at most \(a-1\) cycles: retain one
if each uses one port, and otherwise delete all cycles meeting the
port set. Replace the \(q\) two-cycles by \(d-1\) two-cycles. The gain
\(d-1-q\ge a-1\) pays for the deletions.

\paragraph{Gadget with one private vertex.}
If the private vertex already lies on a two-cycle, keep that cycle.
If it lies on a longer cycle, delete the longer cycle. This frees the
private vertex and every port on that cycle. Add a two-cycle through
the private vertex and one of the ports just freed. We delete one
cycle and add one cycle.

If the private vertex is unused and a port is free, add a two-cycle
through them. If the private vertex is unused and all ports are
occupied, choose a port and delete the cycle containing it. Add a
two-cycle through the newly freed port and the private vertex. In
this case, too, we delete one cycle and add one cycle.

In every case, the packing has at least as many cycles as before.
The two-cycle occupies one port, so the other cycles use at most
\(d-1\) ports.

The steps cannot increase the size of a maximum packing, so the
final packing is maximum and satisfies all capacities.
\end{proof}

\subsubsection{Host graph}
We now build the host graph \(H\), in which the cycles encode a
permutation by pairing rows with columns. For each row and each column
of the grid, we build a path, called a rail. Each input set \(S_s\)
gives one stage along the rails. At that stage, we add connections
between the rails of row \(i\) and column \(j\) for each cell
\((i,j)\in S_s\). We then attach the capacity gadgets to designated
vertices of \(H\).

To index the connections at stage \(s\), let
\(E_s=\{(i,j,s):(i,j)\in S_s\}\). Distinct names designate distinct
vertices.

For each row \(i\), introduce endpoints \(P_i,Q_i\), and for each
column \(j\), introduce endpoints \(A_j,B_j\). The selector vertices
allow a packing to pair a row with a column at either end. For each
\(i,j\), add the two-edge paths
\[
 P_i-p_{ij}-A_j
 \qquad\text{and}\qquad
 Q_i-q_{ij}-B_j.
\]

Between its endpoints, each row has a rail that visits the stages in
order. The columns have analogous rails. Their vertex sequences are
\[
 \begin{aligned}
 &P_i,u^L_{i,1},v^L_{i,1},u^L_{i,2},\ldots,v^L_{i,M},Q_i,\\
 &A_j,u^R_{j,1},v^R_{j,1},u^R_{j,2},\ldots,v^R_{j,M},B_j.
 \end{aligned}
\]
That is, each rail contains its \(u\)-vertex and then its \(v\)-vertex
at every stage, with edges between consecutive vertices.

For each occurrence \(e=(i,j,s)\), introduce vertices
\(a_e,b_e,t_e,d_e,w_e\). Add the two paths
\[
 u^L_{i,s}-a_e-t_e-u^R_{j,s}
 \qquad\text{and}\qquad
 v^L_{i,s}-b_e-d_e-v^R_{j,s},
\]
and add the triangle on \(a_e,b_e,w_e\). These paths allow the packing
to cross between row \(i\) and column \(j\) at stage \(s\).

\begin{figure}[t]
\centering
\begin{tikzpicture}[x=1.0cm,y=0.95cm,
    zv/.style={fv,minimum size=5.5pt,fill=figgray!55},
    pv/.style={fv,minimum size=4.2pt}]
  \coordinate (P) at (0,2);  \coordinate (Q) at (11,2);
  \coordinate (A) at (0,-2); \coordinate (B) at (11,-2);
  \coordinate (p) at (-0.9,0); \coordinate (q) at (11.9,0);
  \foreach \s/\x in {1/2,2/7} {
    \coordinate (xL\s) at (\x,2);  \coordinate (yL\s) at (\x+2,2);
    \coordinate (xR\s) at (\x,-2); \coordinate (yR\s) at (\x+2,-2);
  }
  \coordinate (a) at (2,0.75); \coordinate (c) at (2,-0.75);
  \coordinate (b) at (4,0.75); \coordinate (d) at (4,-0.75);
  \coordinate (z) at (3,0.0);
  \coordinate (a2) at (7,3.2); \coordinate (c2) at (7,4.3);
  \coordinate (b2) at (9,3.2); \coordinate (d2) at (9,4.3);
  \coordinate (z2) at (8,3.75);
  \draw (P)--(xL1)--(yL1)--(xL2)--(yL2)--(Q);
  \draw (A)--(xR1)--(yR1)--(xR2)--(yR2)--(B);
  \draw (P)--(p)--(A); \draw (Q)--(q)--(B);
  \draw (xL1)--(a)--(c)--(xR1); \draw (yL1)--(b)--(d)--(yR1);
  \draw (a)--(b)--(z)--(a);
  \draw (xL2)--(a2)--(c2); \draw (yL2)--(b2)--(d2);
  \draw (a2)--(b2)--(z2)--(a2);
  \draw[densely dashed] (c2) -- ++(0,0.6); \draw[densely dashed] (d2) -- ++(0,0.6);
  \node[flab] at (8,5.15) {to the rail of column \(1\)};
  \begin{scope}[on background layer]
    \draw[figA!45,line width=5pt,line join=round] (P)--(xL1)--(a)--(c)--(xR1)--(A)--(p)--cycle;
    \draw[figB!45,line width=5pt,line join=round] (yL1)--(b)--(d)--(yR1)--(xR2)--(yR2)--(B)--(q)--(Q)--(yL2)--(xL2)--cycle;
    \draw[figC!50,line width=5pt,line join=round] (a2)--(b2)--(z2)--cycle;
  \end{scope}
  \node[zv,label={[flab]above:\(P_1\)}] at (P) {};
  \node[zv,label={[flab]above:\(Q_1\)}] at (Q) {};
  \node[pv,label={[flab]below:\(A_2\)}] at (A) {};
  \node[pv,label={[flab]below:\(B_2\)}] at (B) {};
  \node[pv,label={[flab]left:\(p_{1,2}\)}] at (p) {};
  \node[pv,label={[flab]right:\(q_{1,2}\)}] at (q) {};
  \foreach \s in {1,2} {
    \node[pv,label={[flab]above left:\(u^L_{1,\s}\)}] at (xL\s) {};
    \node[pv,label={[flab]above right:\(v^L_{1,\s}\)}] at (yL\s) {};
    \node[pv,label={[flab]below:\(u^R_{2,\s}\)}] at (xR\s) {};
    \node[pv,label={[flab]below:\(v^R_{2,\s}\)}] at (yR\s) {};
  }
  \node[zv,label={[flab]left:\(a_e\)}] at (a) {};
  \node[zv,label={[flab]right:\(b_e\)}] at (b) {};
  \node[pv,label={[flab]left:\(t_e\)}] at (c) {};
  \node[pv,label={[flab]right:\(d_e\)}] at (d) {};
  \node[pv,label={[flab]below:\(w_e\)}] at (z) {};
  \node[zv,label={[flab]left:\(a_{e'}\)}] at (a2) {};
  \node[zv,label={[flab]right:\(b_{e'}\)}] at (b2) {};
  \node[pv,label={[flab]left:\(t_{e'}\)}] at (c2) {};
  \node[pv,label={[flab]right:\(d_{e'}\)}] at (d2) {};
  \node[pv,label={[flab]above:\(w_{e'}\)}] at (z2) {};
  \draw[fbrace] (4.3,-3.0) -- node[flab,below=4pt]{stage \(1\), \(e=(1,2,1)\)} (1.7,-3.0);
  \draw[fbrace] (9.3,-3.0) -- node[flab,below=4pt]{stage \(2\), \(e'=(1,1,2)\)} (6.7,-3.0);
  \node[flab] at (5.5,1.6) {rail of row \(1\)};
  \node[flab] at (5.5,-1.6) {rail of column \(2\)};
\end{tikzpicture}
\caption{Part of \(H\) before the capacity gadgets are attached, for the instance of \Cref{fig:ph-instance} and \(\pi(1)=2\). The drawing shows the rails of row~\(1\) and column~\(2\) with their selector ports, the occurrence \(e=(1,2,1)\), and the occurrence \(e'=(1,1,2)\), whose cross paths lead to the rail of column~\(1\). Large gray vertices belong to the set \(Z\) used in the proof of \Cref{lem:cover-permutation-sound}. Since row~\(1\) is selected at stage~\(1\) only, its two rails carry two cycles of the packing (red and blue). The triangle of the occurrence \(e'\), which is not selected, is packed as well (green).}
\label{fig:host}
\end{figure}
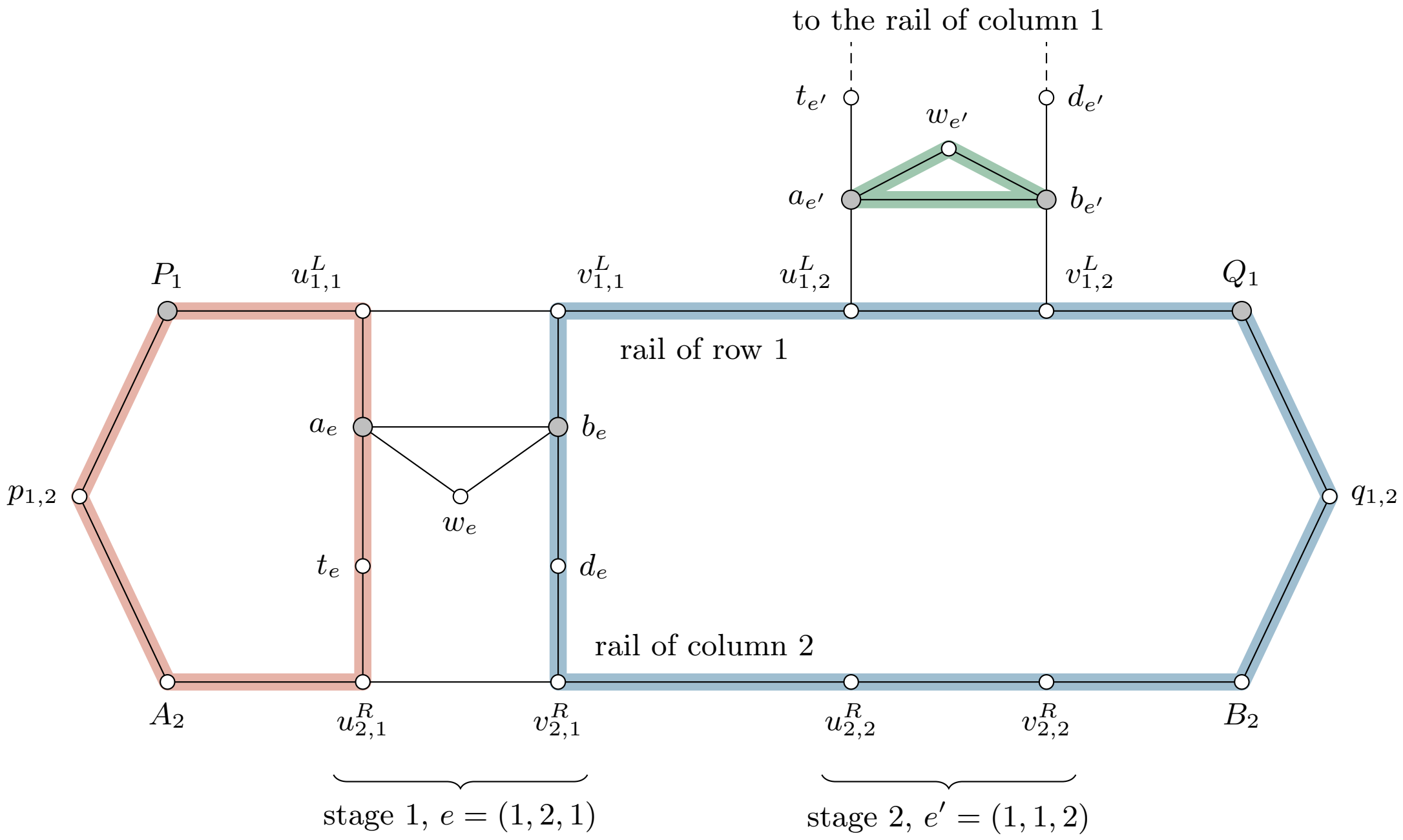

For each row \(i\), attach a sparse selector to each of
\(\{p_{ij}:j\in[D]\}\) and \(\{q_{ij}:j\in[D]\}\).
At each stage \(s\), fix an order of the occurrences in \(E_s\).
Attach a sparse selector to each of \(\{t_e:e\in E_s\}\) and
\(\{d_e:e\in E_s\}\), using that order for both port sets.
Attach a gadget with one private vertex to \(\{w_e:e\in E_s\}\),
allowing the host cycles to use at most \(|E_s|-1\) of those ports.
These port sets are pairwise disjoint.
\Cref{fig:host} shows part of \(H\) for the instance of \Cref{fig:ph-instance}.

Let $m_{\mathrm{cap}}=2D(D-1)+\sum_s(2|S_s|-1)$ be the total number of private vertices in the capacity gadgets. The threshold is $r=m_{\mathrm{cap}}+D+\sum_s|S_s|$.

We first prove that the multigraph has a packing of size at least
\(r\) exactly when a permutation hits every input set.

\begin{lemma}[Completeness]\label{lem:cover-permutation-complete}
If a permutation hits every set, the constructed multigraph has a
cycle packing of size at least \(r\).
\end{lemma}

\begin{proof}
Fix a permutation \(\pi\) hitting every \(S_s\). At each stage \(s\)
choose one occurrence \(e_s=(i_s,\pi(i_s),s)\).
Pack the triangles on \(a_e,b_e,w_e\) for the remaining occurrences.
In \Cref{fig:host}, the occurrence \(e'\) is not chosen, so its triangle is packed.
They contribute \(\sum_s(|S_s|-1)\) cycles.

For a row \(i\) paired with column \(\pi(i)\), use the selector paths through $p_{i,\pi(i)}$ and $q_{i,\pi(i)}$, the two rails, and the cross paths of its selected occurrences $e_s$. If this row is selected at stages \(s_1<\cdots<s_\ell\), the start
selector and first \(u\) cross-path close one cycle. Each \(v\)
cross-path and the next \(u\) cross-path close an intermediate
cycle. The last \(v\) cross-path and end selector close the final
cycle. If \(\ell=0\), the two selectors and the entire rails form one
cycle. Each row thus contributes \(\ell+1\) cycles, for a total of
\(D+M\). In \Cref{fig:host}, row~\(1\) is selected at stage~\(1\) only. The start
selector and the \(u\) cross-path of the chosen occurrence \(e\) close one
cycle, and the \(v\) cross-path of \(e\) and the end selector close the other. These cycles are vertex-disjoint because \(\pi\) is a permutation
and the \(u\) and \(v\) endpoints of a stage are distinct.

Together these host cycles contribute \(D+\sum_s |S_s|\) cycles. They use one \(p\)-port and one \(q\)-port per row, one \(t\)-port
and one \(d\)-port per stage, and all but one \(w\)-port per stage. By the matching property extend them by all \(m_{\mathrm{cap}}\) gadget two-cycles, reaching $r$.
\end{proof}

For soundness, we read a permutation from the way the cycles pair row
rails with column rails at the start of \(H\). The main step is to show
that the same pairing persists through every stage and selects a cell
from the set tested at that stage. To see why the pairing cannot change,
we track how many vertices from a designated set each cycle uses.

Let \(Z\) consist of all \(P_i,Q_i,a_e,b_e\). To follow the pairing from
one stage to the next, cut the rails before each stage, with the start
selectors and all earlier stages on the processed side. The cut after the last
stage has the end selectors on the unprocessed side. At every such
cut, the only crossing edges are the \(2D\) rail edges.

\begin{lemma}[Stage invariant]\label{lem:cp-stage}
Let \(C\) be a cycle packing in \(H\) that respects the capacities,
covers every vertex of \(Z\), and uses exactly two vertices of \(Z\)
on each cycle. Suppose that at the first cut all rail edges are used
by \(C\), and that the parts of \(C\) on the processed side meeting
these edges are \(D\) paths. Suppose path \(i\) joins row \(i\) to
column \(\pi(i)\) and contains exactly one vertex of \(Z\), where \(\pi\)
is a permutation. This property holds at every subsequent cut, with
the same \(\pi\), and \(\pi\) hits every set.
\end{lemma}

\begin{proof}
The first cut satisfies the invariant by hypothesis. Suppose it
holds at the cut before stage \(s\). Each stage-entry vertex
\(u^L_{i,s}\) and \(u^R_{j,s}\) is already incident with a used rail
edge from the processed side. Its cycle therefore already contains
a vertex of \(Z\) from that side.

\paragraph{Finding a hit.}
Capacity on the \(w\)-ports leaves \(w_e\) uncovered for some
\(e=(i,j,s)\). The covered vertex \(a_e\) has only three neighbors
other than \(w_e\), namely \(u^L_{i,s},t_e,b_e\). If its cycle uses
the edge to \(b_e\), its other edge at \(a_e\) leads either directly
to \(u^L_{i,s}\) or through \(t_e\) to \(u^R_{j,s}\). At either entry
vertex the incoming rail edge is already used. This cycle would
therefore contain \(a_e,b_e\) and an earlier vertex of \(Z\), which
is impossible. This also excludes the local four-cycle through
\(a_e,u^L_{i,s},v^L_{i,s},b_e\).

Thus the cycle through \(a_e\) uses the cross-path through \(t_e\)
and joins the row and column entries. If these entries belonged to
different paths from the processed side, the cycle would contain
\(a_e\) and the two distinct vertices of \(Z\) on those paths. Hence
the two entries belong to the same path. This gives \(j=\pi(i)\),
so \(\pi\) hits \(S_s\). The cross-path closes that path into a cycle.

\paragraph{Advancing the cut.}
The selector allows only one \(t\)-port at this stage, and the cycle
through \(a_e\) uses it. For each other occurrence \(e'\), the vertex
\(a_{e'}\) is covered and its \(t\)-port is unavailable. If its cycle
used the edge to its row entry, it would already contain the earlier
vertex of \(Z\) from the incoming rail. Its other edge at \(a_{e'}\)
would then lead to \(b_{e'}\) directly or through \(w_{e'}\), which
has just its two triangle neighbors in \(H\). That would give three
vertices of \(Z\) on this cycle. Thus the cycle must be the local
triangle on \(a_{e'},b_{e'},w_{e'}\).

The covered vertex \(b_e\) cannot use \(a_e\) or \(w_e\), so its cycle
uses the cross-path through \(d_e\). At its two exit vertices
\(v^L_{i,s}\) and \(v^R_{j,s}\), the rail edges back to the entries
are unavailable because those entries already use their incoming
rail edges and the cross-path through \(t_e\). All other occurrences
at this stage use their local triangles, so the exit vertices must
use the two forward rail edges. The resulting path on the processed
side pairs the same row and column, with \(b_e\) as its only vertex
of \(Z\). Every other pair continues along its two rails because all
its attached occurrence vertices are on local triangles. This uses
all the rail edges at the next cut and proves the invariant there.
The induction applies through the last stage, so \(\pi\) hits all sets.
\end{proof}

It remains to show that a packing of size at least \(r\) gives a
packing in \(H\) to which \Cref{lem:cp-stage} applies. We use the
threshold to force exactly two vertices of \(Z\) on each host cycle.
We then read the initial pairing from the cycles through the start
selectors.

\begin{lemma}[Soundness]\label{lem:cover-permutation-sound}
If the constructed multigraph has a cycle packing of size at least
\(r\), some permutation hits every set.
\end{lemma}

\begin{proof}
Apply \Cref{lem:cover-normalization} to a maximum packing and remove
its \(m_{\mathrm{cap}}\) gadget two-cycles. The remaining packing \(C\)
in the host respects the capacities and has at least
\(D+\sum_s |S_s|\) cycles.

The graph \(H-Z\) is a forest. Its components are the interiors of
the row rails, the column rails with selector ports and \(t\)- and
\(d\)-leaves attached, and the isolated \(w\)-vertices. Each vertex
of \(Z\) has at most one neighbor in any component of \(H-Z\).
Consequently a host cycle contains at least two vertices of \(Z\).
Since \(|Z|=2D+2\sum_s |S_s|\) and \(C\) has at least \(|Z|/2\)
cycles, every cycle contains exactly two vertices of \(Z\) and
every vertex of \(Z\) is covered.

The cycle through \(P_i\) uses the first row-rail edge and exactly
one selector port \(p_{i,\pi(i)}\), since capacity forbids using two
such ports. When the cycle reaches \(A_{\pi(i)}\), it must take the
first column-rail edge. If it instead used another selector port,
it would reach another row endpoint \(P_{i'}\) and have to leave
along that row's rail. The two row interiors are in distinct
components of \(H-Z\). The rest of the cycle could join them only
through a third vertex of \(Z\), which is impossible. In particular,
no cycle contains two start endpoints \(P_i\): each such endpoint
uses one row-rail edge, and the row interiors for distinct endpoints
cannot be joined without a third vertex of \(Z\). Thus the cycles
through the \(P_i\) are distinct and use distinct column endpoints
\(A_{\pi(i)}\). This makes \(\pi\) a permutation. The paths through the
\(P_i\) on the processed side of the first cut use all the crossing
rail edges and each contains exactly one vertex of \(Z\).
\Cref{lem:cp-stage} now shows that \(\pi\) hits every set.
\end{proof}

\subsubsection{Explicit path decomposition and simple graphs}
The correctness argument above applies to the multigraph. To obtain
the final instance, we first construct a path decomposition of the
multigraph with width \(O(D)\). We follow the stages in order and
process the cells of each set one at a time, so each bag needs vertices
from only a constant number of cells. We then subdivide each edge
once to obtain a simple graph and extend the decomposition to include
the subdivision vertices.

Let \(B_{\mathrm{glob}}\) consist of the \(4D\) endpoints
\(P_i,Q_i,A_j,B_j\), and let \(R_s\) consist of the \(4D\) rail
vertices at stage \(s\). Keep \(B_{\mathrm{glob}}\) in every bag.
Use the remaining vertices in the following order.
\begin{enumerate}
\item For the start selectors, use one bag per row containing
\(B_{\mathrm{glob}}\), the \(D\) ports of that row's start selector,
and its \(D-1\) private vertices. This covers the edges to the
endpoints and all doubled edges of the start selectors.
\item Start stage \(1\) with \(B_{\mathrm{glob}}\cup R_1\), which
covers all starting rail edges. Between stages \(s\) and \(s+1\)
use \(B_{\mathrm{glob}}\cup R_s\cup R_{s+1}\) to cover the
interstage rail edges, then use \(B_{\mathrm{glob}}\cup R_{s+1}\).
\item Within stage \(s\), process its occurrences in the order
used by the sparse selectors. For each occurrence \(e\), use
a bag containing \(B_{\mathrm{glob}}\cup R_s\), the private
vertex joined to the stage's \(w\)-ports, and the five vertices
\(a_e,b_e,t_e,d_e,w_e\). If \(e\) has a predecessor in this order,
include its \(t\)- and \(d\)-ports and the two private selector
vertices joining the predecessor's ports to the current ports.
This covers the triangle, the cross-paths, the doubled edges
at \(w_e\), and the selector edges to the preceding occurrence.
These are consecutive bags within the stage. Each \(t\)- or
\(d\)-port occurs in its own bag and, if there is a successor, in
the next bag. Every other occurrence-specific vertex is needed
in just one bag, and the private vertex for the \(w\)-ports
occurs throughout the occurrence bags of the stage.
\item End stage \(M\) with \(B_{\mathrm{glob}}\cup R_M\), covering
the rail edges to the end endpoints. Then use one bag per row
for the end selectors, with the same contents as in the start
selector bags but using the end ports and private vertices.
\end{enumerate}
The stage construction and the two-stage transition bags give
consecutive intervals for the rail vertices. The selector bags
at the ends use at most \(6D-1\) vertices, the interstage bags
use \(12D\), and the occurrence bags use at most \(8D+10\):
five current vertices, two preceding ports, two private vertices
of the sparse selectors, and one private vertex for the \(w\)-ports
beyond \(B_{\mathrm{glob}}\cup R_s\). The maximum bag size is thus
at most \(\max\{12D,8D+10\}\), independent of \(|S_s|\) and \(M\).
The graph and the bag sequence have polynomial size in
\(D+M+\sum_s |S_s|\).

Let \(G_0\) be the graph obtained by subdividing every edge of the
loopless multigraph once, using a distinct new vertex for each edge
copy.

\begin{lemma}\label{lem:cp-simple-host}
The graph \(G_0\) is connected, simple, and bipartite. It has a cycle
packing of size at least \(r\) if and only if some permutation hits
every set. The construction supplies a path decomposition of width
at most \(\max\{12D,8D+10\}\).
\end{lemma}

\begin{proof}
The original vertices form one side of the bipartition and the
subdivision vertices form the other. Each subdivision vertex has two
distinct neighbors, and different edge copies use different
subdivision vertices, so \(G_0\) is simple. The start and end
selector paths join all row and column rails, and every other vertex
of the multigraph attaches to a rail or a selector. Subdivision
preserves this connectivity.

Expanding or contracting the subdivided edges preserves the size of
a vertex-disjoint cycle packing. In particular, two parallel edges
become a four-cycle. The equivalence now follows from
\Cref{lem:cover-permutation-complete,lem:cover-permutation-sound}.

For each subdivision vertex, insert a copy of a bag containing the
endpoints of its edge and add that vertex to the copy. The maximum
bag size increases by at most one, so the resulting decomposition
has width at most \(\max\{12D,8D+10\}\).
\end{proof}

\subsection{Putting the reduction together}

We apply the multigraph construction and subdivision to each output
of the reduction from \problem{Clique} to \problem{Permutation Hitting}.
The grid-order bound then gives the required bound on the width of the
supplied path decomposition.

\begin{proof}[Proof of \Cref{cycle-packing-degree-six}]
By \Cref{lem:cp-simple-host}, the \problem{Permutation Hitting}
instance is a yes-instance exactly when the connected simple bipartite
host \(G_0\) has a packing of size at least \(r\). We output \(G_0\),
threshold \(r\), and the constructed path decomposition.

\paragraph{Parameter bound.}
By \Cref{lem:cp-simple-host}, the supplied path decomposition has
width \(p=O(D)\). Since \(D=O(m/\log m)\) by
\Cref{lem:permutation-hitting}, we have \(\Phi(p)=O(m)\).

\paragraph{Construction time.}
Constructing the subdivided host takes polynomial time per
output of \Cref{lem:permutation-hitting}. The total construction takes
fixed-parameter time. Apply \Cref{lem:time-transfer} with \(\varphi=\Phi\).
\end{proof}

\subsection{\problem{Maximum Cycle Cover}}\label{sec:maximum-cycle-cover}

The \problem{Cycle Packing} construction also gives a lower bound
for \problem{Maximum Cycle Cover}. We apply the undirected conversion
of Cygan et al.~\cite[Lemma~C.22]{ccpre} to the host \(G_0\) from
\Cref{lem:cp-simple-host}. This increases the cycle threshold by
\(|V(G_0)|\) and the width of the supplied path decomposition by at
most three.

\problembox{\problem{Maximum Cycle Cover}}%
{A simple undirected graph \(G\), a path decomposition of \(G\),
and a nonnegative integer \(r\).}%
{Can the vertices of \(G\) be partitioned into at least \(r\) simple cycles?}%
{The width \(p\) of the supplied path decomposition.}

\hypertarget{maximum-cycle-cover-supplied-pathwidth-w1}{}
\begin{theorem}\label{maximum-cycle-cover-supplied-pathwidth-w1}
Assume \(\mathrm{FPT}\ne\mathrm{W}[1]\).
There is no deterministic \(2^{o_{\mathrm{eff}}(p\log p)}N^{O(1)}\)-time algorithm
for \problem{Maximum Cycle Cover} parameterized by the width \(p\)
of a supplied path decomposition. This holds on connected graphs.
\end{theorem}

The conversion attaches a private triangle to each host vertex.
In the cover, those private vertices form either a triangle of
their own or a four-cycle with the host vertex. Either choice adds
one cycle to the packing.

\begin{proof}[Proof of \Cref{maximum-cycle-cover-supplied-pathwidth-w1}]
Let \(D\) be the grid order of the \problem{Permutation Hitting}
instance from \Cref{lem:permutation-hitting}, and let \(m\) be the
number of variables in the CNF encoding of the source
\problem{Clique} instance. By \Cref{lem:cp-simple-host}, the
\problem{Permutation Hitting} instance is a yes-instance exactly when
\(G_0\) has a packing of size at least \(r_{\mathrm{pack}}\), the
threshold given by that lemma. The conversion gives an equivalent
instance of \problem{Maximum Cycle Cover} with threshold
\(r=r_{\mathrm{pack}}+|V(G_0)|\). The graph is simple and connected,
since \(G_0\) is simple and connected and each private triangle
attaches to its host vertex.

\paragraph{Parameter bound.}
The conversion constructs a path decomposition of width
\(p\le\max\{12D,8D+10\}+3\), using the decomposition from
\Cref{lem:cp-simple-host}. We supply the resulting decomposition.
Since \(D=O(m/\log m)\) by \Cref{lem:permutation-hitting},
we have \(\Phi(p)=O(m)\).

\paragraph{Construction time.}
The host construction and the conversion take polynomial time per
output of \Cref{lem:permutation-hitting}, so the total construction
takes fixed-parameter time. Apply \Cref{lem:time-transfer} with
\(\varphi=\Phi\).
\end{proof}

\section{\problem{Line Distortion} on bipartite graphs}\label{sec:line-distortion}

\problembox{\problem{Line Distortion}}{A connected simple graph \(G\) and a nonnegative integer \(d\).}{Does the shortest-path metric of \(G\) admit a noncontracting embedding into the line with distortion at most \(d\)?}{The distortion threshold \(d\).}

Let \(\operatorname{dist}_G\) denote the shortest-path metric of \(G\).
An embedding \(f\) into the real or integer line is \emph{noncontracting}
if \(|f(v)-f(w)|\ge \operatorname{dist}_G(v,w)\) for every pair of
vertices \(v,w\). A noncontracting embedding has distortion at most \(d\)
if also \(|f(v)-f(w)|\le d\operatorname{dist}_G(v,w)\) for every pair.
By the triangle inequality, it suffices to check this upper bound on
edges.

The following lemma follows from the normalization to pushing
embeddings of Fellows et al.~\cite[Observation~2.2]{ffllrs-distortion}.

\begin{lemma}\label{lem:ld-integer-coordinates}
A connected graph admits a noncontracting embedding into the real
line with distortion at most \(d\) if and only if it admits one into
the integer line with distortion at most~\(d\).
\end{lemma}

\begin{proof}
Given a noncontracting real-line embedding, keep its vertex order,
place the first vertex at zero, and set each consecutive gap to the
graph distance between its endpoints. The new coordinates are
integers. Each gap is at most its original value, so every upper
bound on distance is preserved. The distance between any two new
coordinates is the sum of the intervening graph distances, which is
at least the graph distance between the two vertices by the triangle
inequality. The converse follows because the integer line is
contained in the real line.
\end{proof}

\hypertarget{graph-distortion-line-w1}{}
\begin{theorem}\label{graph-distortion-line-w1}
Assume \(\mathrm{FPT}\ne\mathrm{W}[1]\).
There is no deterministic \(2^{o_{\mathrm{eff}}(d\log d)}N^{O(1)}\)-time algorithm
for \problem{Line Distortion}.
This holds on connected simple bipartite graphs for embeddings
into either the integer line or the real line.
\end{theorem}

We reduce from \problem{Clique} via \problem{Permutation Hitting}.
We use the consecutive-pair encoding of Lokshtanov, Marx, and
Saurabh~\cite[Theorem~4.1]{lms}. For a \problem{Permutation Hitting}
instance on a \(D\times D\) grid, we introduce a symbol for each row
and each column. We arrange these symbols together with auxiliary
symbols in a single sequence. The sequence encodes a permutation
\(\pi\) by placing the symbol for row \(i\) next to the symbol for
column \(\pi(i)\).

For each input set, we create \(D\) tests, one per row. These tests
must use distinct consecutive pairs in the sequence. A chain of
auxiliary symbols provides only \(D-1\) pairs that can serve any test,
so at least one test must use a row-column pair corresponding to a
cell in the set.

To realize these tests in a line embedding, we give each layer its
own copy of the symbols. As in the reduction of Lokshtanov, Marx,
and Saurabh~\cite[Theorem~4.2]{lms}, edges between corresponding
symbols enforce a common order across layers. Lokshtanov, Marx, and
Saurabh attach a clique at each end of their construction to control
how the layers are spaced on the line. In our construction, a complete
bipartite graph at each end fixes the spacing at distortion \(10D+5\)
and keeps the whole graph bipartite. Within each layer, symbol
vertices alternate with test vertices, so each
test sits between two symbols to which it is adjacent. The common
order therefore represents a single permutation that must hit every set.

\subsection{Orders and consecutive pairs}\label{sec:ld-orders}

We first express \problem{Permutation Hitting} as a problem of ordering
symbols. Consecutive row and column symbols encode the permutation,
and a separate layer of tests checks whether it hits each input set.

Let $D$ be the order of a \problem{Permutation Hitting} instance, with sets $S_1,\ldots,S_H\subseteq[D]\times[D]$. We construct layers of tests on the same universe of $s_{\mathrm{sym}}=5D+2$ symbols. Given an order of the symbols, a test is a subset of symbols that must contain both endpoints of a consecutive pair in the order. The assignments to pairs must be injective within each layer.

Use symbols \(r_i,R_i,\chi_i,C_i\) for \(i\in[D]\) and the
\emph{guarded chain} \(\gamma_L,\zeta_1,\ldots,\zeta_D,\gamma_R\), with
\(\gamma_L\) and \(\gamma_R\) as its two guards. Impose every pair $\{r_i,R_i\}$, $\{\chi_i,C_i\}$, and every consecutive pair of the chain as a separate layer containing a single test. Add a layer with tests $\{R_i\}\cup\{\chi_j:j\in[D]\}$ for all $i\in[D]$.

For each set $S_\lambda$ add a layer with the $D$ tests
\[
 \{\zeta_1,\ldots,\zeta_D\}\cup\{R_i\}\cup\{\chi_j:(i,j)\in S_\lambda\},
 \qquad i\in[D].
\]
The forced consecutive pairs place the guards at the two ends of the
chain. Neither guard occurs in these \(D\) tests, so a test that uses a
\(\zeta\) symbol must use one of the \(D-1\) pairs between consecutive
\(\zeta\) symbols. Pad each layer to exactly \(s_{\mathrm{sym}}-1\) tests by adding
copies of the full symbol universe.

\begin{lemma}\label{lem:ld-order-tests}
There is a permutation hitting every set of the
\problem{Permutation Hitting} instance if and only if the
constructed layers admit a common order with valid test assignments.
\end{lemma}

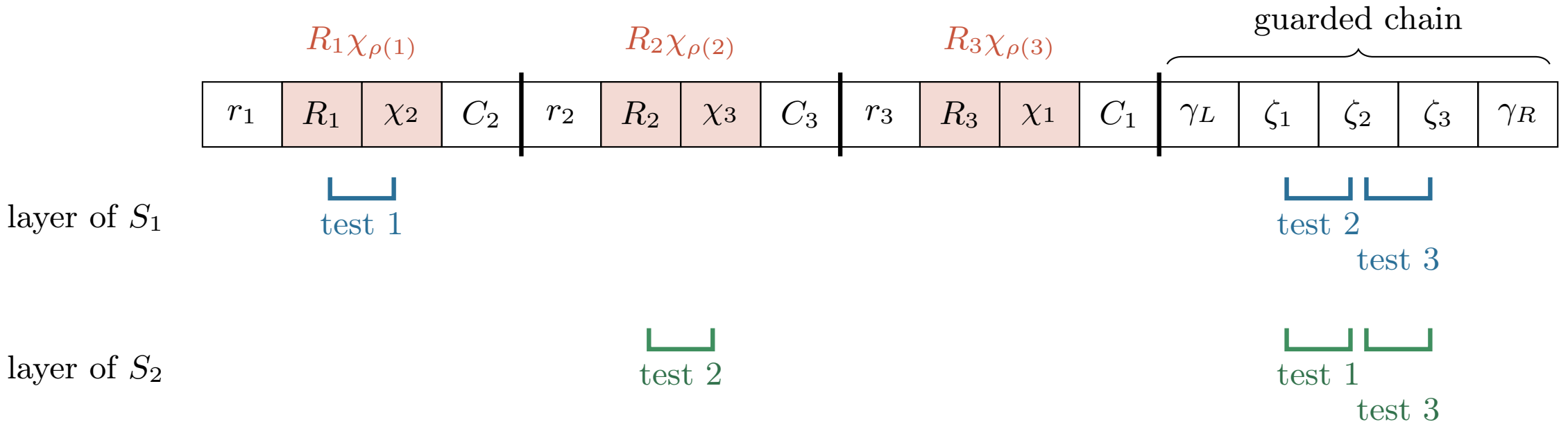
\begin{figure}[t]
\centering
\begin{tikzpicture}[x=0.76cm,y=0.8cm,
   sym/.style={draw,minimum width=0.76cm,minimum height=0.62cm,inner sep=0pt,flab}]
  \foreach \l [count=\c from 0] in {r_1,R_1,\chi_2,C_2,r_2,R_2,\chi_3,C_3,r_3,R_3,\chi_1,C_1,\gamma_L,\zeta_1,\zeta_2,\zeta_3,\gamma_R}
    \node[sym] (s\c) at (\c,0) {\(\l\)};
  \foreach \c in {1,5,9} \node[sym,fill=figA!22] at (\c,0) {\(R_{\the\numexpr(\c+3)/4\relax}\)};
  \foreach \c/\l in {2/\chi_2,6/\chi_3,10/\chi_1} \node[sym,fill=figA!22] at (\c,0) {\(\l\)};
  \foreach \c in {3.5,7.5,11.5} \draw[very thick] (\c,-0.5) -- (\c,0.5);
  \foreach \c/\i in {1.5/1,5.5/2,9.5/3}
    \node[flab,text=figA] at (\c,0.85) {\(R_{\i}\chi_{\rho(\i)}\)};
  \draw[fbrace] (11.6,0.6) -- node[flab,above=4pt]{guarded chain} (16.4,0.6);
  \node[flab,anchor=east] at (-0.8,-1.25) {layer of \(S_1\)};
  \foreach \a/\b/\l/\dy in {1/2/{test 1}/0,13/14/{test 2}/0,14/15/{test 3}/-0.42} {
    \draw[hiB] (\a+0.1,-0.75) -- (\a+0.1,-1.0) -- (\b-0.1,-1.0) -- (\b-0.1,-0.75);
    \node[flab,text=figB] at (\a/2+\b/2,-1.3+\dy) {\l};
  }
  \node[flab,anchor=east] at (-0.8,-3.05) {layer of \(S_2\)};
  \foreach \a/\b/\l/\dy in {5/6/{test 2}/0,13/14/{test 1}/0,14/15/{test 3}/-0.42} {
    \draw[hiC] (\a+0.1,-2.55) -- (\a+0.1,-2.8) -- (\b-0.1,-2.8) -- (\b-0.1,-2.55);
    \node[flab,text=figC!80!black] at (\a/2+\b/2,-3.1+\dy) {\l};
  }
\end{tikzpicture}
\caption{A satisfying order of the \(s_{\mathrm{sym}}=17\) symbols for the instance of \Cref{fig:ph-instance} and \(\rho=\pi\). The shaded pairs join \(R_i\) to \(\chi_{\rho(i)}\). Below the order, the brackets show the consecutive pairs assigned to the three tests of the layer of \(S_1\) and of the layer of \(S_2\) before padding. In each layer, one test uses a partner pair that corresponds to a hit, and the other two tests use the internal pairs of the chain.}
\label{fig:ld-order}
\end{figure}

\begin{proof}\leavevmode
\paragraph{Completeness.}
Let \(\rho\) be a permutation that hits every set. Concatenate
the blocks \(r_i,R_i,\chi_{\rho(i)},C_{\rho(i)}\) and append the
guarded chain. This gives the forced pairs and pairs each \(R_i\)
with \(\chi_{\rho(i)}\). In the layer for a set, assign the test
for one hit to its partner pair and assign the other \(D-1\)
tests to the \(D-1\) internal pairs of the guarded chain.
In each layer, assign the added tests to the unused consecutive
pairs. \Cref{fig:ld-order} shows such an order.

\paragraph{Soundness.}
Suppose the layers admit a common order with valid test assignments.
Discard the added tests and their assignments. The forced consecutive
pairs make the guarded chain contiguous in one of its two
orientations. The $D$ distinct assigned pairs each consume at least one $\chi_j$. Each $\chi_j$ can be used at most once, since the pair with $C_j$ already consumes one of its neighbors and cannot be assigned in this layer. Hence each assigned pair uses one $\chi_j$ and the other endpoint must be $R_i$ for test $i$. This defines a permutation $\rho$ by pairing $R_i$ with $\chi_{\rho(i)}$. In particular, no pair of symbols of the form $\chi_j$ is consecutive.

In the layer for \(S_\lambda\), each assigned pair must be internal to the $\zeta$ chain or be the pair joining $R_i$ to $\chi_{\rho(i)}$, provided $(i,\rho(i))\in S_\lambda$. The guards prevent a pair between a $\zeta$ symbol and a symbol outside the chain. There are only $D-1$ internal pairs, so the permutation hits~$S_\lambda$.
\end{proof}

\subsection{Encoding the order by line distortion}

We now turn each padded test layer into a graph layer so that an
embedding at the target distortion realizes the test assignments.

Let $M$ denote the number of padded layers, let $r=s_{\mathrm{sym}}+1$, and let $d=2r-1=2s_{\mathrm{sym}}+1$. We construct two parallel paths with vertices $v_0,\ldots,v_M$ and $w_0,\ldots,w_M$ and add edges $v_iw_i$ for each $i$. In layer $a\in[M]$ use one vertex $u_{a,s}$ per symbol $s$, adjacent to $v_{a-1}$ and $w_a$. Join corresponding symbol vertices between consecutive layers. A test in layer $a$ has its own vertex adjacent exactly to the layer's symbol vertices belonging to its subset. Call this connected graph the core.

\begin{figure}[t]
\centering
\begin{tikzpicture}[x=1cm,y=1cm,
   pv/.style={fv,minimum size=4.4pt},
   sv/.style={pv,fill=figB!55},
   tv/.style={fterm,minimum size=4pt,fill=figgray!55}]
  \node[flab,anchor=west] at (-3.3,2.3) {\textbf{(a)}};
  \foreach \i in {0,1,2} {
    \node[pv,fill=black,label={[flab]above:\(v_{\i}\)}] (v\i) at (4.2*\i,1.7) {};
    \node[pv,label={[flab]below:\(w_{\i}\)}] (w\i) at (4.2*\i,-1.7) {};
    \draw (v\i) -- (w\i);
  }
  \draw (v0) -- (v1) -- (v2); \draw (w0) -- (w1) -- (w2);
  \foreach \a in {1,2} {
    \foreach \s/\y in {1/0.8,2/0,3/-0.8}
      \coordinate (x\a\s) at (4.2*\a-2.6,\y);
    \foreach \t/\y in {1/0.4,2/-0.4}
      \coordinate (t\a\t) at (4.2*\a-1.5,\y);
  }
  \foreach \a/\vv/\ww in {1/v0/w1,2/v1/w2}
    \foreach \s in {1,2,3} { \draw[figgray] (\vv) -- (x\a\s); \draw[figgray] (x\a\s) -- (\ww); }
  \foreach \s in {1,2,3} \draw[hiB,line width=0.8pt] (x1\s) -- (x2\s);
  \foreach \a in {1,2} {
    \draw (t\a1) -- (x\a1); \draw (t\a1) -- (x\a2);
    \draw (t\a2) -- (x\a2); \draw (t\a2) -- (x\a3);
    \foreach \s in {1,2,3} \node[sv] at (x\a\s) {};
    \foreach \t in {1,2} \node[tv] at (t\a\t) {};
  }
  \node[flab] at (1.6,-2.45) {layer \(1\)};
  \node[flab] at (5.8,-2.45) {layer \(2\)};
  \node[fnode,minimum width=1.5cm,minimum height=1.5cm] (KL) at (-2.3,0) {\(K_{r,r+1}\)};
  \node[fnode,minimum width=1.5cm,minimum height=1.5cm] (KR) at (10.7,0) {\(K_{r,r+1}\)};
  \draw[hiA,line width=0.8pt] (w0) -- node[flab,below left=-1pt,text=black]{\(r\) edges} (KL.south east);
  \draw[hiA,line width=0.8pt] (v0) -- node[flab,above left=-1pt,text=black]{\(r-1\) edges} (KL.north east);
  \draw[hiA,line width=0.8pt] (v2) -- node[flab,above right=-1pt,text=black]{\(r\) edges} (KR.north west);
  \draw[hiA,line width=0.8pt] (w2) -- node[flab,below right=-1pt,text=black]{\(r-1\) edges} (KR.south west);
  \begin{scope}[shift={(-2.75,-5.3)},x=0.43cm,y=0.43cm]
    \node[flab,anchor=west] at (-1.3,3.6) {\textbf{(b)}};
    \draw[figgray] (-0.6,0) -- (33.6,0);
    \fill[hypothesisfill,draw=hypothesisink!45,rounded corners=2pt] (-0.5,-0.8) rectangle (8.5,0.8);
    \fill[hypothesisfill,draw=hypothesisink!45,rounded corners=2pt] (24.5,-0.8) rectangle (33.5,0.8);
    \foreach \x in {0,2,4,6,8,25,27,29,31,33} \node[pv,fill=figgray] at (\x,0) {};
    \foreach \x in {1,3,5,7,26,28,30,32} \node[pv,fill=figgray!25] at (\x,0) {};
    \foreach \i/\x in {0/10,1/17,2/24} \node[pv,fill=black] (ev\i) at (\x,0) {};
    \foreach \i/\x in {0/9,1/16,2/23} \node[pv] (ew\i) at (\x,0) {};
    \foreach \a/\o in {1/10,2/17} {
      \foreach \s in {1,2,3} \node[sv] (ex\a\s) at (\o+2*\s-1,0) {};
      \foreach \t in {1,2} \node[tv] at (\o+2*\t,0) {};
    }
    \draw (ev0) to[out=60,in=120] node[flab,above,pos=.5]{\(d\)} (ev1);
    \draw (ev1) to[out=60,in=120] node[flab,above,pos=.5]{\(d\)} (ev2);
    \draw[hiB,line width=0.8pt] (ex12) to[out=-60,in=-120] node[flab,below,pos=.5,text=black]{\(d\)} (ex22);
    \draw[hiA,line width=0.8pt] (ew0) to[out=120,in=60] node[flab,above,pos=.5,text=black]{\(d\)} (2,0);
    \draw[hiA,line width=0.8pt] (ew2) to[out=-60,in=-120] node[flab,below,pos=.5,text=black]{\(d\)} (30,0);
    \foreach \x/\l in {9/{w_0},10/{v_0},16/{w_1},17/{v_1},23/{w_2},24/{v_2}}
      \node[font=\scriptsize,inner sep=0pt] at (\x,-3.9) {\(\l\)};
    \foreach \x/\l in {9/{-1},10/{0},17/{d},24/{2d}}
      \node[font=\scriptsize,inner sep=1pt] at (\x,-5.0) {\(\l\)};
    \node[flab,anchor=east] at (8.2,-5.0) {coordinate};
    \node[flab,anchor=east] at (8.2,-3.9) {vertex};
  \end{scope}
\end{tikzpicture}
\caption{The reduction to \problem{Line Distortion}, drawn schematically for \(s_{\mathrm{sym}}=3\) symbols and \(M=2\) layers, so that \(r=4\) and \(d=7\). (a)~The core between the two copies of \(K_{r,r+1}\). Blue circles are symbol vertices, gray squares are test vertices, and the blue edges join corresponding symbols of consecutive layers. (b)~The embedding. Each copy of \(K_{r,r+1}\) occupies \(2r+1\) consecutive positions. In each layer, the symbols take the odd offsets and the tests the even offsets. The drawn edges are examples of edges that span exactly \(d\).}
\label{fig:ld-embed}
\end{figure}
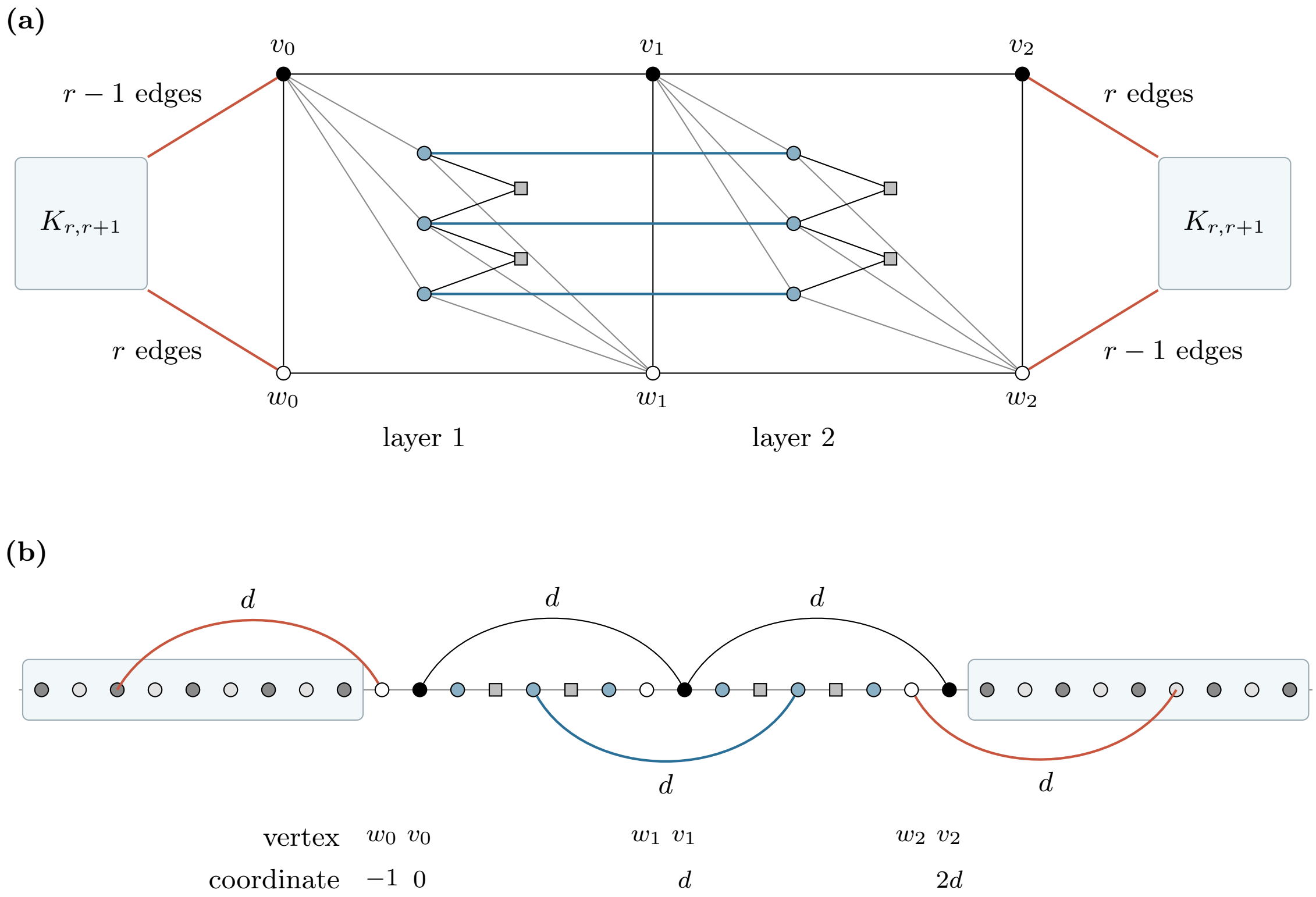

Attach two disjoint copies of $K_{r,r+1}$ at opposite ends of the core. For the first copy, join $w_0$ to $r$ vertices of the larger side and $v_0$ to $r-1$ vertices of the smaller side. For the second copy, do the same with $v_M$ and $w_M$, respectively. The graph is bipartite. Give $v_i$ color $i\bmod 2$ and $w_i$ the other color. The vertices for symbols of layer $a$ have the opposite color to $v_{a-1}$, and the vertices for tests in that layer have its color. Each added complete bipartite graph has its larger side opposite its neighbor $w_0$ or $v_M$. All test sets are nonempty, so the resulting simple graph is connected. \Cref{fig:ld-embed}(a) shows the graph.

We use a common order of the symbols to place vertices in each layer.

\begin{lemma}[Completeness]\label{lem:ld-embedding-complete}
If the layers admit a common order with valid test assignments,
then the constructed graph has a noncontracting embedding into
the integer line of distortion at most \(d\).
\end{lemma}

\begin{proof}
Suppose there is a satisfying order and assignment of tests to consecutive pairs. Set \(f(v_i)=id\) and \(f(w_i)=id-1\). In layer \(a\), place the \(s_{\mathrm{sym}}\) symbol vertices at the odd offsets from \(f(v_{a-1})\) in order and the tests at the even offsets of their assigned pairs. At each end, immediately outside the core, place the vertices of the attached complete bipartite graph so they alternate between sides, with larger-side vertices at both ends. Within each side choose the order so the vertices attached to the core are nearest the core. Consecutive positions are adjacent, so the embedding is noncontracting. Edges along either path and edges joining corresponding symbols span $d$. The other core edges have span at most $d$, as do all edges within either complete bipartite graph since its opposite-side vertices have opposite parity. The farthest of the $r$ larger-side neighbors of the first attached core vertex is at distance $1+2(r-1)=d$. The farthest of the $r-1$ smaller-side neighbors of the second attached core vertex is at distance $2+1+2(r-2)=d$. Thus every edge respects the threshold. See \Cref{fig:ld-embed}(b).

\end{proof}

For the reverse direction, we recover the order from the coordinates
forced by the two attachments.

\begin{lemma}[Soundness]\label{lem:ld-embedding-sound}
If the constructed graph has a noncontracting embedding into
the real line of distortion at most \(d\), then the layers
admit a common order with valid test assignments.
\end{lemma}

\begin{proof}
Suppose there is a noncontracting embedding \(f\) of distortion at most \(d\).
By \Cref{lem:ld-integer-coordinates}, we may assume that \(f\) has integer coordinates. Each attached complete bipartite graph spans at least $2r$ since it has $2r+1$ vertices. Its extremes must be on the same side, as otherwise their edge would exceed $d$. If they were both on the smaller side, the $r+1$ larger-side vertices would lie within distance $d$ of both extremes and would be separated by gaps of at least two. The span could then be at most $2d-2r=2r-2$, which is impossible. With extremes on the larger side, applying the same estimate to the $r$ smaller-side vertices bounds the span by $2d-2(r-1)=2r$. Thus every integer position in the interval is occupied and the smaller-side vertices fill the alternating interior positions.

No edge between exterior vertices can cross an attached interval because it would span at least $2r+2$. Since the core is connected it lies on one side of each such interval. Its vertex adjacent to $r$ larger-side vertices is at distance at most one from the nearest extreme, since its furthest larger-side neighbor is at least $2(r-1)$ farther away than that extreme. The second attached core vertex is at distance at most two, using the analogous lower bound $1+2(r-2)$ on its furthest smaller-side neighbor's offset from the extreme. The two core vertices therefore occupy the first two exterior coordinates. As $w_0$ and $v_M$ are distinct and are each a core extreme, they occupy opposite extremes. Reflect and translate to set \(f(w_0)=-1\) and \(f(v_0)=0\) at the left end. The core has
\[
 2(M+1)+M(2s_{\mathrm{sym}}-1)=Md+2
\]
vertices. Its span is at least $Md+1$. Its span is at most $Md+1$ by following the edge $w_0v_0$ and the $M$ edges of the $v$ path to $v_M$. The core therefore occupies every integer position between its extremes. Equality along the path from \(w_0\) to \(v_M\) forces \(f(v_i)=id\) for each \(i\). The second attachment vertex at the right end has \(f(w_M)=Md-1\). Equality along the path from \(w_0\) to \(w_M\) then forces \(f(w_i)=id-1\) for each \(i\).

Every symbol vertex in layer \(a\) has its coordinate in \([f(v_{a-1})-1,f(w_a)+1]\) by its two attachments. The first two and last two positions are occupied by \(w_{a-1},v_{a-1},w_a,v_a\). The $s_{\mathrm{sym}}$ independent symbol vertices must fit in the remaining $2s_{\mathrm{sym}}-1$ positions, with gaps of at least two. They therefore occupy the odd offsets. The edges between corresponding symbols force their offsets to be nonincreasing over the layers. Because the sum of all offsets is fixed, all layers have the same order. The vertex at each even offset is adjacent to both flanking symbol vertices, since it is at distance one from them in the embedding and the embedding is noncontracting. Their only common neighbors are test vertices of that layer and the two shared attachment vertices. The latter already occupy their fixed positions outside the test slots. Thus every even offset contains a test vertex of the layer, and every test is assigned to a valid consecutive pair. This gives an order with valid test assignments.
\end{proof}

The construction has $O((M+1)s_{\mathrm{sym}})$ vertices and $O((M+1)s_{\mathrm{sym}}^2)$ edges. It takes time polynomial in the total size of the listed sets and $D$, and its threshold is $d=10D+5$.

\subsection{Putting the reduction together}

We combine the ordering construction with its graph realization to
obtain the lower bound for embeddings into both the integer and the
real line. The distortion threshold is linear in the grid order.

\begin{proof}[Proof of \Cref{graph-distortion-line-w1}]
By \Cref{lem:ld-order-tests}, the layers admit a common order
with valid test assignments exactly when the corresponding
\problem{Permutation Hitting} instance is a yes-instance.
By \Cref{lem:ld-embedding-complete}, such an order gives an
integer-line embedding at the prescribed threshold. By
\Cref{lem:ld-embedding-sound}, any real-line embedding at that
threshold gives such an order. Since the integer line is
contained in the real line, the equivalence holds for both
target spaces. The constructed graph is connected, simple, and bipartite.

\paragraph{Parameter bound.}
The distortion threshold is \(d=10D+5\), where \(D\) is the grid
order. By \Cref{lem:permutation-hitting}, \(D=O(m/\log m)\) for
\(m=k\lceil\log_2 n\rceil\), where \(k\) is the requested clique
size and \(n\) is the source graph order. Thus \(\Phi(d)=O(m)\).

\paragraph{Construction time.}
The construction takes polynomial time per output of
\Cref{lem:permutation-hitting}, hence fixed-parameter time overall.
Apply \Cref{lem:time-transfer} with \(\varphi=\Phi\).
\end{proof}

\section{Conclusion}

Earlier work already provides several ways to recover historical ETH
lower bounds under \(\mathrm{FPT}\ne\mathrm{W}[1]\). Some recoveries
follow from hardness for a logarithmically rescaled parameter, while
others require an additional reduction or a new analysis of a published
construction. In our own reductions, the common step is to encode
clique choices in few bits and use that encoding to bound the target
parameter, even when the target instance is much larger. For
\problem{Vertex-Disjoint Paths}, \problem{Cycle Packing}, and
\problem{Maximum Cycle Cover}, the reductions also provide path
decompositions within the required parameter bounds.

The comparison with historical ETH results also identifies where the
transfer loses something. Several exclusions recover the same exponent
scale with computable cutoffs. For \problem{Chromatic Number} by vertex
cover number, the exponent differs from the historical benchmark by a
logarithmic factor. Graphs of small treewidth can have arbitrarily large
vertex cover number, so the treewidth lower bound leaves this gap open.
Two questions are
whether the computable-cutoff requirement can be dropped in the
scale-matching results and whether different encodings can recover the
larger exponent for this parameterization. Both questions ask how much
quantitative information about running time can be obtained from the
separation \(\mathrm{FPT}\ne\mathrm{W}[1]\).

Another direction is to ask which running-time lower bounds follow
from \(\mathrm{NP}\nsubseteq\mathrm{QP}\), where \(\mathrm{QP}\) is
deterministic quasipolynomial time. One example is \problem{Edge Clique
Cover}, parameterized by the number \(k\) of cliques used to cover all
edges. Cygan, Pilipczuk, and Pilipczuk exclude
\(2^{k^{O(1)}}N^{O(1)}\) time under this premise~\cite{ecc}. Their
reduction uses \(k=O(\log n)\) for a formula on \(n\) variables, so
such a running time would give a quasipolynomial algorithm for
\problem{3-SAT}. Which other parameterized problems admit reductions
that make the parameter small enough for this argument?

\subsection*{AI usage disclosure}

OpenAI's ChatGPT was used to develop all
proofs in this manuscript, including the adaptations and combinations
of previously published constructions. It also assisted with checking the literature, checking
arguments, comparing the results with the literature, and writing and
revising the text. Anthropic's Claude was used to create figures and to check proofs. The author checked and revised all proofs and takes responsibility for the manuscript.

\end{document}